\documentclass[letterpaper,twocolumn,10pt]{article}
\usepackage{usenix}
\usepackage{majorrev}

\usepackage{tikz}
\usepackage{amsmath}
\usepackage{graphicx}
\usepackage{xcolor,colortbl}
\usepackage{adjustbox}
\usepackage{multirow}
\usepackage{soul}
\usepackage{filecontents}

\begin{document}

\date{}

\title{``I just hated it and I want my money back'': Data-driven Understanding\\ of Mobile VPN Service Switching Preferences in The Wild}

\def\plainauthor{Author name(s) for PDF metadata. Don't forget to anonymize for submission!}

\author{
{\rm Rohit Raj}\\
IIT Kharagpur\\
rrohit2901@gmail.com
\and
{\rm Mridul Newar}\\
IIT Kharagpur\\
mridulnewar2000@gmail.com
\and
{\rm Mainack Mondal}\\
IIT Kharagpur\\
mainack@cse.iitkgp.ac.in
} %

\maketitle

\begin{abstract}
Virtual Private Networks (VPNs) are a crucial Privacy-Enhancing Technology (PET) leveraged by millions of users and catered by multiple VPN providers worldwide; thus, understanding the user preferences for the choice of VPN apps should be of importance and interest to the security community. To that end, prior studies looked into the usage, awareness and adoption of VPN users and the perceptions of providers. However, no study so far has looked into the user preferences and underlying reasons for switching among VPN providers and identified features that presumably enhance users' VPN experience. This work aims to bridge this gap and shed light on the underlying factors that drive existing users when they switch from one VPN to another. 
\newtext{In this work, we analyzed over 1.3 million reviews from 20 leading VPN apps, identifying 1,305 explicit mentions and intents to switch. Our NLP-based analysis unveiled distinct clusters of factors motivating users to switch. An examination of 376 blogs from six popular VPN recommendation sites revealed biases in the content, and we found ignorance towards user preferences. We conclude by identifying the key implications of our work for different stakeholders. The data and code for this work is available at \href{https://github.com/Mainack/switch-vpn-datacode-sec24}{https://github.com/Mainack/switch-vpn-datacode-sec24}.}

\oldtext{In this work, we collected more than 1.3 M reviews from 20 popular VPN apps on the Google Play Store and Apple App Store. Out of these reviews, using a semi-automated analysis, we identified 1,305 reviews where users explicitly mentioned switching between VPNs. Leveraging this dataset, we used Natural Language Processing methods to identify a total of 185,399 reviews where the user expressed intent to switch VPNs. Our systematic qualitative analysis (assisted by machine learning techniques) of these reviews reveals an underlying hierarchy of themes describing user-specified (often VPN-agnostic) factors for switching---ranging from technical aspects of VPN (e.g., connection latency) to nontechnical aspects (geographic location or cost). We further found that multiple factors considered while switching VPNs often co-occur in reviews and form distinct clusters, identifying that different users look for different \textit{sets} of requirements from VPN services. Finally, our analysis of 376 blogs from six popular VPN recommendation websites identifies that these websites are often biased towards specific VPN features and services---they do not consider the sets of user requirements, which likely necessitates VPN-switching for users. We conclude by identifying the key implications of our work for different stakeholders. }

\end{abstract}

\section{Introduction}\label{sec:intro}

Virtual Private Networks (VPNs) are a Privacy Enhancing Technology (PET) that has become increasingly popular over the last two decades. With the advent of commercial VPN services, VPNs are adopted by millions of users~\cite{vpn-usage-2020}. Today, VPN usage ranges from accessing geo-restricted content to accessing the web securely through public wifi~\cite{why-vpn-1, why-vpn-2}.

Prior work looked into the end-user's awareness and usage of VPNs~\cite{VPNPET, VPNapp}. They found that end-users have varied expectations from VPN services. People who are emotionally invested in protecting their privacy are generally resilient VPN users. Some users might expect that a VPN can help them access Netflix US while they are on vacation in China, and they may not care much about privacy, while others might want to use a VPN to securely connect to their offices, protecting them from potential privacy threats. Intuitively, this variation in expectation, combined with massive growth in the user base and ecosystem with multiple key players~\cite{vpn-growth}, provides users with the choice to use the VPN service that best caters to their needs. In fact, VPN users today have the obvious option of switching between VPN providers~\footnote{Since VPNs are often used as applications (or apps), in this work, we use VPN, VPN-provider and VPN-app interchangeably.}.

However, a user who is adopting a VPN and a user who is switching between VPN service providers are potentially different in terms of enhanced awareness resulting from previous VPN usage (and perhaps due to facing associated problems). Thus, recent work has reviewed and investigated the technical aspects of VPN infrastructure (e.g., for finding vulnerabilities) and the adoption of VPNs~\cite{VPNapp, VPNPET}. However, it is unclear whether users switch between VPN providers and what are their key motivators (e.g., specific VPN usage preferences). This issue is further complicated due to the unclear impact of VPN-review websites on this VPN-switching behaviour. According to recent work,  most of the users consider the VPN review blogs available online to choose from a large number of VPNs during VPN adoption~\cite {royaPaper}. These prior works also noted that these websites could be biased towards particular VPN service providers.
However, it is unclear how these websites might cater to a more experienced VPN user who wants to switch between VPNs. This question of the requirements of existing VPN users and whether the current VPN apps and VPN review blogs cater to those desired preferences has serious implications for VPN providers, VPN-review blogs, and the general research community. Unfortunately, there is not much work focusing on understanding end-users' behaviour and preferences regarding VPN switching. Thus, in this work, we fill this gap and investigate VPN-switching behaviour and preferences of end-users in the wild.

We focus on mobile VPN apps (due to the ubiquity of smartphones and additional metadata like reviews and \# downloads provided by the app stores). We systematically selected 20 popular VPN apps on the Android and iOS platforms. Then, we collected 1,333,160 user-generated reviews for these apps from the Google Play Store and Apple App Store. Next, we created a systematic keyword search, manual labelling and automated text-classification-driven approach\footnote{The dataset and code for this work can be found at \href{https://github.com/Mainack/switch-vpn-datacode-sec24}{https://github.com/Mainack/switch-vpn-datacode-sec24}.} to identify 185,399 reviews where users intended to switch between VPN providers (often with reason). Next, we took a deep dive into the why question---our in-depth qualitative analysis of these reviews uncovered a hierarchy of reasons for VPN-switching---we synthesized these reasons into the most desirable and undesirable properties of VPNs as indicated by users in the wild.
Furthermore, we demonstrate that these reasons for VPN switching have interesting differences from prior work on VPN adoption reasons. Finally, we collected 376 VPN-review blogs and checked in detail if they potentially covered user-desired features of VPNs during switching. Specifically, we answer five key research questions. 

\vspace{1mm}
\textit{\textbf{RQ1} Do users switch between VPNs? Is the switching uniform across different VPN providers?}

\noindent \oldtext{We found that 241,252 reviews (58.6\% of all meaningful reviews) contain user intent of VPN-switching. These reviews were posted by 219,160 unique usernames (average of 1.1 reviews), signifying a non-trivial number of users expressing intent to switch VPNs. Furthermore, we show that the percentage of VPN-switching is quite diverse across different VPN apps (Section~\ref{sec:doswitch}). }

\vspace{1mm}
\textit{\textbf{RQ2} What are the key reasons driving VPN-switching?}

\noindent \oldtext{Using our review dataset, we found a hierarchy of reasons motivating users to switch between VPNs. Perhaps due to the number of VPNs and the increase in VPN usage for non-privacy-related reasons, the problems that users face with their VPNs are not limited to technical, security, and privacy-related aspects. In fact, our findings show a mix of technical and non-technical reasons which motivate users to switch between VPNs (Section~\ref{sec:reasons}).}

\vspace{1mm}
\textit{\textbf{RQ3} Are the features desired during VPN-switching, by existing VPN users, different from features desired during VPN adoption?}

\oldtext{\noindent We found important differences between features desired by users during VPN adoption (reported in previous work) and the desired features during VPN switching (unearthed in this work). During VPN-switching, end-users often desired advanced security features like kill-switch and customized geography-specific solutions to evade censorship (Section~\ref{sec:prior}).}

\vspace{1mm}
\textit{\textbf{RQ4} What are the concrete features desired by users during VPN-switching?}

 \noindent \oldtext{Leveraging themes from the qualitative analysis, we identified the most popular desired features in our reviews ranging from Network speed to Security and UI/UX. However, we also performed a deeper analysis of the co-occurrence of these individual features using community detection of a feature co-occurrence graph. We found that there are ten concrete subsets of features that co-occurred in reviews---in these sets, most frequent VPN properties are accompanied by other (non-so frequent) properties, and they provide a roadmap of customised VPN solutions to users' needs (Section~\ref{sec:sem}).}

\vspace{1mm}
\textit{\textbf{RQ5} Does information by VPN review websites align with desired requirements of users during switching VPNs?}

\noindent \oldtext{We perform an NLP-technique-driven topical, entity and word similarity analysis of 376 distinct blog articles from six popular VPN-review websites. Our analysis reveals that current VPN blog articles are often biased towards specific VPN apps (praising them on all fronts) while not providing any information on a large number of user-desired features. This finding identified a key user need to have better VPN-related information, which will provide information aligned with their desired feature(Section~\ref{sec:blogs}).}

\vspace{1mm}
\noindent
\grumbler{REVIEW A.1.2}{During our endeavour to answer the above-listed questions, we found that VPN switching is a very common phenomenon among users, and there is a clear hierarchy of factors that primarily contribute to user switching between VPNs. We also found that users have different preferences while \textit{choosing} VPNs vs. \textit{switching} between VPNs. This can be clearly attributed to the fact that with the usage of VPN, users' expectations and requirements change, and hence, the factors they consider. Thus, we tried to enlist and group users based on factors they primarily consider while deciding to switch VPNs. Finally, to complete the study of user switching, we looked into primary sources which users refer to aid in their VPN switching decisions, i.e. VPN recommendation blogs and found that there are multiple ways in which VPN review websites fail to address the themes which are most relevant to users when switching between VPNs. In line with earlier literature, we also found that these websites show bias towards particular VPN providers and misguide the users in their search for the right VPN service (Section~\ref{sec:blogs}).}

\vspace{1mm}
\textbf{Limitations.} One of our study's key limitations is the potential of our approach to miss users who indeed switched VPNs but did not write a review. Furthermore, \grumbler{Review A.3}{we used keyword search to identify reviews relevant to user switching context for manual analysis, but during this, we will miss some reviews due to the limited number of VPN names used as keywords and not using terms like ``switching'' for the search, as that might lead to a large number of false positives making the manual process more tedious. However, this limitation won't affect the results significantly since we added ML models in the pipeline to further identify relevant reviews.}
However, some inaccuracy is associated with models used to identify potential VPN-switching reviews and themes mentioned in any review. We have tried to reduce this error through hyperparameter tuning and model selection. We also established the correctness and suitability of our VPN-switching review dataset with additional manual analysis. \grumbler{Review A.4}{We also acknowledge that ML models used for the identification of themes in reviews will not be able to identify new themes, but thematic saturation attained during thematic analysis suggests that there will be a very small number of such unidentified themes and potentially have a small impact on the validity of presented results.} Furthermore, we did not conduct a survey---rather, we preferred collecting and analysing an ecologically valid dataset posted by users of their own volition. However, this limitation of data also indicates that our reported results are a lower bound on the identified reasons. \grumbler{REVIEW B}{Another limitation is posed by fake reviews on the App Store or Play Store, but as indicated in \cite{fake-reviews}, the fraction of such reviews is too small to impact our results.} Yet another limitation is that we will miss the users who only used desktop VPN apps.
\oldtext{However, in general, VPN providers provide both desktop and mobile apps, signifying at least, in some cases, the user review considered their experiences for both desktop and mobile environments.} 

We uncovered interesting and potentially generalizable reasons for VPN-switching using our ecologically valid dataset despite these limitations. We also acknowledge the lack of generalizability of results associated with review blogging websites, as we have only considered 6 popular websites, but since these are top recommendation websites used by users, the results are still relevant. \grumbler{REVIEW B}{In fact, our blog post analysis results might be impacted by SEO techniques employed by VPN recommendation websites \cite{SEO}, but we still used these blogs as a data source since a regular user looking for VPN recommendation will also be subject to these SEO techniques and hence, will see these same websites.}

\vspace{1mm}
\textbf{Ethical considerations.} In this work, we collected data from the Google Play and Apple App Store and conducted a qualitative (thematic) analysis of these reviews. Since we analysed user-generated data from the wild, we tried our best to conduct our study ethically and protect users' privacy in our dataset. \grumbler{REVIEW D}{Our Institute does not have a formal IRB, so we carefully reviewed and followed best ethical practices while conducting this study}. In line with earlier work, we noted that no registration was necessary to view App Store or Google Play store reviews~\cite{eysenbach_ethical_2001}, signifying it's an open forum. Our analyzed reviews did not divulge any personally identifiable information. Also, as advised by earlier work to protect user privacy, we further hashed user names (thus anonymizing them) in our dataset to protect user privacy~\cite{cook-2018-ethics}.

Next, we discuss related work regarding adopting and using VPN apps. Then, in section~\ref{sec:datacollection}, we discuss our methodology involved in the collection and filtering of data, followed by details of thematic analysis (Section~\ref{sec:analysismethod}). Section~\ref{sec:reasons} describes the results of our thematic analysis,  and Section~\ref{sec:prior} compares desired features during VPN-switching and VPN adaptation. Section~\ref{sec:sem} describes the in-depth set of desirable features expressed by users during VPN-switching. Section \ref{sec:blogs} describes our analysis of VPN review websites. We discuss the implications in Section~\ref{sec:disussion} and conclude in Section~\ref{sec:conclusion}.

\section{Related Work}\label{sec:relwork}
We discuss related work on the user-centric view of VPNs in two broad directions. 

\vspace{1mm}
\noindent \textbf{Prior research of VPN awareness and usage.} Virtual Private Network (VPN) has been studied rigorously by prior research because of their vital role as Privacy Enhancing Technology (PET) and for other purposes like censorship circumvention, bypassing geo-blocking, etc. Most studies have focused on people's awareness about the existence/usage of VPN~\cite{MobUsers, ISP, FB}  and on security-related concerns associated with VPN like anonymization, DNS leaks, etc~\cite{VitalRole, Permission, Commercial, EmpCommercial, ClientCommercial}. A recent study by Dutkowska-Zuk et al.~\cite{feamster-2020-safe} conducted a study of 349 university students. They found that the students were mainly motivated to use VPNs to access restricted content rather than privacy. These studies about awareness and usage of VPNs were conducted through interviews~\cite{MobUsers} and surveys~\cite{FB} while later were studied either by developing a pipeline for testing VPN apps~\cite{Commercial} or by installing the app on a target device and then running tests on it~\cite{IOS} (an inherently invasive approach). Our study is complementary to this prior research as we focus on existing VPN users and identify their desired VPN characteristics that resulted in VPN-switching. Furthermore, contrary to the prior studies, we did not leverage user interviews or surveys~\cite{FB}. Rather, we leveraged app reviews of popular VPN apps from the Google Play and Apple App stores, where users organically present their opinions and attitudes without researcher intervention, enhancing the ecological validity of results.

\vspace{1mm}

\noindent \textbf{Prior research of VPN adoption.} On the user side, data leaks, geoblocking services and awareness of privacy risks gave way to the wide adoption of VPNs~\cite{mcdonald-2018-geoblocking}. Namara et al.~\cite{VPNPET} conducted a study about the adoption and usage of VPN apps and the barriers users encountered towards their adoption. Their survey of 90 participants found users' emotional reasons to adopt VPNs---like surveillance or desire for privacy. A closely related study by Sombatruang et al.~\cite{VPNapp} checked factors influencing users' decision to adopt a particular VPN application. They found user review ratings and pricing significantly affected the user's decision to choose a specific VPN app. Their results align with ours, with financial incentives and user reviews also playing a role in users' VPN-switching decisions. However, other factors like Geographic location and interaction of the VPN providers with users also played a significant role in our case. Overall, our work is complementary to these earlier studies. They investigated why users adopt VPNs. On the other hand, our study %
we investigated what factors motivate users of a current VPN app to switch to another VPN app. In fact, we demonstrate that these users seek more advanced and customized features than those uncovered in previous work. Our work also builds upon work by Ramesh et al.\cite{royaPaper}, which conducted end-user surveys and interviews with VPN service providers. They investigated users' dependence on VPN review websites for selecting VPNs and how these websites are not good sources of information. In that line, we unearth underlying reasons for VPN-switching and investigate VPN service recommendation websites to find the misalignment of user requirements and content provided by these websites. Next, we present our semi-automated approach of creating a review dataset mentioning VPN-switching in the wild.

\section{Collecting VPN App Reviews Corresponding to VPN Service Switching}\label{sec:datacollection}

Our study's first step was collecting app reviews of the most popular VPN apps. To that end, we turn our attention towards the Google Play Store and Apple App Store, which host all available VPN mobile apps for the Android and iOS platforms, respectively and also contain a user review section where users post their perceptions and attitudes towards these apps. 
This section describes the stages associated with collecting and filtering reviews in detail.

\subsection{Collecting reviews for popular VPN apps}

\vspace{1mm}

\noindent{\bf Selecting VPN apps.} We started with Google Play Store and  \grumbler{REVIEW AC}{we first searched for ``VPN'' in app stores. However, the search results repeatedly contained the same VPN providers (e.g., VPN app, VPN-enabled browser app, lite versions). On the contrary, searching for ``best VPN 2023'' gave the most popular versions of individual apps.} So, we searched for ``best VPN 2023'' and noted the top 50 VPN apps (in terms of downloads) which appeared in the search results. To ensure the diversity of the user base of VPN apps,  
We bucketed these apps by their number of downloads into 5 buckets---50K to 100K, 100K to 500K, 500K to 1M, 1M to 5M and 5M to 100M or more. \grumbler{REVIEW AC}{and in order to study the variation of VPN-switching with popularity, instead of checking only top VPNs, we did stratified sampling from popularity buckets} and ended up with 20 VPNs; they are---Hotspot Shield, Nord, Express, Vypr, Freedome, Tunnelbear, Windscribe, HideMyAss, Surfshark, Surfeasy, VPN Unlimited, Zenmate, Proton, Hola, Cyberghost, IP Vanish, Strong VPN, Private Internet Access (PIA), Super VPN, Turbo VPN. To further test the popularity of these VPN apps, we compiled a list of the most discussed VPNs from VPN market report statistics~\cite{Socia-Media-Article1, Socia-Media-Article2}---our set of 20 VPNs have a 75\% overlap with this list.

\vspace{1mm}

\noindent{\bf Collecting reviews for the VPN apps.} We used open-source tools to collect public reviews written in English (for simplifying analysis) from the Google Play store and Apple app store~\cite{google-play-scraper, app-store-scraper}. In total, we collected 1,333,160 unique reviews during our data collection. 
The selected VPN apps, the number of downloads on Google Play Store and the number of reviews (on both platforms) are in Table~\ref{tab:bestvpnapp} (Appendix~\ref{sec:vpn-details}).

\subsection{Identifying reviews mentioning VPN-switching}\label{sec:keyword-switching}

First, we pre-processed the review dataset (to retain only meaningful ones)  and then used a simple heuristic to identify reviews explicitly revealing VPN app-switching.

\begin{table}[h!]
\footnotesize
\centering
\caption{Number of VPN app reviews identified after our preprocessing step and application of heuristic.}\label{tab:reviewstat}
\vspace{1mm}
    \begin{tabular}{|l | c|}
        \hline
        \textbf{Number of VPNs used} & 20\\
        \hline
        \textbf{\#reviews} & 1,333,160\\
        \hline
        \textbf{\#reviews with \#words} $>$ 5 & 411,751\\
        \hline
        \textbf{\#reviews containing multiple VPN names} & 4,403\\
        \hline
    \end{tabular}
\end{table}

\noindent{\bf Pre-processing reviews.} Guzman et al.~\cite{length} shows that short reviews are often non-informative and contain praise or dispraise without reason. Thus, we removed all reviews with fewer than five words to retain only meaningful reviews. Finally, we ended up with 266,895 reviews from the Google Play Store and 144,856 reviews from the Apple App Store. 

\vspace{1mm}
\noindent \textbf{Detecting reviews possibly mentioning VPN-switch.} Note that most of the 411,751 reviews are likely written about a particular VPN app. To that end, to uncover potential reviews, we used a simple idea: \emph{the reviews that contain names of multiple VPNs might indicate that the user is comparing them}. Thus, they might have used or considered using both apps (signalling switching). We leverage this idea to filter the reviews via a simple keyword search---we extracted keywords from each VPN app name (Table~\ref{tab:keywords} of Appendix~\ref{sec:review-keywords} contains the full list of keywords) identified reviews with keywords from multiple VPN apps. We identified 4,403 such reviews. Table~\ref{tab:reviewstat} summarizes the number of VPN app reviews at each step of our process.

\begin{table}[]
\footnotesize
\centering
\caption{\#reviews stating Actual switch, potential switch and neither (irrelevant) after our manual analysis of reviews mentioning multiple VPN app names.}
\label{tab:final-reviews-property}
\vspace{1mm}
\begin{tabular}{|p{2cm}|l|ll|ll|}
\hline
\multirow{2}{*}{\bf Review type} & \multirow{2}{*}{\bf \# Reviews} & \multicolumn{2}{l|}{\bf \#words}     & \multicolumn{2}{c|}{\bf \#characters}  \\ \cline{3-6} 
                                &                     & \multicolumn{1}{l|}{Avg.} & s.d  & \multicolumn{1}{l|}{Avg.}  & s.d   \\ \hline
\textbf{Actual switch}             & 454                 & \multicolumn{1}{l|}{59.6} & 45.8 & \multicolumn{1}{l|}{318.7} & 206.1 \\ \hline
\textbf{Potential switch}          & 851                 & \multicolumn{1}{l|}{47.8} & 38.2 & \multicolumn{1}{l|}{257.9} & 209.1 \\ \hline
\textbf{Irrelevant}              & 295                 & \multicolumn{1}{l|}{25.3} & 39.1 & \multicolumn{1}{l|}{132.8} & 203.2 \\ \hline
\end{tabular}
\end{table}

\vspace{1mm}

\noindent \textbf{Manual analysis to create final review dataset mentioning VPN-switch.} Since mentioning multiple VPN app names might not be enough to actually mention VPN-switching, at this point, we resorted to manual analysis. We randomly sampled 1,600 reviews out of these 4,403, and one researcher manually reviewed each of these reviews to classify the reviews into three categories---Actual switch, Potential Switch and Irrelevant. For \textit{Actual switch} reviews, a user explicitly mentioned that they switched between VPNs. For example, P661 mentioned \emph{ ``I left [VPN1] for these guys! [VPN1] was slow.. and i mean slow... tried [VPN2]...and never looked back. love the speed and security!!''.} \grumbler{REVIEW E}{P2496 also mentioned \emph{`` Not able to connect to campus from dubai... bye to incompetent customer team..''}}. In \textit{potential switch} reviews, users showed intent of switching by comparing two VPNs but did not confirm switching VPN providers (however, we consider that if these users switched, they would switch from the VPN, which they are talking negatively about in the review). \grumbler{REVIEW E}{This comparison can be there without mention of a second VPN; for example, P1459 mentioned \emph{ ``This is so stupit vpn I'll connect from Pakistan... unable to Netflix... frnds able to watch!! useless!!''}}. Irrelevant reviews are the ones that neither compared multiple VPNs nor mentioned switching. Out of 1,600 reviews, 1,305 (81.3\%) indicated either an actual or potential switch. \grumbler{REVIEW B}{To check the correctness of annotation, we randomly sampled 100 reviews out of 1600 reviews and had them re-annotated by another independent annotator, and found Cohen Kappa between these two annotations to be 0.74, indicating a substantial agreement}. Table~\ref{tab:final-reviews-property} summarizes the number of reviews in each category as well as the average number of words and characters for each of these types of reviews.
Interestingly, both the actual and potential reviews have a considerably higher number of words and characters than the irrelevant reviews. This observation hints that users provide detailed and potentially better explanations in actual-switch and potential-switch reviews. Next, we perform a multi-stage qualitative analysis to identify the reasons for VPN-switching revealed in these 1,305 reviews. 

\section{Identifying Reasons of VPN-switching Using Qualitative Analysis}\label{sec:analysismethod}

\begin{table*}[!t]
\scriptsize
\centering
\caption{First three levels of our hierarchical themes explaining the reasons for switching VPNs.}
\vspace{1mm}
\label{tab:themes-three-levels}
\begin{tabular}{|l|l|l}
\hline
\cellcolor[HTML]{A8AAC2}A Network & \cellcolor[HTML]{D5D5D5}B.5 Deciet & \multicolumn{1}{l|}{F.1.1 Payment medium} \\ \hline
\cellcolor[HTML]{D5D5D5}A.1 Network speed & B.5.1 Removal of app reviews & \multicolumn{1}{l|}{F.1.2 Subscription options} \\ \hline
A.1.1 Network speed & B.5.2 Financial Fraud & \multicolumn{1}{l|}{\cellcolor[HTML]{D5D5D5}F.2 Price of VPN} \\ \hline
A.1.2 Ping value & B.5.3 False claims & \multicolumn{1}{l|}{\cellcolor[HTML]{D5D5D5}F.3 Refund} \\ \hline
\cellcolor[HTML]{D5D5D5}A.2 Server count & B.5.4 User data usage & \multicolumn{1}{l|}{\cellcolor[HTML]{D5D5D5}F.4 Premium version of app} \\ \hline
\rowcolor[HTML]{D5D5D5} 
A.3 Server switching & B.6 Reliability & \multicolumn{1}{l|}{\cellcolor[HTML]{D5D5D5}F.5 Free version of app} \\ \hline
A.3.1 Ease of switching server & \cellcolor[HTML]{D5D5D5}B.7 Reputation & \multicolumn{1}{l|}{\cellcolor[HTML]{A8AAC2}G Dynamic data services} \\ \hline
A.3.2 Manual switching required & B.7.1 Branding & \multicolumn{1}{l|}{\cellcolor[HTML]{D5D5D5}G.1 OTT} \\ \hline
A.3.3 Issues with server selection & B.7.2 Online reviews & \multicolumn{1}{l|}{\cellcolor[HTML]{D5D5D5}G.2 Games} \\ \hline
A.3.4 Issues with country selection & B.7.3 History & \multicolumn{1}{l|}{\cellcolor[HTML]{D5D5D5}G.3 Ads} \\ \hline
\rowcolor[HTML]{D5D5D5} 
A.4 Network Protocol & \cellcolor[HTML]{A8AAC2}C Connection & \multicolumn{1}{l|}{\cellcolor[HTML]{D5D5D5}G.4 Streaming} \\ \hline
A.4.1 Port forwarding & \cellcolor[HTML]{D5D5D5}C.1 Connection speed & \multicolumn{1}{l|}{\cellcolor[HTML]{A8AAC2}H Software} \\ \hline
A.4.2 Support for IPv6 & \cellcolor[HTML]{D5D5D5}C.2 Connection stability & \multicolumn{1}{l|}{\cellcolor[HTML]{D5D5D5}H.1 Updates} \\ \hline
A.4.3 Wireguard & \cellcolor[HTML]{D5D5D5}C.3 Split tunneling & \multicolumn{1}{l|}{\cellcolor[HTML]{D5D5D5}H.2 Features} \\ \hline
A.4.4 Issues with TCP/UDP & \cellcolor[HTML]{D5D5D5}C.4 Ease of reconnection & \multicolumn{1}{l|}{H.2.1 Number of features} \\ \hline
\cellcolor[HTML]{A8AAC2}B User Interaction & \cellcolor[HTML]{D5D5D5}C.5 Auto-connect feature & \multicolumn{1}{l|}{H.2.2 Presence/Absence of features} \\ \hline
\rowcolor[HTML]{D5D5D5} 
B.1 Assistance & \cellcolor[HTML]{A8AAC2}D Geography & \multicolumn{1}{l|}{\cellcolor[HTML]{D5D5D5}H.3 Devices} \\ \hline
B.1.1 Technical support & \cellcolor[HTML]{A8AAC2}E Security and privacy & \multicolumn{1}{l|}{\cellcolor[HTML]{D5D5D5}H.4 Bypassing capabilities} \\ \hline
B.1.2 Customer support & \cellcolor[HTML]{D5D5D5}E.1 Security & \multicolumn{1}{l|}{\cellcolor[HTML]{D5D5D5}H.5 Resource consumption} \\ \hline
\cellcolor[HTML]{D5D5D5}B.2 Services Provided & E.1.1 Encyption standard & \multicolumn{1}{l|}{H.5.1 Battery consumption} \\ \hline
B.2.1 Amount of data available & E.1.2 App security & \multicolumn{1}{l|}{H.5.2 System resources consumption} \\ \hline
B.2.2 Support for multiple devices & E.1.3 Server security & \multicolumn{1}{l|}{\cellcolor[HTML]{D5D5D5}H.6 Self sufficiency of software} \\ \hline
B.2.3 Cross OS support & E.1.4 DNS leaks & \multicolumn{1}{l|}{\cellcolor[HTML]{D5D5D5}H.7 Software stability} \\ \hline
B.2.4 Compatibility with environment & E.1.5 Headquater country of VPN &  \\ \cline{1-2}
B.2.5 Issues with third-party apps & E.1.6 User activity logs &  \\ \cline{1-2}
B.2.6 Support for Always-On-VPN & E.1.7 Kill switch &  \\ \cline{1-2}
\cellcolor[HTML]{D5D5D5}B.3 UI/UX & \cellcolor[HTML]{D5D5D5}E.2 Privacy &  \\ \cline{1-2}
B.3.1 Mobile app & E.2.1 RTC leaks &  \\ \cline{1-2}
B.3.2 General UI issues & E.2.2 Usage access Permission &  \\ \cline{1-2}
B.3.3 General UX issues & E.2.3 Privacy policy of VPN &  \\ \cline{1-2}
\cellcolor[HTML]{D5D5D5}B.4 User friendliness & E.2.4 SSH issues &  \\ \cline{1-2}
B.4.1 Ease of setup & E.2.5 Obfuscated servers present &  \\ \cline{1-2}
B.4.2 Ease of use & \cellcolor[HTML]{A8AAC2}F Financial Aspects &  \\ \cline{1-2}
B.4.3 Accessibility & \cellcolor[HTML]{D5D5D5}F.1 Payment &  \\ \cline{1-2}
\end{tabular}
\end{table*}

We leveraged our dataset of 1,305 VPN-switching reviews to investigate the reasons for VPN-switching via qualitative analysis. 
To that end, two researchers together extracted explanatory quotes from these 1,305 reviews---in total, there were 2,808 final quotes for our analysis, which at least one researcher deemed explanatory for VPN-switching. Next, we use open coding and affinity diagram~\cite{saldana-qualitative} to develop a hierarchy of themes explaining reasons for VPN-switching. 

\vspace{1mm}

\noindent \textbf{Open coding.} First, we applied open coding on the quotes. We randomly sampled 200 quotes, and using these quotes, two researchers collaboratively created a codebook. \grumbler{REVIEW C}{Apart from this, we set aside 50 quotes initially to check the saturation of themes.} Next, using this codebook, those two researchers independently coded all the reviews. Inter-rater agreement (Cohen's kappa) at the end of the open coding round was 0.81, indicating almost perfect agreement. At the end of the open coding round, the two researchers met to discuss the specific user quotes whose codes were mismatched and assign a final code to each quote (if necessary, creating new codes). \grumbler{REVIEW C}{After completing coding all quotes (apart from the random 50 quotes we set aside earlier), one researcher verified that the identified themes are sufficient to code those 50 quotes, indicating thematic saturation of our open coding}. Finally, the coders assigned a total of 217 codes to 1,305 user quotes. 

\noindent \textbf{Affinity diagramming to identify the hierarchy of themes.} After this open coding round, the two researchers collaboratively analyzed the uncovered codes using  \textit{affinity diagramming} where they together looked into the set of quotes for each code, not only the code itself~\cite{harboe-2015-affinity}. \grumbler{REVIEW C}{First, we set aside 10 random codes to check the saturation of hierarchy.} Then, the coders collaboratively created a set of higher-level themes from the rest of the open codes. They continued this process on the newly developed higher-level themes for two more rounds when the coders felt no new higher-level theme could emerge.
At the end of the process, we identified a 4-level hierarchy where Level-1 contains abstract highest-level themes created in the third (final) round of affinity diagramming, and Level-4 contains the very specific codes specified by the open coding step. \grumbler{REVIEW C}{Finally, we checked that including the 10 random codes did not add any new Level-1 and Level-2 themes, indicating thematic saturation of affinity diagramming.}
Our hierarchy of themes (first three levels) explaining the themes for reasons of VPN-switching is shown in Table~\ref{tab:themes-three-levels}. Level 1 contains the eight highest-level themes---Network, User interaction, Connection, Geography, Security and Privacy, Financial aspects, Dynamic data services and Software. However, we note that, so far, our analysis primarily considered only 1,305 semi-automatically filtered reviews (the complete hierarchy of themes is in Table~\ref{tab:frequencies} of Appendix~\ref{sec:frequencies-theme}). Now, we scale our VPN-switching review dataset with themes using automated classification.

\section{Scaling VPN-switching Review Dataset and Theme Annotation}\label{sec:scale}

\begin{figure}[]
    \centering
    \includegraphics[width=7cm]{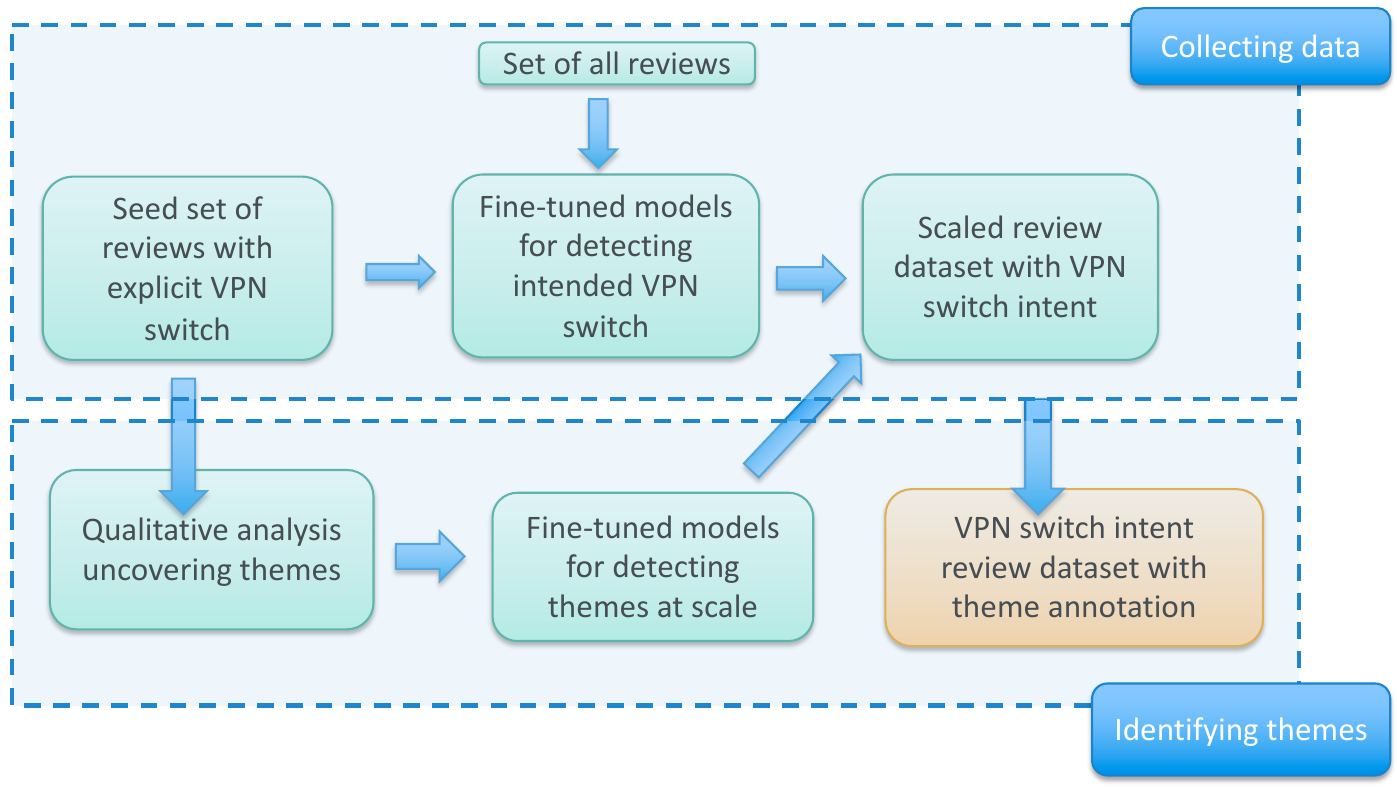}
    \caption{Pipeline for filtering review and identifying themes occurring in the reviews.}
    \label{fig:pipeline}
\end{figure}

We note that our VPN-switching review dataset comprises only 1,305 reviews (0.3$\%$ of a total of 411,751 reviews). To that end, we hypothesize that we might be missing a large-number of VPN-switching reviews which do not mention names of multiple VPNs. One such review in our corpus is \textit{``[VPN1] had been working great for me in the past but after last update it sucks...will look for other vpn if bug not fixed''}. Thus, to identify such reviews, we designed a machine learning pipeline presented in figure~\ref{fig:pipeline}. This section will describe the pipeline's working and the resultant dataset.

\subsection{Leveraging text-classification to extend VPN-switching review dataset}
Originally, we manually divided 1,600 reviews into three categories---Actual switching, Potential switching and Irrelevant (Section~\ref{sec:keyword-switching}). We note that we can consider these as three distinct classes assigned to the reviews---based on this idea, we model the problem of finding reviews mentioning VPN switching as a three-class classification problem. 
\oldtext{To that end, we tried two separate classifiers---a sentence transformer model~\cite{Reimers2019SentenceBERTS} and a pre-trained BERT \cite{bertbase} based language model fine-tuned on our initial dataset of 1,600 reviews. The sentence transformer model performed poorly (F1-score of 45\%), so we only report the results from BERT model.}
\public{To that end, we tried multiple models as classifiers, including recent models like LLAMA2 \cite{llama2}, and FLAN-T5 \cite{flant5}. The performance of individual models has been shown in Table \ref{tab:val-performance}. We used the fine-tuned DeBERTa model for further analysis since it offered best performance (F1-score of 0.85).}

\vspace{1mm}
\noindent\textbf{Model Architecture. } Given the small number of labelled data points, 
\oldtext{we used the pre-trained BERT model (weights taken from 'bert-base-uncased'~\cite{bertbase}) with a linear feed-forward network on the top as the classifier. We fine-tuned the model weights on the downstream task of review classification.}
\public{we primarily used following pre-trained models, LLAMA-2\cite{llama2}, FLAN-T5\cite{flant5}, BERT\cite{bertbase}, DeBERTa\cite{deberta} and GAN-BERT\cite{gan-bert}. We fine-tuned these model weights on the dataset for the downstream task of review classification. We used Optuna framework\cite{optuna} for systematic hyper-parameter tuning. Given the varying spectrum of models, the usage method also varied among the models (details in appendix \ref{sec:review-model-appendix}).}

\vspace{1mm}
\noindent\textbf{Model Performance. }We randomly split the dataset into 80\% training and 20\% validation subsets and post-training observed \newtext{$85\%$ accuracy and 84.73\% F1-score on the validation dataset using DeBERTa model} (presented in Table~\ref{tab:val-performance}). 

To further investigate the performance and misclassification of our model, 
we randomly selected an additional 100 reviews mentioning multiple VPN names (excluding the 1,600 annotated reviews), and one annotator manually annotated the selected reviews to create ground truth. Next, we used our text classification on these 100 reviews and constructed the confusion matrix (Table-\ref{tab:manual-model-1}). We note that, in this confusion matrix, in only 9 cases, a review mentioning an Actual or potential switch is misclassified as irrelevant. Thus, we identify actual or potential VPN-switching reviews in 91\% of cases. Many misclassified reviews with strong negative/positive sentiments were not directed towards VPN features. Finally, we used this validated text-classified model on our dataset of 411,751 reviews to identify a total of 185,399 reviews mentioning actual or potential VPN-switching as presented in Table \ref{tab:classModelShift}.

\begin{table}[]
\footnotesize
\centering
{\color{\mytablecolor}
\caption{Performance of our fine-tuned models on the validation dataset for classifying reviews into Actual/Potential switch and irrelevant. Since the problem of review classification was a multi-class classification problem so metrics are computed using the methodology described in \cite{metricsMultiClass}, where we compute model prediction using argmax among scores assigned by the model to all classes.}
\label{tab:val-performance}
\vspace{1mm}
\begin{tabular}{|c|c|c|}
\hline
\rowcolor[HTML]{C0C0C0} 
\textbf{Model}   & \textbf{Accuracy} & \textbf{\begin{tabular}[c]{@{}c@{}}Macro avg F1\end{tabular}} \\ \hline
FLAN T5          & 0.77              & 0.77                                                                    \\ \hline
\textbf{DeBERTa} & \textbf{0.85}     & \textbf{0.85}                                                           \\ \hline
LLAMA2           & 0.69              & 0.65                                                                    \\ \hline
BERT             & 0.81              & 0.79                                                                    \\ \hline
GANBERT          & 0.83              & 0.75                                                                    \\ \hline
\end{tabular}
}
\end{table}

\begin{table}[]
\footnotesize
\centering
\caption{Confusion matrix for our DeBERTa text-classifier annotation and ground truth for previously unseen 100 reviews. For only 9 reviews Actual/Potential switch is misclassified as irrelevant or vice-versa (marked in gray).}
\label{tab:manual-model-1}
\vspace{1mm}
\begin{tabular}{cl|ccc|}
\cline{3-5}
\multicolumn{1}{r}{}                                                                           & \multicolumn{1}{r|}{} & \multicolumn{3}{c|}{\bf Predicted}                                                                                                                                                                    \\ \cline{3-5} 
\multicolumn{1}{l}{}                                                                           &                       & \multicolumn{1}{l|}{\begin{tabular}[c]{@{}l@{}}{\bf Actual} \\ {\bf switch}\end{tabular}} & \multicolumn{1}{l|}{\begin{tabular}[c]{@{}l@{}}{\bf Potential}\\ {\bf switch}\end{tabular}} & \multicolumn{1}{l|}{{\bf Irrelevant}} \\ \hline
\multicolumn{1}{|c|}{\multirow{3}{*}{\begin{tabular}[c]{@{}c@{}}{\bf Ground} \\ {\bf Truth}\end{tabular}}} & {\bf Actual switch}         & \multicolumn{1}{c|}{33}                                                       & \multicolumn{1}{c|}{{3}}                                                 & \cellcolor{gray!30}{1}                               \\ \cline{2-5} 
\multicolumn{1}{|c|}{}                                                                         & {\bf Potential switch}      & \multicolumn{1}{c|}{13}                                                       & \multicolumn{1}{c|}{20}                                                         & \cellcolor{gray!30}{2}                               \\ \cline{2-5} 
\multicolumn{1}{|c|}{}                                                                         & {\bf Irrelevant}            & \multicolumn{1}{c|}{\cellcolor{gray!30}{2}}                                               & \multicolumn{1}{c|}{\cellcolor{gray!30}{4}}                                                 & {22}                     \\ \hline
\end{tabular}
\end{table}

\begin{table}[]
\footnotesize
\centering
{\color{\mytablecolor}
\caption{Final size of our extended dataset for VPN-switching reviews.}
\label{tab:classModelShift}
\vspace{1mm}
\begin{tabular}{|l|l|l|l|}
\hline
                          & \textbf{\begin{tabular}[c]{@{}l@{}}Apple App Store \\ reviews\end{tabular}} & \textbf{\begin{tabular}[c]{@{}l@{}}Google Play \\ Store reviews\end{tabular}} & \textbf{Total reviews} \\ \hline
\textbf{Actual switch}    & 38,139                                                                & 52,166                                                                        & 90,305                 \\ \hline
\textbf{Potential switch} & 35,614                                                                & 59,480                                                                       & 95,094                \\ \hline
\textbf{Irrelevant}       & 71,103                                                                & 119,685                                                                       & 190,788                \\ \hline
\end{tabular}
}
\end{table}

\subsection{Identifying themes mentioned in extended dataset}

Next, we wanted to assign themes to the reviews in our extended dataset of 185,399 reviews. To that end, we leveraged a similar idea as before---we modelled this problem as a \newtext{theme identification problem and multi-label classification problem}. Specifically, we model the Level-2 themes as labels assigned to each review 
\oldtext{(as Level-1 themes might be too broad)}. 
\grumbler{REVIEW C}{We chose to build our analysis based on L-2 themes---L-1 themes appeared too broad for granular analysis. Also, L-3 and L-4 themes might be more granular than L-2 themes, but we have a relatively low amount of labelled data per L-3/L-4 theme. Thus automated theme identification with high accuracy would have been very difficult for them.}

\vspace{1mm}
\noindent\textbf{Model Architecture. }\public{We again used the following pre-trained models,  LLAMA-2\cite{llama2}, BART\cite{bart}, FLAN-T5\cite{flant5}, BERT\cite{bertbase}, DeBERTa\cite{deberta} and GAN-BERT\cite{gan-bert}. Given the different architectures of these models, we fine-tuned BERT and GAN-BERT with multi-head classification layers on top of the pre-trained model. FLAN-T5 and BART were fine-tuned on the seq2seq task, and LLAMA2 was fine-tuned on the CausalLM task. We used Optuna framework\cite{optuna} for systematic hyper-parameter tuning (details in appendix \ref{sec:theme-identification-appendix}).}

\vspace{1mm}
\noindent\textbf{Model Performance. }
\oldtext{This model was a multi-label classifier which took review as input and produced a list of themes mentioned in that review.}
To measure the performance of the fine-tuned models, we split the labelled dataset into 80\% training and 20\% validation subsets. 
\public{We found that the BART model outperformed other models, identifying more than 80\% of themes correctly in the reviews}. \newtext{The accuracy of all models on the validation set is shown in Table \ref{tab:theme-model-performance} of Appendix~\ref{sec:theme-identification-appendix}. To further ensure the correctness of the model, we also performed a manual analysis---we selected 100 random reviews (excluding the original 1,305 annotated) and manually annotated them with Level-2 themes. Then, we compared the correctness of our predicted themes against this ground truth. The result of the manual analysis is detailed in Table \ref{tab:classModelTheme} of Appendix~\ref{sec:theme-identification-appendix}. 
Our classifier was correct for the majority of the reviews (more than 80\%) in both the validation set and additional ground truth dataset.} Next, we will leverage this extended dataset of theme-annotated reviews to explore our research questions.

\section{Do Users Switch Between VPN Apps? (RQ1)}\label{sec:doswitch}

In our extended dataset, we found that 185,399 reviews contain user intent of VPN-switching out of 411,751 reviews, i.e.,  for a significant fraction of 45\% of reviews. We further ask if unique users contribute these reviews and if the VPN-switching behaviour is uniform across different VPN apps.

\vspace{1mm}

\noindent{\bf How many users are switching VPN apps?} We consider usernames as proxies for user accounts. Our collected dataset shows that the \newtext{185,399} VPN-switching reviews were posted by \newtext{169,201} unique usernames. The only exception is the ``A google user'' tag in place of the name, which represents user accounts that have either been deleted or anonymized due to the user's privacy settings. Thus, these actual or potential VPN-switching reviews are posted mainly by individual user accounts (as opposed to a small subset of userbase).

\vspace{1mm}

\noindent{\bf Are VPN apps uniformly affected by VPN-switching?} We further checked if the phenomena of VPN-switching is uniform across all VPN apps. To investigate, we checked what fraction of reviews per VPN app mention VPN-switching. Interestingly, there is a wide variation. Super VPN has only \newtext{2.0\%} of the reviews mentioned VPN-switching, and for VPN unlimited \newtext{61.9\%} of the reviews mentioned VPN-switching---showing a 30 times increase. The percentage of VPN-switching reviews for each of our VPN apps is in Table~\ref{tab:switches} (Appendix~\ref{sec:vpn-details}). To that end, we computed the Pearson product-moment correlation coefficient between the number of downloads for each VPN and the percentage of reviews mentioning VPN-switch. We found there is a statistically significant negative correlation \newtext{($R = -0.5842$, $p = 0.006858$)} between these two variables. Thus, VPN apps with smaller user bases have a higher chance of their users switching to other apps. Next, we investigate why users switch VPNs.

\section{Reasons of VPN-Switching (RQ2)}\label{sec:reasons}

In this section, we delve into each of the eight level-1 themes (i.e., highest-level themes) to understand how the switching of users can be explained based on that theme. According to our study, the reason for the switch can be among a large number of causes ranging from Geography, i.e. the country in which the user resides or visits, to User interaction, i.e. how VPN service provider interacts with users through channels like customer service or through software/app. In this section, we will not use the names of VPNs in this section, as our goal is to understand the high-level reasons for users to switch VPNs, not to identify the specific shortcomings/superiority of any VPN. \newtext{In this section, we will discuss the themes which have been identified in the manual analysis.}

\subsection{Security and privacy}
VPN userbase comprises users who want to protect their privacy against hackers and government agencies. Thus, switching from one VPN to another seems evident for users who think their security is under threat. \newtext{In our manual analysis, security and privacy appeared in 196 reviews out of 1,305 manually coded reviews. %
} In this theme, we described particular aspects of security and privacy that were most dominant in motivating users to switch from one VPN to another. 

\vspace{1mm}
\noindent{\bf Hiding IP address: }Commercial VPNs typically function by tunnelling the user’s Internet traffic through a trusted remote server before it is forwarded to its final destination. This achieves the goal that the destination server does not learn the real IP address of the client~\cite{ClientCommercial}. Still, sometimes this fails to happen, and the actual IP address becomes visible to the destination server making the client vulnerable. This seemed to be a very relevant thing motivating users to switch from one VPN to another. For example, \emph{P148 mentions, ``Doesn't work. Currently in Germany, and prime video knows I'm in Germany. After selecting US- New York or Washington DC, still knows I'm in Germany... [VPN1] had been the only one that works for me''}, i.e. user feels their current VPN fails to hide their IP as Prime Video knows their correct location; thus, this user may switch to VPN1, which seems to work fine. \newtext{We found quotes related to this theme in 40 reviews during our manual analysis}.

\vspace{1mm}
\noindent{\bf User logs: }User activity logs were another essential thing which users seemed to be concerned about because it was observed by users that VPN stored logs even if VPN service providers claimed that no logs were kept. \newtext{In our manual analysis, user activity logs-related quotes appeared in 19 reviews}. Also, many users expressed concerns about cases of selling logs, which threatened their privacy. One such example can be seen when \emph{P10 mentioned, ``It (VPN1) was caught selling user's data and logs in 2017. Use at ur own risk. Use [VPN2] instead of this''}, clearly warning users against VPN1 as it has a history involving the sale of user logs and data.

\vspace{1mm}
\noindent{\bf Headquarter country of VPN: }One surprising yet recurring reason \newtext{(found in 20 reviews during the manual analysis} motivating users to switch from one VPN to another was the country where the VPN is headquartered. Many users felt that companies with headquarters in some countries were more likely to betray them than other companies. For example, \emph{P899 mentioned, ``...I let it run out. Why? Because [VPN1] is AMERICAN, and I'm done trusting Americans with my data...I've switched to [VPN2], a Swiss company. So far so good''}, where the user switched from VPN1 to VPN2 just because VPN1 is an American country while VPN2 is not. In our corpus, users were often against companies headquartered in ``14 eyes countries''.

\vspace{1mm}
\noindent{\bf App usage access: }Permission control is one of the major Android privacy mechanisms. When an 
application is to be installed, a user has a choice whether to allow specific permissions or not~\cite{Permission2}. 
While getting installed, many apps ask for ``Usage Access'' permission, allowing the app to track what other apps you're using and how often, as well as your carrier, language settings, and other details~\cite{appUsage}. This appeared as a concern to some users. \newtext{We found quotes related to this theme in 6 reviews}. For example, \emph{P30 mentioned, ``Why does it need ``App Usage Access'' permission? [VPN1] doesnt ask for it...''}.

\vspace{1mm}
\noindent{\bf DNS leaks: }The Domain Name System (DNS) is an essential Internet service that translates domain names into IP addresses~\cite{dns1}. Information leakage of DNS servers may cause the disclosure of users' behaviour or network zone structure and may even induce phishing or intranet penetration attacks~\cite{dns2}. DNS leaks were also raised as concerns in reviews mentioning them as reasons for switching from VPN. Other fundamental issues raised by users were related to issues with SSH, webRTC leak, etc. \newtext{During our manual analysis, we found this concern in 10 reviews.}

\subsection{User demographics}
Now we check how user demographics, such as place of residence, financial constraints, etc., are also responsible for making users switch between VPNs.  

\subsubsection{\bf Geographic location of user}

Due to the increase in awareness about the usage of VPNs as a Privacy Enhancing Technique and the rise in geoblocking imposed by various governments, there has been a worldwide increase in usage of VPNs, but this is sometimes constrained due to poor performance of VPNs in certain countries. In our study, we found this lack of performance in certain locations to be one of the reasons why users switch from one VPN to another \newtext{(expressed in 88 out of 1,305 manually-coded reviews)}. For example, \emph{P818 mentioned, ``[VPN1] hasn't worked in China for a year. The app is useless on the phone and computer. The back up open listings are inadequate... Will never subscribe again. [VPN2]'s free service has worked better''}. Apart from the lack of customer care service, the failure of [VPN1] to work properly in China was an important reason why users were willing to switch to [VPN2]. %

\vspace{1mm}
\noindent{\bf User residing in China: }Among the censorship regimes on the Internet, China is one
of the most notorious, having developed an advanced filtering
system known as the Great Firewall (GFW) to control the
the flow of online information~\cite{chinaFirewall}. The strength of this firewall was also reflected in our study, as most reviews related to this theme were related to China \newtext{(50 out of 88 Geography-related reviews)}, followed by Middle East countries like Saudi Arabia and UAE. In China, users complained of VPNs not upgrading their services as their servers were blocked as reflected in \emph{review of P835 mentioning ``anywhere else, hong kong for instance, connect right away. Now in nanchang, china, seldom succeed, because censors keep attacking and cutting access. [VPN1] stays ahead by highly frequent updates. perhaps [VPN2] could do something similar''}, where the user is currently using [VPN2] and expects [VPN2] to introduce updates frequently, like [VPN1], to fight censorship. 

\vspace{1mm}
\noindent{\bf User residing in Middle East countries: }Similar patterns were observed in Middle East countries, according to ~\cite{MiddleEast}, blocked content often contains messaging that opposes the censoring government’s own narrative of the conflicts, and that governments block content that could potentially spark internal dissent.%
\newtext{We found quotes mentioning these countries in 12 reviews under this theme.}
Thus, VPN users in these regions face the unavailability of proper services and switch from one VPN to another. For example, \emph{P139 mentioned in their review, ``Cannot get the app to connect from UAE to USA. The chrome extensions says unable to load. Tried [VPN1] and that one worked. It even made [VPN1] run. Looks like [VPN1] is blocked in UAE. Otherwise it worked great in Europe''}.

\vspace{1mm}
\noindent{\bf Other locations: }Other than these two regions, reviews also mentioned problems faced by users in Australia, Europe and South East Asia. One reason was the inactivity of VPN after government intervention, e.g., in Turkey the government blocked some servers as expressed by \emph{P529 mentioning ``Works Flawlessly. I'm in Turkey and the government blocked [VPN1] so I tried [VPN2]. It is fast and always works. Highly recommend''}. Within this theme, we noted that this inaction of VPNs towards government policies (e.g., blocking specific services) was the main motivation to switch VPNs.

\subsubsection{\bf Financial constraints of user}

Most VPN services worldwide provide either premium or free service, or both leaving the choice to the user about the type of service they want to procure. Companies that provide users with the option of both free and premium versions of their VPN incentivize users to buy the premium version by increasing the daily usage limit or the number of simultaneous connections users can have. Some VPNs also come with a free trial period where users can try the premium features for a limited period (typically 7 days) and then evaluate if the user's needs are satisfied by this VPN. The sub-themes for this theme cover the contribution of these aspects of VPN service in motivating the user to switch. \newtext{In our manual analysis, we found quotes related to this theme in 294 reviews out of 1,305 manually-coded reviews. %
} Specifically, users mentioned the following reasons. 

\vspace{1mm}
\noindent{\bf Pricing of VPN: }High pricing of VPN was a recurring theme for some VPNs and sometimes served as the primary reason for users switching to another VPN. \newtext{During our manual analysis, we found quotes related to this theme in 182 reviews}. \emph{P559 mentioned, ``Good service but just too expensive. I have switched to [VPN1] and saved a fortune. I'm afraid that with such capable competitors offering such good value, [VPN2] will have to tidy up their pricing''}. Here, it is clear that the user switched from VPN2 to VPN1 even though the user feels that the service provided by VPN2 is good. While the pricing of VPN was a big issue for some users, a proportion of users cared more about the service provided than the price, for example, in \emph{``This is probably the best VPN on the market. It costs more than other 'budget' VPNs but you get what you pay for. The speeds are phenomenal, especially for someone like me coming from [VPN] where I was paying to sit and look at the buffering symbol on YouTube...''}, where the user is ready to pay more and believes the service justifies the cost.

\vspace{1mm}
\noindent{\bf Free trial: }Companies often provide free trial to lure users towards buying a premium version of VPN. However, it sometimes backfires---\emph{review of P1478 mentioned ``I tested this as an alternative to [VPN1] when my subscription ended, however they did not give the full 7 day trial, which instead ended after 5 days. The performance was okay, but that trial bug madee trust them less and in the end I went back to [VPN1]''}, here the bug in the free trial was disappointing for the user; thus, the user moved back to the previous VPN. Some companies ask users to enter payment information or make some initial payment for accessing a free trial but it also has some repercussions, as in, \emph{review of P145 mentioning ``Expensive and the Free Trial isnt a free trial. It still makes you put in billing information. That's just wrong. Never try this. They shouldnt have your info until you're ready to commit. Thats not a free trial idc if it allows you 7 days to cancel. Most people forget and thats what they're counting on''}. Here, we can see that the user is just switching due to distrust that VPN3 might be a fraud, and that's why they are asking for payment, which isn't what these payments are intended for.  \newtext{We found quotes related to this theme in 41 reviews during our manual analysis}.

\vspace{1mm}
\noindent{\bf User finding no perk of paying for premium: }One interesting theme observed from reviews is some users, after buying premium features, felt no difference from the free version of the VPN.  \newtext{During our manual analysis, we found quotes related to this theme in 10 reviews}. \emph{P745 mentioned, ``from the first day I buy a one month premium subscription I felt like it was the same as the free version of the app. now I regret buying this piece of junk and thinking that I have to deal with it for the whole month. makes me angry and I feel like a stupid not going for the [VPN1]''}. Thus some VPN services are either not able to convey all features of the premium version to the users or it is not providing anticipated features to users. %

\vspace{1mm}
\noindent{\bf General issues related to finances: }Problems with a payment portal, lack of subscription options, lack of offers for existing users, etc. were other factors noted in our study as visible in reviews, for example, \emph{P340 mentioned, ``The description states that it has in app purchases but there is no way to pay from your Google play account. I had been saving money in my account for this purchase and now I find out I can't use it for this app. if you want to use your Google Play balance look into ``[VPN1]'' they will let you pay with your Google play balance''}. Apart from these, some positive factors also motivate users to switch from their current VPN to that particular VPN, for example, the availability of more subscription options or lower price as in their review, \emph{P258 mentioned ``Love it. Cheaper than [VPN], more subscription options than [VPN]. So easy to use''}. \newtext{During our manual analysis, we found quotes related to this theme in 57 reviews}.

\subsection{Software usability}
While the primary purpose of VPNs is to protect the user from privacy attacks while maintaining anonymity but usability is also an important factor considered by users. Our study found that many users switched among VPNs due to usability issues like poor UI, lack of required features, etc. \newtext{In our manual analysis, we found quotes related to this theme in 524 reviews out of 1,305 reviews. %
} Out of many recurring reasons, we have described a few of the most frequently appearing reasons responsible for users switching between VPNs.

\vspace{1mm}
\noindent{\bf UI/UX: }User experience covered under UI/UX also appears as an essential aspect of this theme. Better UI serves as an important reason motivating users to switch between VPNs. \newtext{During our manual analysis, we found quotes related to this theme in 129 reviews.} Also, we observed that many users admired the UI features provided by other VPNs. For example, \emph{P360 said, ``Pretty basic app but it works well. I would like a quick toggle in the notification to pull down like [VPN1]''}, where the user admires the feature provided by VPN1. In some cases, we also observed that this admiration is so extreme that users tend to switch just because of them. Bugs in the user interface also seemed to matter for users motivating them to switch if not solved. For example, \emph{P491 mentioned, ``I just installed your VPN on my phone. However, I am uninstalling it now. There is no way to go back to the home screen once you connect to [VPN1]. If you go back, you get disconnected. This is disgusting. Take some cue from [VPN2] where there is an on or off switch...''}, where this bug in UI makes the VPN useless for the user and hence the switching. 

\vspace{1mm}
\noindent{\bf Presence of too many features: }While the lack of some basic features bothered users, the presence of too many features was also mentioned as the reason for switching between VPNs by some users, as with the number of features available in a software increase, resource consumption of software is also generally observed to increase. For example, \emph{P54 said, ``So sick of this VPN, they keep adding intrusive and unwanted features. I never wanted a file cleaner or task killer in my VPN app. Uninstalling and switching to [VPN2]''}, here the user is switching to VPN2 primarily because the current VPN has a lot of overhead in terms of features like file cleaner, task cleaner, etc. A similar pattern was observed in User Interface, where we found both very simple and very complex UI motivating users to switch VPNs. \newtext{We found quotes related to this theme in 17 reviews.}

\vspace{1mm}
\noindent{\bf Connection speed: }We found speed is a significant factor for users switching between VPNs. While network speed is related to the latency experienced by users when the user is connected with a VPN, the connection speed is the time the VPN takes to connect to the desired service (e.g., a webpage). Some users mentioned slow connection speed as a primary motivation for VPN-switching. For example, \emph{P769 mentioned, ``You better take this review because I have used [VPN1] for three years with friends who use the same...[VPN1] is the worst at connecting \& also log in...Connecting to country take hours. Other vpn connects in 5 sec. Just downgrade ur app, man''} Clearly, the user mentioned [VPN1] has issues with logging in, and the connection speed was not up to the mark. \newtext{During our manual analysis, we found quotes pertaining to this theme in 39 reviews.}

\vspace{1mm}
\noindent{\bf Network speed: }Network speed has emerged as the most commonly cited metric for characterizing the
quality of network offerings~\cite{speed}. Numerically speaking, this was the most common theme that arose from our corpus study. For example, \emph{P103 mentioned in their review ``Nice app. The one thing that this app(VPN1) is lacking is the speed of the vpn. I've used [VPN2] for quite a while but this speed is pretty slow...''}---the user was directly comparing speeds provided by VPN1 and VPN2. \newtext{During our manual analysis, we found quotes pertaining to this theme in 375 reviews.} 

\vspace{1mm}
\noindent{\bf Software updates: }Software updates are essential to keeping systems and programs up-to-date. These updates fix bugs and improve performance and usability, but arguably their most important function is enhancing system security by fixing vulnerabilities~\cite{updates}. In our study, we observed that infrequent updates are criticised by users as \emph{P963 mentioned in their review ``...It's been unreliable for weeks and the last mobile update was more than a fortnight ago. During the China blockages, [VPN2] had version updates and information for customers on their website''}. Here the user felt  VPN2 is superior because of their frequent updates.

Another important aspect arising in our study related to updates is when users think that updates introduced in apps make service worse for them. For example, in \emph{P1 says, ``There are better VPN Apps. [VPN1] become dead. Try [VPN2]...[VPN1] they still old school and now these frequent updates, its get worse''}, where users feel that the updates introduced in the app are making their user experience even worse. Also, we found certain reviews where, according to users, no persisting problems were solved by updates---\emph{P227 mentioned, ``Users in China, don't bother. I've had [VPN1] for 2 years, no idea why I renewed my subscription. You can't connect more than half of the time. Constant updates that never seem to fix anything...''}, where constant updates don’t seem to solve the network issues faced in China by the user. \newtext{During our manual analysis, we found quotes related to this theme in 54 reviews.}

\subsection{Other frequently appearing factors motivating users to switch}
In this sub-section, we will discuss other factors that frequently pop up in many user reviews. These themes are mostly related to the specific purposes for which users use VPNs, like accessing OTT platforms from restricted regions. The frequent appearance of these factors showed the extent to which VPN technology has been democratized.

\vspace{1mm}
\noindent{\bf OTT Platforms: }The OTT service providers deliver audio, video and other
media over the internet and bypass the traditional operator’s network. Since the OTT players do not require any
business or technology affiliations with network operators to provide such services, they are often known by
the term `Over-The-Top' (OTT). Some popular OTT platforms were mentioned quite frequently in our study---
Netflix, Amazon Prime Video, Hulu and Hotstar~\cite{netflix, amazon, hulu, hotstar}. \newtext{During our manual analysis, we found quotes pertaining to this theme in 51 reviews}.

Many users whose primary purpose was accessing the content of another country on these OTT platforms, which are generally blocked due to policies of government or OTT service providers, switch between VPNs if their current VPN does not provide expected access. One such example is seen when \emph{P1400 mentions in their review, ``There is no streaming Netflix server for another country than the US as US don't 
necessary have all title. [VPN1] has that..and able to connect to any 
region might consider jumping back to [VPN1]''}, where the user is considering switching from the current VPN to VPN1 because of restricted access to Netflix.

\vspace{1mm}
\noindent{\bf Gaming issues: }By encrypting the internet connection and hiding the IP address, using a VPN for gaming can potentially help in several ways — granting access to different servers, improving ping and reducing lag, or even protecting you against DDoS attacks~\cite{gamers}. Thus, many gamers nowadays look forward to VPNs for a better and more secure gaming experience. Our study also found that many users switch between VPNs to get the expected gaming experience. For example, \emph{P1395 mentioned in their review, ``I was happy with this service until I realize that this service dosen't work with online games. The games keep disconecting. I though it was game issues until I switch to [VPN1]. taa daa, no more online games disconections...''}---the user switched VPNs because of disconnections in online games.  \newtext{During our manual analysis, we found quotes related to this theme in 8 reviews}.

\vspace{1mm}
\noindent{\bf Streaming issues: }Under this theme (\newtext{included in 34 reviews}), we also studied the importance of ease of streaming using VPNs in motivating users to switch from one VPN to another. Many streamers use a VPN to protect their actual IP address for privacy and performance reasons. Some use it to access Twitch or unblock content and games~\cite{streamers}. From the collected reviews, we found that many users considered switching to another VPN just because their current VPN doesn't perform well with streaming. For example, \emph{P126 mentioned, ``...I found it unreliable for video streaming. It was extremely slow and videos often just stopped playing and I couldn't get them to restart. I've switched to [VPN1]...''}, where the user switched to VPN1 because of issues with streaming videos. %

\vspace{1mm}
\noindent{\bf Technical and customer support offered: }Assistance provided by the company through technical support and customer support are the most common ways through which a company interacts with its current and potential customers. In our study, we discovered that poor knowledge of technical support, which made it difficult for users to solve the problems they faced while using VPNs, was an important reason for motivating users to switch. \newtext{During our manual analysis, we found quotes related to this theme in 151 reviews}. As \emph{P146 mentioned, ``I used [VPN1] for around 7 years. However, during the middle of 2020 I was unable to view TV anymore. After five months and numerous emails and promises that technicians were working on the issue, I gave up! I had read positive reviews of [VPN2] and decided to give it a go...''}, the inefficiency of technical support was a reason why the user switched from VPN1 to VPN2. 

This section discussed various relevant features for users switching between VPNs. In the following sections, we will discuss how these factors differ from those identified in the previously published literature on factors influencing the choice/usage of VPNs. We will analyze the impact of these factors on users' decisions to choose VPNs while switching.

\section{Comparison of Features influencing VPN-adaption and VPN-switching (RQ3)}\label{sec:prior}

Our study identified reasons for VPN-switching, however, it is not obvious if these reasons for VPN-switching are different from the general VPN adoption motivation of users. Especially since Sombatruang et al. examined factors which users consider while choosing a VPN app which suits their purpose and preferences~\cite{VPNapp}, to that end, we contrasted those factors from earlier work with the ones identified in our study to find how user requirements and concerns regarding the choice of VPN change with prior usage of VPNs.

Sombatruang et al. found that factors like device battery consumption, change in internet speed due to VPN, connection stability, ease of connection-establishment of connection, etc., are primary considerations of users while choosing a VPN~\cite{VPNapp}. They also highlighted the importance of app store reviews and star ratings of apps in the user app selection procedure. The subscription price and User personality are also crucial in the VPN adaption process~\cite{VPNPET}. These themes are also present in reviews exhibiting intent to switch, but there are some key differences, presumably due to users' experience. 

\vspace{1mm}
\noindent{\bf Advanced user requirements for VPN: }Unlike prior work, in our study, one recurring theme was that users complained about the lack of advanced features like split tunnelling, kill switch, better UI, streaming support while playing specific games, etc. Out of these, kill switch and split tunnelling security features are more common---we can infer that existing users understand the importance of these features after facing associated problems in their existing VPN. This informed understanding strongly influenced the VPN selection criteria during switching. 

\vspace{1mm}
\noindent{\bf Dependence on user Demographics: }Prior work found that the purpose of using a VPN is an important factor in the VPN selection process. We found that while this was still true, user demographics (specifically the geographical location of users) also played an important role in VPN switching. This was particularly visible in reviews from users residing in China and the Middle East, where web censorship is very high and VPNs are often blocked. In those situations, users are forced to switch between VPNs frequently if the app prevents them from evading geographical constraints. 

\vspace{1mm}
\noindent{\bf Reduced weightage to ratings and reviews: }Sombatruang et al. identified that ratings and reviews are most significantly correlated with VPN adaption, indicating herding attitude among users\cite{VPNapp}. In contrast, our dataset had only 4.15\% of reviews in which users talked about app ratings and user reviews of the VPN app, while other factors like geography, security, etc., are mentioned in significantly more reviews. Thus, our results hint that user experience with existing VPN apps often reduces the weightage of ``ratings and reviews'' as a factor during VPN-switching. Next, we will explore the answer to our RQ4 posted in Section~\ref{sec:intro} and check the key features desired by users during VPN-switching. 

\section{VPN Features Desired By Users During VPN-switching (RQ4)}\label{sec:sem}

In this section, we check what features are desired by existing users during VPN-switching and if users desire multiple features together while switching their VPN providers. 

\vspace{1mm}

\noindent{\bf Top desirable features in a VPN.} We use our identified level-2 themes as features desired by users while switching their VPNs. To that end, we simply searched for these themes in the reviews and took the frequency of occurrence as importance for these features. \newtext{The top-5 most mentioned features in VPN-switching reviews are Connection, Network speed, User-friendliness, Service Provided, and UI/UX appearing in 41,020, 28,406, 22,402, 21,689, 19,406 reviews, respectively (top-10 of such features with their frequency is in Table~\ref{tab:most-frequent-indiv-feature} of Appendix~\ref{sec:desired-features-appendix})}. We noted that privacy is not among the top 10 mentioned features in VPN-switching reviews. This finding aligns with earlier work~\cite{VPNapp}. Next, we moved beyond only the most popular features and checked if these popular features also co-occurred with not-so-popular ones for specific VPN-switching reviews. 

\vspace{1mm}
\noindent \textbf{Co-occurrence of desired features in VPN-switching reviews.} 
For checking the co-occurrence of multiple desired factors, in line with prior work, which examined the co-occurrence of tokens in the same posts~\cite{fearspeech}, we created a co-occurrence network~\cite{network} of level-2 themes where themes served as nodes of the network and weights of edges were calculated using the formula: $W_{ij} = F_{ij}/(F_i * F_j)$
where $F_{ij}$ are \#reviews in which both theme i and j occur, $F_i$ is the number of reviews in which theme $i$ occurred, and $F_j$ are \#reviews in which theme $j$ occurred.
After constructing the co-occurrence network, we used the clique-based Louvain algorithm~\cite{cbla, lov} to find 10 overlapping communities in the graph. %

\vspace{1mm}

\noindent \textbf{Specific subsets of desired features co-occurred in VPN-switching reviews.} In our detected communities, we found that the most frequent factors in user reviews about switching VPN and network speed co-occurred with non-frequent factors like streaming. This shows that user requirements often encompass multiple features, and furthermore, there are distinct feature sets that are desired together in VPN-switching reviews, signifying a concrete spectrum of needs which led to VPN-switching. \newtext{One interesting observation is that 7 out of 10 communities have ``Security'' (3 communities) and ``Privacy'' (5 communities)} in them even though privacy did not occur most frequently in reviews---signifying at least a non-trivial subset of users still consider better security and privacy as a valid reason for switching VPNs. We provide the complete list of the feature co-occurrence communities for interested readers in Table~\ref{tab:comms} (Appendix~\ref{sec:desired-features-appendix}). Having uncovered these sets of distinct requirements during VPN-switching, we finally check if the current VPN review blogs identify and cater to these user requirements. 

\section{Alignment of Information in VPN Review Websites with User Requirement During VPN-switching (RQ5)}\label{sec:blogs}

We previously noted in Section~\ref{sec:prior} that users switching among VPNs put lesser emphasis on reviews and ratings while switching. However, as previous work found, general users still rely on VPN review blog websites for information regarding VPNs~\cite{royaPaper}. 
To that end, we ask if these VPN-review blog articles can help the users who want to switch VPNs.

\subsection{Collecting data from VPN-review blogs}
In order to identify popular VPN-review websites and their relevant blog articles, first, we queried Google with the term ``best VPN'' and collected the top 10 results. Four out of these ten websites prohibit automatically collecting their articles, so we focus on the rest. We searched each website domain using Google's site search with the keywords ``Best VPN recommendation blogs'', ``VPN review blogs'', and ``VPN comparison blogs'', which resulted in 583 blog articles. We manually reviewed these articles to identify 376 blog articles related to VPN or its uses. Out of these, 176 are VPN review blogs, and 200 are VPN best-pick blogs (i.e. blogs which either recommend a ranked list of best VPNs or compare VPNs). For ethical reporting of results, this section will refer to individual sites with alphabets A to F. %

\subsection{Topic analysis of VPN-review blogs}

\noindent\textbf{Topic Modelling: }To identify the content available in the blogs, we started with Topic Modelling implemented using LDA algorithm\cite{lda}. Firstly, we applied LDA to the text of the entire 376 blog corpus, followed by applying it to the corpus formed by only the VPN review blogs and VPN best-pick blogs separately. 

\vspace{1mm}
\noindent\oldtext{
\textbf{Blogs do not give information on important and desired VPN features: }We started by applying the LDA topic modelling algorithm on our blog dataset corpus\cite{lda}. LDA output word clusters as topics. Following the method described in prior work, we identify the optimal number of topics and assigned semantic labels to these word clusters via manually checking the blogs where words from a cluster appear~\cite{fakenews-lda}. Finally, we identified overall ten topics in the blogs, out of which four topics were related to VPN features. Two topics primarily contained terms related to streaming services like ``Hulu'', ``Netflix'', etc. The other two topics were related to ``Security and Privacy'' and ``Connection and Network speed''. These seemed in line with the themes identified in our qualitative analysis. Still, we also noted a lack of coverage of other important aspects mentioned by users in reviews like financial aspects (like price, free version availability, etc.) and Geography. We also applied topic modelling on subsets of blog articles---VPN review blogs and VPN best-pick blogs separately. Topic analysis of VPN reviews and best-pick blogs yielded 10 topics each. Most of the topics in VPN-review blogs were related to Streaming and Speed, while topics in VPN best-pick blogs were mainly related to the Speed and Pricing of VPNs. However, these topics did not touch upon a plethora of features mentioned in Section~\ref{sec:sem}---these blogs might provide information about a significant number of desired features which users consider during VPN-switching. The topics (semantic labels with word cluster) are in Table~\ref{tab:topics-complete-corpus}, Table~\ref{tab:topics-review} and Table~\ref{tab:topics-best-pick} of Appendix~\ref{sec:blog_appendix}. 
}

\vspace{1mm}
\noindent\grumbler{REVIEW E}{\textbf{Identifying themes in reviews:  }Apart from topic modelling, we also applied the BART model fine-tuned for identifying themes in reviews on the blogs corpus. Given the input length constraint of the BART model, we fed the blogs to the model in small chunks followed by a union of identified themes. This provided us with the themes covered by blogs from these VPN recommendation websites.}

\vspace{1mm}

\noindent\textbf{Blogs do not give information on important and desired VPN features: }We started by applying the LDA topic modelling algorithm on our blog dataset corpus\cite{lda}. LDA output word clusters as topics. Following the method described in prior work, we identify the optimal number of topics and assigned semantic labels to these word clusters via manually checking the blogs where words from a cluster appear~\cite{fakenews-lda}. Finally, we identified ten topics in the blogs, out of which four topics were related to VPN features. Two topics primarily contained terms related to streaming services like ``Hulu'', ``Netflix'', etc. The other two topics were related to ``Security and Privacy'' and ``Connection and Network speed''. These are in line with the themes identified in our qualitative analysis. 

We also noted a lack of coverage of other important aspects mentioned by users in reviews like financial aspects (like price, free version availability, etc.) and Geography. We also applied topic modelling on subsets of blog articles---VPN review blogs and VPN best-pick blogs separately. Topic analysis of VPN reviews and best-pick blogs yielded 10 topics each. Most of the topics in VPN-review blogs were related to Streaming and Speed, while topics in VPN best-pick blogs were mainly related to the Speed and Pricing of VPNs. However, these topics did not touch upon a plethora of features mentioned in Section~\ref{sec:sem}---these blogs might provide information about a significant number of desired features which users consider during VPN-switching. The topics (semantic labels with word cluster) are in Table~\ref{tab:topics-complete-corpus},~\ref{tab:topics-review} and~\ref{tab:topics-best-pick} of Appendix~\ref{sec:blog_appendix}.

\newtext{
Then, we identified themes covered in the blogs using the BART model fine-tuned to identify themes in reviews to understand the coverage of user requirements by these blogs. The top themes identified by the model are listed in Table \ref{tab:themes-blogs}. Among the top themes, we can see ``Privacy'' and ``Features'', which, according to our findings, are also the top aspects users look at when making the decision to switch VPNs, but we see no coverage of important themes like ``Geography'' or ``User Friendliness'' or ``Reliability'' which is also important for users. We also checked the variation of topics covered in VPN review blogs and VPN best-pick blogs and saw a similar trend as before. Best-pick blogs mostly focused on Speed and Pricing, and VPN review blogs focused on features but lacked coverage of other important aspects related to VPN switching.
}

\subsection{Entities mentioned in VPN-review blogs}

We Performed an off-the-shelf Named Entity Recognition technique (NER) \cite{ner} on individual blogs and then calculated the frequency of these entities in blogs from different websites as well as in the whole blog corpus. The full list of the top 10 most frequently occurring entities for each of the websites is presented in Table~\ref{tab:ner} of Appendix~\ref{sec:blog_appendix}.

\vspace{1mm}

\noindent\textbf{Information in blogs is heavily biased towards specific VPN apps and services: } Our frequency analysis of entities reveals that there are 1 to 5 specific VPN app names in the top 10 most occurring entities across websites. Out of these two entities occur very frequently---``Netflix'' and ``Surfshark'' (Table~\ref{tab:ner} of Appendix~\ref{sec:blog_appendix}). We make two observations from this analysis. First, these websites focus disproportionately more on streaming services like Netflix, backing our observation in Topic modelling. Second, these websites primarily mention a few VPNs like Nord or SurfShark while ignoring other VPNs. However, interestingly, a Pearson correlation coefficient analysis did not find any statistically significant correlation between the frequency of the entities (which are VPN apps) and the number of downloads. Finally, we do a word association analysis to check (i) if these mentions of specific entities in blogs (i.e., co-occurred) are biased (e.g., overwhelming praise or criticism) and (ii) if the themes identified in Section~\ref{sec:sem} (which are also user-desired features) are at all mentioned in any form in these blogs.

\section{Bias towards specific entities and coverage of features in VPN-review blogs}

\noindent\textbf{Word association analysis:} We leverage a technique developed in prior work to perform a word association analysis, i.e., given a set of words, what are similar or associated words appear in a corpus?~\cite{wordvec}. We first created two sets of keywords. The first set contains keywords that are not related to any specific themes, and rather, they contain general terms describing desired VPN characteristics like ``best'', ``fastest'' and ``Cheapest''. The second set contained a set of keywords which are related to themes (i.e., features) desired in VPN-switching reviews. 
\oldtext{Then, we trained a Word2Vec model\cite{word2vec} on the corpus generated by blog posts of individual websites. Finally, we analysed embedding of which words are similar (as described in prior work~\cite{wordvec}) to embedding words in our sets of keywords. The full sets of keywords are in Table~\ref{tab:ref-terms} of Appendix~\ref{sec:blog_appendix}.}
\grumbler{REVIEW E}{Next, to capture words used in similar contexts, we used a pre-trained BERT model\cite{bertbase} and fine-tuned it using the blog corpus on a Masked Language Modelling task to get a model acquainted with the context used in blogs. Then, we generated embeddings for keywords using this BERT model. Finally, we analysed embedding of which words are similar (as described in prior work~\cite{wordvec}) to embeddings of our sets of keywords. This approach helped us to capture the context of input keywords within the blogs while also taking the general meaning of words into context due to the nature of the pre-trained model. The full sets of keywords are in Table~\ref{tab:ref-terms} of Appendix~\ref{sec:blog_appendix}.}

\vspace{1mm}

\noindent\textbf{Bias towards specific VPN apps in blogs:} For the keyword ``best'', in most websites, the names of popular VPNs like \newtext{Nord and Proton} are among the most similar words. We made similar observations when we tried to find similar words for terms like ``fastest'' or ``cheapest'', indicating a bias towards selected VPNs corresponding to all words in our first set of keywords---signifying desired VPN characteristics are always associated with these VPN apps in the blogs. 

\vspace{1mm}

\noindent \textbf{Lack of coverage for user-desired features in VPN-blogs:} Via our second set of theme-related keywords, we tried identifying if the VPN blogs mention the user-desired themes (in VPN-switching reviews) in any form. For each such keyword in our set (e.g., ``logs'', ``free'', or ``price''), we checked the top 10 most similar words in the corpus \newtext{(in terms of our fine-tuned BERT model embedding)}. We found that for most themes, these words are not relevant, hinting that most of these themes are not discussed in depth in the VPN blogs.

\section{Implications}\label{sec:disussion}

Our first-ever study of VPN-switching study unearths a number of important observations. To that end, in this section, we examine the implications of our study for VPN providers, VPN-review blogs and the security community in general. 

\vspace{1mm}
\noindent{\bf VPN-switching is intended by a non-trivial fraction of users: } We note that VPN-switching, although not studied before, is a significant phenomenon. More specifically, out of 411,751 reviews from 20 popular VPN apps, 185,399 (45\%) indicated towards VPN-switching. This observation heavily emphasizes the need to assist these users from the VPN-review blogs as well as from VPN providers by providing them services that cater to these users' desired preferences.  

\vspace{1mm}

\noindent{\bf Users desire advanced and more customized features while switching VPN: }As opposed to VPN adoption, the users who want to switch VPNs are more knowledgeable. Thus, they seek more advanced security features and customized options (e.g., for evading the censorship employed in specific countries). Thus, VPN providers might want to provide customized solutions to these experienced users. 

\vspace{1mm}

\noindent{\bf Users desire sets of features while switching VPN: }Our cluster analysis revealed that, interestingly, the VPN-switching reviews often desire multiple features for their new VPN app. In fact, these features form distinct clusters, hinting at separate user bases (since mostly unique users posted these reviews) with distinct requirements. In fact, often during VPN-switching, security and privacy is an important factor, but it is accompanied by additional requirements in a VPN. This finding provides a roadmap for VPN providers and researchers to create customized VPN solutions which can cater to these distinct user bases with different sets of requirements. 

\vspace{1mm}

\noindent{\bf Current VPN-review blogs are insufficient for VPN-switching for bias toward specific VPNs and low coverage of user-desired features: }Finally, our analysis reveals a huge gap and opportunity for researchers and the general security education community. Currently, VPN review blogs paint a biased picture of VPN apps and manipulate users to switch to specific VPNs. In fact, presumably, due to this reason, the users who want to switch VPNs did not mention these reviews and ratings as significant factors for choosing a new VPN. Furthermore, these blogs are insufficient to provide information to users regarding their desired privacy, security and usability features. Hence, our findings imply that the security community should develop more tools like Vpnalyzer~\cite{vpnalyzer} and make them user-friendly so that users can check the VPNs and switch to a VPN according to their requirements.

\subsection{Recommendations for stakeholders}

\noindent\grumbler{REVIEW CD}{Finally, we synthesize a list of recommendations for important stakeholders in the context of VPN switching---VPN providers, VPN recommendation websites, the Security Research Community, and finally, the users themselves.}

\noindent\newtext{
\textbf{Recommendations for VPN service providers: }
Our findings indicate that there is no one-size-fits-all thing in the case of VPN. Hence, service providers need to provide users with options to customize VPN features based on users' priorities and requirements. The providers can also aid users by customizing their products according to user geography and demographic groups. In our study, we found that there is a large section of VPN users who expressed ``Potential Shift'' in their reviews. These users are basically giving their existing VPNs a chance to satisfy their need. Thus sensing these needs and establishing a smoother communication channel between user and providers (e.g., via customer or technical support) can help the company retain these users.
}

\noindent\newtext{
\textbf{Recommendations for VPN recommendation websites: } We identify that VPN recommendation websites are biased towards particular VPN service providers and do not cover important themes that users consider while switching VPNs. Therefore, coverage of such topics in blogs can help users make more informed decisions. We also noted a difference in factors responsible for VPN adoption and switching, which can guide these websites to tailor their blogs according to users' VPN usage experience. These websites can also provide automated VPN recommendation systems according to users' experience and requirements.
}

\noindent\newtext{
\textbf{Recommendations for the security research community: }Our results show that users sometimes compromise their security for their requirements, like watching Netflix. Thus the security community might consider developing more tools like Vpnalyzer~\cite{vpnalyzer} and make them usable so that users can test the VPNs by themselves. %
}

\noindent\newtext{
\textbf{Recommendations for the users: }Finally, from previous literature, we found that users rely a lot on VPN recommendation websites\cite{royaPaper}, which is a problem. In our study, we found that these websites are biased towards and selected VPNs and hence, users should rely on more reliable sources like academic tools or materials to select the VPN. Also, users need to acknowledge that their peers or even families will have different requirements compared to theirs; hence, just relying on word-of-mouth might not work while selecting VPNs.
}

\section{Conclusion}\label{sec:conclusion}
\oldtext{In this study, we aimed to explore the VPN-switching behaviours and preferences of users in the wild. Using a semi-automated approach, we collected a total of 185,399 VPN app reviews (from 20 popular VPN apps and two popular app stores from Google and Apple).
Leveraging this data, we found a non-negligible number of users (identified by distinct usernames) who intend to switch between VPNs. In fact, different VPN providers have non-uniform rates of users migrating from them--- the size of the VPN user base shows a statistically significant negative correlation with the fraction of reviews mentioning VPN-switching. 
To that end, our qualitative analysis revealed a hierarchy of reasons which motivate users to switch between VPNs. These reasons include technical factors, like network speed, etc., and non-technical factors, like financial considerations. These factors also helped us to identify properties that matter to VPN users. Finally, using our identified themes for switching VPNs and our dataset, we identified communities of themes by applying a community detection algorithm on the co-occurrence graph of themes. We identified that users seek advanced and often customized features while switching VPNs. To that end, our novel automated analysis of VPN review blogs revealed a significant issue. We found that these websites prefer some popular VPNs over others, showing bias. Moreover, these blogs significantly miss providing information about desired user requirements.
Our research is the first to investigate the reasons for retaining/losing users for VPN providers and provide a glimpse of users' inherent reasons for switching their VPN providers. We are also the first to investigate VPN review websites, which are vital stakeholders in the VPN ecosystem from the user's point of view. We strongly believe our research provides VPN providers, VPN-review blogs and the security community, in general, a roadmap to help the end users by creating better VPN services which can cater to users' desired preferences and provide more relevant VPN-related information to the end users. 
}

\noindent\newtext{
This study delved into the VPN-switching behaviours of users through an analysis of \newtext{185,399} app reviews from 20 popular VPN apps across Google and Apple stores. Our findings revealed a significant number of users intending to switch between VPNs, with varied rates among providers. Notably, the size of a VPN provider's user base exhibited a statistically significant negative correlation with the frequency of VPN-switching mentions in reviews. Our qualitative analysis identified a hierarchy of reasons motivating users to switch, encompassing technical factors such as network speed and non-technical considerations like financial aspects.
}

\newtext{
Furthermore, our research introduced a novel exploration of VPN review blogs, exposing biases towards certain VPNs and a notable absence of information regarding user requirements. By shedding light on the reasons behind user retention and loss for VPN providers and unveiling patterns in user preferences, this study offers valuable insights for enhancing VPN services and addressing user needs more effectively.
}

\section{Acknowledgement}\label{sec:acknowledgement}
We thank the anonymous reviewers and our shepherd for their valuable feedback. We also thank Sankalp Ramesh for his help with an earlier iteration of this work. This research was (partially) funded by a Google India Faculty Research Award.

\bibliographystyle{plain}
\bibliography{usenix2022_SOUPS.bib}

\begin{thebibliography}{10}

\bibitem{why-vpn-1}
Susan Alexandra.
\newblock 5 popular reasons for using a vpn.
\newblock \url{https://www.globalsign.com/en-in/blog/5-popular-reasons-using-vpn}, 2020.

\bibitem{amazon}
Amazon.
\newblock Amazon prime video.
\newblock \url{https://www.primevideo.com/}, 2023.

\bibitem{speed}
Steven Bauer, David Clark, and William Lehr.
\newblock Understanding broadband speed measurements.
\newblock 08 2010.
\newblock \url{https://groups.csail.mit.edu/ana/Publications/Understanding_broadband_speed_measurements_bauer_clark_lehr_TPRC_2010.pdf}.

\bibitem{lov}
Vincent~D Blondel, Jean-Loup Guillaume, Renaud Lambiotte, and Etienne Lefebvre.
\newblock Fast unfolding of communities in large networks.
\newblock {\em Journal of Statistical Mechanics: Theory and Experiment}, 2008(10), Oct 2008.
\newblock \url{https://arxiv.org/abs/0803.0476}.

\bibitem{MobUsers}
Bram Bonné, Gustavo Ruiz, Peter Quax, and Wim Lamotte.
\newblock Insecure network, unknown connection: Understanding wi-fi privacy assumptions of mobile device users.
\newblock {\em Information (Switzerland)}, 8, 07 2017.
\newblock \url{https://www.mdpi.com/2078-2489/8/3/76/htm}.

\bibitem{fakenews-lda}
Ceren Budak.
\newblock What happened? the spread of fake news publisher content during the 2016 u.s. presidential election.
\newblock In {\em Proceedings of the Web Conference 2019}, WWW '19, page 139–150, 2021.

\bibitem{ClientCommercial}
Thanh Bui, Siddharth~Prakash Rao, Markku Antikainen, and Tuomas Aura.
\newblock {\em Client-Side Vulnerabilities in Commercial VPNs}, pages 103--119.
\newblock 11 2019.
\newblock \url{https://arxiv.org/pdf/1912.04669.pdf}.

\bibitem{flant5}
Hyung~Won Chung, Le~Hou, Shayne Longpre, Barret Zoph, Yi~Tay, William Fedus, Yunxuan Li, Xuezhi Wang, Mostafa Dehghani, Siddhartha Brahma, Albert Webson, Shixiang~Shane Gu, Zhuyun Dai, Mirac Suzgun, Xinyun Chen, Aakanksha Chowdhery, Alex Castro-Ros, Marie Pellat, Kevin Robinson, Dasha Valter, Sharan Narang, Gaurav Mishra, Adams Yu, Vincent Zhao, Yanping Huang, Andrew Dai, Hongkun Yu, Slav Petrov, Ed~H. Chi, Jeff Dean, Jacob Devlin, Adam Roberts, Denny Zhou, Quoc~V. Le, and Jason Wei.
\newblock Scaling instruction-finetuned language models, 2022.

\bibitem{vpn-usage-2020}
Sarah Coble.
\newblock Vpn usage in us quadruples.
\newblock \url{https://www.infosecurity-magazine.com/news/vpn-usage-in-us-quadruples/}, 2020.

\bibitem{cook-2018-ethics}
Natalie Cook, Susan Ayers, and Antje Horsch.
\newblock Maternal post traumatic stress disorder during the perinatal period and child outcomes: A systematic review.
\newblock {\em J Affect Disord .}, 2018(225):18--31, 2018.

\bibitem{gan-bert}
Danilo Croce, Giuseppe Castellucci, and Roberto Basili.
\newblock {GAN}-{BERT}: Generative adversarial learning for robust text classification with a bunch of labeled examples.
\newblock In Dan Jurafsky, Joyce Chai, Natalie Schluter, and Joel Tetreault, editors, {\em Proceedings of the 58th Annual Meeting of the Association for Computational Linguistics}, pages 2114--2119, Online, July 2020. Association for Computational Linguistics.

\bibitem{streamers}
Cyberwaters.
\newblock Do streamers use a vpn (do you really need it?).
\newblock 2021.
\newblock \url{https://cyberwaters.com/do-streamers-use-vpn/}.

\bibitem{bertbase}
Jacob Devlin, Ming{-}Wei Chang, Kenton Lee, and Kristina Toutanova.
\newblock {BERT:} pre-training of deep bidirectional transformers for language understanding.
\newblock {\em CoRR}, abs/1810.04805, 2018.

\bibitem{feamster-2020-safe}
Agnieszka Dutkowska{-}Zuk, Austin Hounsel, Andre Xiong, Marshini Chetty, Nick Feamster, Molly Roberts, and Brandon Stewart.
\newblock Practicing safe browsing: Understanding how and why university students use virtual private networks.
\newblock {\em CoRR}, 2020.

\bibitem{why-vpn-2}
Max Eddy.
\newblock Why you need a vpn, and how to choose the right one.
\newblock \url{https://in.pcmag.com/privacy/113633/what-is-a-vpn-and-why-you-need-one}, 2022.

\bibitem{eysenbach_ethical_2001}
Gunther Eysenbach and James~E. Till.
\newblock Ethical issues in qualitative research on internet communities.
\newblock {\em BMJ}, 323(7321):1103--1105, November 2001.

\bibitem{vpn-growth}
Yahoo Finance.
\newblock Global virtual private network (vpn) market to reach us\$77.1 billion by the year 2026.
\newblock \url{https://finance.yahoo.com/news/global-virtual-private-network-vpn-122300132.html}, 2021.

\bibitem{Permission2}
Erza Gashi and Zhilbert Tafa.
\newblock Permission-based privacy analysis for android applications.
\newblock 10 2017.
\newblock \url{https://knowledgecenter.ubt-uni.net/cgi/viewcontent.cgi?article=1230&context=conference}.

\bibitem{metricsMultiClass}
Margherita Grandini, Enrico Bagli, and Giorgio Visani.
\newblock Metrics for multi-class classification: an overview, 2020.

\bibitem{cbla}
Sumit~Kumar Gupta and Dr. Dhirendra~Pratap Singh.
\newblock Cbla: A clique based louvain algorithm for detecting overlapping community.
\newblock {\em Procedia Computer Science}, 218:2201--2209, 2023.
\newblock International Conference on Machine Learning and Data Engineering.

\bibitem{length}
Emitza Guzman and Walid Maalej.
\newblock How do users like this feature? a fine grained sentiment analysis of app reviews.
\newblock {\em 2014 IEEE 22nd International Requirements Engineering Conference, RE 2014 - Proceedings}, pages 153--162, 09 2014.
\newblock \url{https://ieeexplore.ieee.org/stamp/stamp.jsp?tp=&arnumber=6912257}.

\bibitem{harboe-2015-affinity}
Gunnar Harboe and Elaine~M. Huang.
\newblock {Real-World Affinity Diagramming Practices: Bridging the Paper-Digital Gap}.
\newblock In {\em Proc.\ CHI}, 2015.

\bibitem{deberta}
Pengcheng He, Xiaodong Liu, Jianfeng Gao, and Weizhu Chen.
\newblock Deberta: Decoding-enhanced bert with disentangled attention, 2021.

\bibitem{chinaFirewall}
Nguyen~Phong Hoang, Arian~Akhavan Niaki, Jakub Dalek, Jeffrey Knockel, Pellaeon Lin, Bill Marczak, Masashi Crete-Nishihata, Phillipa Gill, and Michalis Polychronakis.
\newblock How great is the great firewall? measuring china’s dns censorship.
\newblock In {\em USENIX Security Symposium}, pages 3381--3398, 2021.
\newblock \url{https://www.usenix.org/system/files/sec21-hoang.pdf}.

\bibitem{hotstar}
Hotstar.
\newblock Watch tv shows, movies, specials, live cricket \& football.
\newblock \url{https://www.hotstar.com/}, 2023.

\bibitem{optuna}
{\em Optuna: A hyperparameter optimization framework}.

\bibitem{hulu}
Hulu.
\newblock Hulu: Stream tv and movies live and online.
\newblock \url{https://www.hulu.com/}, 2023.

\bibitem{Permission}
Muhammad Ikram, Narseo Vallina-Rodriguez, Suranga Seneviratne, Mohamed~Ali Kaafar, and Vern Paxson.
\newblock An analysis of the privacy and security risks of android vpn permission-enabled apps.
\newblock pages 349--364, 11 2016.
\newblock \url{https://dl.acm.org/doi/10.1145/2987443.2987471}.

\bibitem{FB}
Asela Jayatilleke and Parakum Pathirana.
\newblock Smartphone vpn app usage and user awareness among facebook users.
\newblock pages 1--6, 10 2018.
\newblock \url{https://ieeexplore.ieee.org/stamp/stamp.jsp?arnumber=8550081}.

\bibitem{lda}
Hamed Jelodar, Yongli Wang, Chi Yuan, and Xia Feng.
\newblock Latent dirichlet allocation (lda) and topic modeling: models, applications, a survey.
\newblock {\em Multimedia Tools and Applications}, 78:15169--15211, 2017.

\bibitem{VitalRole}
Chamandeep Kaur and Dr.~Yogesh Sharma.
\newblock The vital role of vpn in making secure connection over internet world.
\newblock 8:2336--2339, 03 2020.
\newblock \url{https://www.ijrte.org/wp-content/uploads/papers/v8i6/F8335038620.pdf}.

\bibitem{EmpCommercial}
Mohammad Khan, Joe DeBlasio, Geoffrey Voelker, Alex Snoeren, Chris Kanich, and Narseo Vallina-Rodriguez.
\newblock An empirical analysis of the commercial vpn ecosystem.
\newblock pages 443--456, 10 2018.
\newblock \url{https://dl.acm.org/doi/10.1145/3278532.3278570}.

\bibitem{ISP}
Sheharbano Khattak, Mobin Javed, Syed~Ali Khayam, Zartash Uzmi, and Vern Paxson.
\newblock A look at the consequences of internet censorship through an isp lens.
\newblock {\em Proceedings of the ACM SIGCOMM Internet Measurement Conference, IMC}, pages 271--283, 11 2014.
\newblock \url{https://dl.acm.org/doi/10.1145/2663716.2663750}.

\bibitem{gamers}
Kristina.
\newblock Why do i need vpn for gaming? 10 reasons it makes sense.
\newblock 2023.
\newblock \url{https://www.vpnmentor.com/blog/why-need-a-vpn-for-gaming/}.

\bibitem{bart}
Mike Lewis, Yinhan Liu, Naman Goyal, Marjan Ghazvininejad, Abdelrahman Mohamed, Omer Levy, Ves Stoyanov, and Luke Zettlemoyer.
\newblock Bart: Denoising sequence-to-sequence pre-training for natural language generation, translation, and comprehension, 2019.

\bibitem{network}
Taoying Li, Jie Bai, Xue Yang, Qianyu Liu, and Yan Chen.
\newblock Co-occurrence network of high-frequency words in the bioinformatics literature: Structural characteristics and evolution.
\newblock {\em Applied Sciences}, 8:1994, 10 2018.
\newblock \url{https://www.mdpi.com/2076-3417/8/10/1994}.

\bibitem{app-store-scraper}
Eric Lim.
\newblock app-store-scraper 0.3.5.
\newblock \url{https://pypi.org/project/app-store-scraper/}, 2020.

\bibitem{dns2}
Sui Luo, Heng Li, and Shuyuan Jin.
\newblock Measuring information leakage of dns server.
\newblock In {\em 2021 IEEE 6th International Conference on Computer and Communication Systems (ICCCS)}, pages 796--805, 2021.
\newblock \url{https://ieeexplore.ieee.org/abstract/document/9449155}.

\bibitem{updates}
Arunesh Mathur, Nathan Malkin, Marian Harbach, Eyal P{\'{e}}er, and Serge Egelman.
\newblock Quantifying users' beliefs about software updates.
\newblock {\em CoRR}, abs/1805.04594, 2018.
\newblock \url{http://arxiv.org/abs/1805.04594}.

\bibitem{mcdonald-2018-geoblocking}
Allison McDonald, Matthew Bernhard, Luke Valenta, Benjamin VanderSloot, Will Scott, Nick Sullivan, J.~Alex Halderman, and Roya Ensafi.
\newblock 403 forbidden: A global view of cdn geoblocking.
\newblock In {\em Proceedings of the Internet Measurement Conference 2018}, IMC '18, page 218–230, 2018.

\bibitem{dns1}
P.~V. Mockapetris.
\newblock {\em RFC1034: Domain Names - Concepts and Facilities}.
\newblock RFC Editor, 1987.
\newblock \url{https://www.ietf.org/rfc/rfc1034.txt}.

\bibitem{Socia-Media-Article1}
Rebecca Moody.
\newblock Vpn market report 2023: Who’s got the biggest vpn market share?
\newblock 2020.
\newblock \url{https://www.comparitech.com/blog/vpn-privacy/vpn-market-share-report/}.

\bibitem{VPNPET}
Moses Namara, Daricia Wilkinson, Kelly Caine, and Bart Knijnenburg.
\newblock Emotional and practical considerations towards the adoption and abandonment of vpns as a privacy-enhancing technology.
\newblock {\em Proceedings on Privacy Enhancing Technologies}, 2020:83--102, 01 2020.
\newblock \url{https://www.researchgate.net/publication/338470351_Emotional_and_Practical_Considerations_Towards_the_Adoption_and_Abandonment_of_VPNs_as_a_Privacy-Enhancing_Technology}.

\bibitem{netflix}
Netflix.
\newblock Netflix-watch tv shows online, watch movies online.
\newblock \url{https://www.netflix.com/}, 2023.

\bibitem{MiddleEast}
Helmi Noman.
\newblock Internet censorship and the intraregional geopolitical conflicts in the middle east and north africa.
\newblock 01 2019.
\newblock \url{bit.ly/36fvwG1}.

\bibitem{Commercial}
Vasile Perta, Marco Barbera, Gareth Tyson, Hamed Haddadi, and Alessandro Mei.
\newblock A glance through the vpn looking glass: Ipv6 leakage and dns hijacking in commercial vpn clients.
\newblock volume~1, 04 2015.
\newblock \url{https://haddadi.github.io/papers/PETS2015VPN.pdf}.

\bibitem{google-play-scraper}
PlanB.
\newblock google-play-scraper 1.0.3.
\newblock \url{https://pypi.org/project/google-play-scraper/}, 2020.

\bibitem{fake-reviews}
Mizanur Rahman, Nestor Hernandez, Ruben Recabarren, Syed~Ishtiaque Ahmed, and Bogdan Carbunar.
\newblock The art and craft of fraudulent app promotion in google play.
\newblock In {\em Proceedings of the 2019 ACM SIGSAC Conference on Computer and Communications Security}, CCS '19, page 2437–2454, New York, NY, USA, 2019. Association for Computing Machinery.

\bibitem{vpnalyzer}
Reethika Ramesh, Leonid Evdokimov, Diwen Xue, and Roya Ensafi.
\newblock Vpnalyzer: Systematic investigation of the vpn ecosystem.
\newblock {\em Proceedings 2022 Network and Distributed System Security Symposium}, 2022.

\bibitem{royaPaper}
Reethika Ramesh, Anjali Vyas, and Roya Ensafi.
\newblock "all of them claim to be the best": Multi-perspective study of vpn users and vpn providers.
\newblock {\em ArXiv}, abs/2208.03505, 2022.

\bibitem{ner}
Alan Ritter, Sam Clark, Mausam, and Oren Etzioni.
\newblock Named entity recognition in tweets: An experimental study.
\newblock In {\em Conference on Empirical Methods in Natural Language Processing}, 2011.

\bibitem{fearspeech}
Punyajoy Saha, Binny Mathew, Kiran Garimella, and Animesh Mukherjee.
\newblock “short is the road that leads from fear to hate”: Fear speech in indian whatsapp groups.
\newblock In {\em Proceedings of the Web Conference 2021}, WWW '21, page 1110–1121, 2021.

\bibitem{saldana-qualitative}
J.~M. Saldana.
\newblock {\em The coding manual for qualitative researchers}.
\newblock SAGE, 2015.

\bibitem{VPNapp}
Nissy Sombatruang, Tan Omiya, Daisuke Miyamoto, M.~Angela Sasse, Youki Kadobayashi, and Michelle Baddeley.
\newblock Attributes affecting user decision to adopt a virtual private network {(VPN)} app.
\newblock {\em CoRR}, abs/2008.06813, 2020.
\newblock \url{https://arxiv.org/abs/2008.06813}.

\bibitem{wordvec}
Fateme Tahmasbi, Leonard Schild, Chen Ling, Jeremy Blackburn, Gianluca Stringhini, Yang Zhang, and Savvas Zannettou.
\newblock “go eat a bat, chang!”: On the emergence of sinophobic behavior on web communities in the face of covid-19.
\newblock {\em Proceedings of the Web Conference 2021}, 2021.

\bibitem{appUsage}
Dallas Thomas.
\newblock 20 privacy \& security settings you need to check on your google pixel.
\newblock 2019.
\newblock \url{bit.ly/36nAtN9}.

\bibitem{llama2}
Hugo Touvron, Louis Martin, Kevin Stone, Peter Albert, Amjad Almahairi, Yasmine Babaei, Nikolay Bashlykov, Soumya Batra, Prajjwal Bhargava, Shruti Bhosale, Dan Bikel, Lukas Blecher, Cristian~Canton Ferrer, Moya Chen, Guillem Cucurull, David Esiobu, Jude Fernandes, Jeremy Fu, Wenyin Fu, Brian Fuller, Cynthia Gao, Vedanuj Goswami, Naman Goyal, Anthony Hartshorn, Saghar Hosseini, Rui Hou, Hakan Inan, Marcin Kardas, Viktor Kerkez, Madian Khabsa, Isabel Kloumann, Artem Korenev, Punit~Singh Koura, Marie-Anne Lachaux, Thibaut Lavril, Jenya Lee, Diana Liskovich, Yinghai Lu, Yuning Mao, Xavier Martinet, Todor Mihaylov, Pushkar Mishra, Igor Molybog, Yixin Nie, Andrew Poulton, Jeremy Reizenstein, Rashi Rungta, Kalyan Saladi, Alan Schelten, Ruan Silva, Eric~Michael Smith, Ranjan Subramanian, Xiaoqing~Ellen Tan, Binh Tang, Ross Taylor, Adina Williams, Jian~Xiang Kuan, Puxin Xu, Zheng Yan, Iliyan Zarov, Yuchen Zhang, Angela Fan, Melanie Kambadur, Sharan Narang, Aurelien Rodriguez, Robert Stojnic, Sergey Edunov, and Thomas
  Scialom.
\newblock Llama 2: Open foundation and fine-tuned chat models, 2023.

\bibitem{IOS}
Jack Wilson, David McLuskie, and Ethan Bayne.
\newblock Investigation into the security and privacy of ios vpn applications.
\newblock pages 1--9, 08 2020.
\newblock \url{https://dl.acm.org/doi/10.1145/3407023.3407029}.

\bibitem{hallucination}
Jia-Yu Yao, Kun-Peng Ning, Zhen-Hui Liu, Mu-Nan Ning, and Li~Yuan.
\newblock Llm lies: Hallucinations are not bugs, but features as adversarial examples, 2023.

\bibitem{Socia-Media-Article2}
Jan Youngren.
\newblock Who is dominating the rising vpn market in 2023?
\newblock 2023.
\newblock \url{https://vpnpro.com/blog/vpn-market-share-overview/}.

\bibitem{SEO}
Qing Zhang, David~Y. Wang, and Geoffrey~M. Voelker.
\newblock Dspin: Detecting automatically spun content on the web.
\newblock In {\em Network and Distributed System Security Symposium}, 2014.

\end{thebibliography}

\appendix

\section{Details of the VPN apps used} \label{sec:vpn-details}

\begin{table*}[t]
\footnotesize
\centering
{\color{\mytablecolor}
\caption{Twenty popular VPN apps were considered in our study, along with their number of downloads and reviews collected from the Google Play Store and Apple App Store.}
\label{tab:bestvpnapp}
\vspace{1mm}
\begin{tabular}{|c|c|c|c|c|c|}
\hline
\rowcolor[HTML]{C0C0C0} 
\textbf{VPN}          & \textbf{URL}                                       & \textbf{No. Of Downloads} & \textbf{No. Of Reviews} & \textbf{\begin{tabular}[c]{@{}c@{}}Available On \\ Apple Store\end{tabular}} & \textbf{\begin{tabular}[c]{@{}c@{}}Available On \\ Google Play Store\end{tabular}} \\ \hline
Hotspotshield         & https://www.hotspotshield.com/                     & 100 M+                    & 257,533                 & YES                                                                          & YES                                                                                \\ \hline
SuperVPN              & https://supervpn.im/                               & 100 M+                    & 419,788                 & YES                                                                          & YES                                                                                \\ \hline
TurboVPN              & https://turbovpn.com/                              & 100 M+                    & 75,926                  & YES                                                                          & YES                                                                                \\ \hline
Nord                  & https://nordvpn.com/                               & 5 M+                      & 120,237                 & YES                                                                          & YES                                                                                \\ \hline
VyprVPN               & https://www.vyprvpn.com/                           & 5 M+                      & 12,465                  & \textbf{NO}                                                                  & YES                                                                                \\ \hline
ExpressVPN            & https://www.tunnelvpn.com/                         & 1 M+                      & 59,284                  & YES                                                                          & YES                                                                                \\ \hline
Tunnelbear            & https://www.tunnelvpn.com/                         & 1 M+                      & 59,303                  & YES                                                                          & YES                                                                                \\ \hline
Cyberghost            & https://www.cyberghostvpn.com/en\_US/              & 1 M+                      & 31,209                  & YES                                                                          & YES                                                                                \\ \hline
Hola                  & https://hola.org/                                  & 1 M+                      & 82,657                  & YES                                                                          & YES                                                                                \\ \hline
Windscribe            & https://windscribe.com/                            & 1 M+                      & 14,710                  & YES                                                                          & YES                                                                                \\ \hline
Privateinternetaccess & https://ww.privateinternetaccess.com/              & 500 K+                    & 25,688                  & YES                                                                          & YES                                                                                \\ \hline
HMA                   & https://www.hidemyass.com/en-in/index              & 500 K+                    & 11.117                  & YES                                                                          & YES                                                                                \\ \hline
Surfshark             & https://surfshark.com/                             & 500 K+                    & 13,739                  & YES                                                                          & YES                                                                                \\ \hline
Zenmate               & https://zenmate.com/                               & 500 K+                    & 9,762                   & YES                                                                          & YES                                                                                \\ \hline
ProtonVPN             & https://protonvpn.com/                             & 500 K+                    & 8,176                   & YES                                                                          & YES                                                                                \\ \hline
FreedomeVPN           & https://www/f-secure.com/en/home/products/freedome & 500 K+                    & 8,985                   & YES                                                                          & YES                                                                                \\ \hline
IPVanish              & https://www.ipvanish.com/                          & 500 K+                    & 11,524                  & YES                                                                          & YES                                                                                \\ \hline
Surfeasy              & https://www.surfeasy.com/                          & 500 K+                    & 94,710                  & YES                                                                          & YES                                                                                \\ \hline
VPNUnlimited          & https://www/vpnunlimited.com/                      & 100 K+                    & 9,501                   & YES                                                                          & YES                                                                                \\ \hline
StrongVPN             & https://strongvpn.com/                             & 50 K+                     & 1,725                   & YES                                                                          & YES                                                                                \\ \hline
\end{tabular}
}
\end{table*}

\begin{table*}[]
\footnotesize
\centering
{\color{\mytablecolor}
\caption{Total number and percentage of VPN-switching reviews for different VPNs where VPNs are sorted by the number of downloads. There is a wide variation.}
\label{tab:switches}
\vspace{1mm}
\begin{tabular}{|c|c|c|c|c|}
\hline
\rowcolor[HTML]{E7E7E7} 
\textbf{VPN}            & \textbf{\begin{tabular}[c]{@{}c@{}}Number of\\ Downloads\end{tabular}} & \textbf{\begin{tabular}[c]{@{}c@{}}Number of\\ VPN-switching Reviews \end{tabular}} & \textbf{\begin{tabular}[c]{@{}c@{}}Total Number \\ of\\ Reviews\end{tabular}} & \textbf{\begin{tabular}[c]{@{}c@{}}\% of \\ VPN-switching Reviews\end{tabular}} \\ \hline
\rowcolor[HTML]{FFFFFF} 
Turbo VPN               & 100M+                                                                  & 2,557                                                                                    & 75,926                                                                        & {3.4}                                                                                \\ \hline
\rowcolor[HTML]{FFFFFF} 
Super VPN               & 100M+                                                                  & 8,570                                                                                 & 419,788                                                                       & {2.0}                                                                                \\ \hline
\rowcolor[HTML]{FFFFFF} 
Hotspot shield           & 100M+                                                                  & 27,828                                                                                   & 257,533                                                                       & {10.8}                                                                                \\ \hline
\rowcolor[HTML]{FFFFFF} 
Nord VPN                & 5M+                                                                    & 41,280                                                                                   & 120,237                                                                       & \textbf{34.3}                                                                                         \\ \hline
\rowcolor[HTML]{FFFFFF} 
Vypr VPN                & 5M+                                                                    & 2,308                                                                                    & 12,465                                                                        & \textbf{18.5}                                                                                         \\ \hline
\rowcolor[HTML]{FFFFFF} 
Hola                    & 1M+                                                                    & 7,310                                                                                   & 82,657                                                                        & {8.8}                                                                                \\ \hline
\rowcolor[HTML]{FFFFFF} 
Express                 & 1M+                                                                    & 21,062                                                                                   & 59,284                                                                        & \textbf{35.5}                                                                                         \\ \hline
\rowcolor[HTML]{FFFFFF} 
Cyberghost              & 1M+                                                                    & 7,546                                                                                    & 31,209                                                                        & 24.2                                                                                         \\ \hline
\rowcolor[HTML]{FFFFFF} 
Windscribe              & 1M+                                                                    & 3,358                                                                                    & 14,710                                                                        & 22.8                                                                                         \\ \hline
\rowcolor[HTML]{FFFFFF} 
Tunnelbear              & 1M+                                                                    & 15,897                                                                                   & 59,303                                                                        & 26.8                                                                                         \\ \hline
\rowcolor[HTML]{FFFFFF} 
IPVanish                & 500K+                                                                  & 5,076                                                                                    & 11,524                                                                        & \textbf{44.0}                                                                                         \\ \hline
\rowcolor[HTML]{FFFFFF} 
HideMyAss               & 500K+                                                                  & 2,034                                                                                    & 11,117                                                                        & 18.3                                                                                        \\ \hline
\rowcolor[HTML]{FFFFFF} 
Surfshark               & 500K+                                                                  & 5,534                                                                                   & 13,739                                                                        & \textbf{40.3}                                                                                         \\ \hline
\rowcolor[HTML]{FFFFFF} 
Private Internet Access & 500K+                                                                  & 10,351                                                                                   & 25,688                                                                        & \textbf{40.3}                                                                                         \\ \hline
\rowcolor[HTML]{FFFFFF} 
Proton VPN              & 500K+                                                                  & 3,164                                                                                    & 8,176                                                                         & \textbf{38.7}                                                                                         \\ \hline
\rowcolor[HTML]{FFFFFF} 
Zenmate VPN             & 500K+                                                                  & 1,892                                                                                   & 9,762                                                                         & 19.4                                                                                         \\ \hline
\rowcolor[HTML]{FFFFFF} 
Surfeasy                & 500K+                                                                  & 10,433                                                                                   & 94,710                                                                        & {11.0}                                                                                \\ \hline
\rowcolor[HTML]{FFFFFF} 
Freedome VPN            & 500K+                                                                  & 2,464                                                                                    & 8,985                                                                         & 27.4                                                                                          \\ \hline
\rowcolor[HTML]{FFFFFF} 
VPN unlimited           & 100K+                                                                  & 5,885                                                                                   & 9,501                                                                         & \textbf{61.9}                                                                                   \\ \hline
\rowcolor[HTML]{FFFFFF} 
Strong VPN              & 50K+                                                                   & 850                                                                                      & 1,725                                                                         & \textbf{49.3}                                                                                   \\ \hline
\end{tabular}
}
\end{table*}

In this section, we have provided the details of the twenty popular VPN apps used in this paper. Table~\ref{tab:bestvpnapp} presents the selected VPN apps, the number of downloads on Google Play Store and reviews (on both Google Play Store and Apple App Store). Table~\ref{tab:switches} presents the percentage of VPN-switching reviews posted for each app.

\section{Keywords used to identify relevant reviews}\label{sec:review-keywords}

The keywords used in identifying reviews potentially corresponding to VPN-switching have been listed in table ~\ref{tab:keywords}.

\begin{table*}[]
\centering
\footnotesize
\caption{Keyword used in filtering the VPNs corresponding cases where users switched from one VPN to another. To filter reviews, we performed a keyword search of these keywords in reviews VPNs other than their corresponding VPN}
\label{tab:keywords}
\vspace{1mm}
\begin{tabular}{|l|l|l|l|}
\hline
\rowcolor[HTML]{E7E7E7} 
\multicolumn{1}{|c|}{\cellcolor[HTML]{E7E7E7}\textbf{Keyword}} &
  \textbf{Corresponding VPN} &
  \multicolumn{1}{c|}{\cellcolor[HTML]{E7E7E7}\textbf{Keyword}} &
  \textbf{Corresponding VPN} \\ \hline
hotspotshield & Hotspotshield & nord                    & Nord          \\ \hline
hola          & Hola          & private internet access & PIA           \\ \hline
expressvpn    & Express VPN   & privateinternetaccess   & PIA           \\ \hline
express       & Express VPN   & cyberghost              & Cyberghost    \\ \hline
strong vpn    & Strong VPN    & surfshark               & Surfshark     \\ \hline
strongvpn     & Strong VPN    & turbo                   & Turbo VPN     \\ \hline
ipvanish      & IPVanish      & supervpn                & Super VPN     \\ \hline
tunnelbear    & TunnelBear    & super vpn               & Super VPN     \\ \hline
hidemyass     & HideMyAss     & vpn unlimited           & VPN Unlimited \\ \hline
hma           & HideMyAss     & vpn\_unlimited          & VPN Unlimited \\ \hline
windscribe    & Windscribe    & proton                  & Proton VPN    \\ \hline
vypr          & Vypr          & surfeasy                & Surfeasy VPN  \\ \hline
freedome      & Freedome VPN  & zenmate                 & Zenmate VPN   \\ \hline
\end{tabular}
\end{table*}

\section{Hierarchy of themes}\label{sec:frequencies-theme}
In table \ref{tab:frequencies}, we summarize the overall hierarchy of reviews for each higher-level theme presented in the section \ref{sec:reasons}.

\begin{table*}[]
\centering
\footnotesize
\color{\mytablecolor}
\caption{The hierarchy of themes derived from open coding of reviews in the category of "Software". Here, F1 refers to the number of reviews in which this theme appeared in our manual analysis, and F2 refers to the number of reviews in which the theme was identified by the automated approach.}
\label{tab:frequencies}
\begin{tabular}{|
>{\columncolor[HTML]{EFEFEF}}l |l|l|l||
>{\columncolor[HTML]{EFEFEF}}l |l|l|l|}
\hline
\cellcolor[HTML]{C0C0C0}\textbf{Theme} & \cellcolor[HTML]{C0C0C0}\textbf{Hierarchy} & \cellcolor[HTML]{C0C0C0}\textbf{F1} & \cellcolor[HTML]{C0C0C0}\textbf{F2} & \cellcolor[HTML]{C0C0C0}\textbf{Theme}                          & \cellcolor[HTML]{C0C0C0}\textbf{Hierarchy} & \cellcolor[HTML]{C0C0C0}\textbf{F1} & \cellcolor[HTML]{C0C0C0}\textbf{F2} \\\hline
Network                                & Table \ref{tab:network-frequency}                                    & 445                                 & 33,301                              & \begin{tabular}[c]{@{}l@{}}Security and \\ privacy\end{tabular} & Table \ref{tab:security-frequency}                                    & 196                                 & 21,425                              \\\hline
User Interaction                       & Table \ref{tab:user-freind-frequency}                                    & 575                                 & 94,263                              & Financial Aspects                                               & Table \ref{tab:finance-frequency}                                    & 294                                 & 29,106                              \\\hline
Connection                             & Table \ref{tab:connection-heirarchy}                                    & 244                                 & 41,020                              & \begin{tabular}[c]{@{}l@{}}Dynamic Data\\ Services\end{tabular} & Table \ref{tab:dds-frequency}                                    & 116                                 & 16,242                              \\\hline
Geography                              & -                                    & 88                                  & 12,487                              & Software                                                        & Table \ref{tab:software-frequency}                                    & 275                                 & 36,241   \\\hline                          
\end{tabular}
\end{table*}

\begin{table*}[]
\color{\mytablecolor}
\caption{The hierarchy of themes derived from open coding of reviews in the category of ``Network''. Here, F1 refers to the number of reviews in which this theme appeared in our manual analysis, and F2 refers to the number of reviews in which this theme was identified by the automated approach.}
\label{tab:network-frequency}
\centering
\footnotesize
\begin{tabular}{|l|l|l|l|l|l|lll}
\hline
\rowcolor[HTML]{C0C0C0} 
\textbf{Theme}                                  & \textbf{F1}              & \textbf{F2}              & \textbf{Theme}                                             & \textbf{F1}              & \textbf{F2}              & \multicolumn{1}{l|}{\cellcolor[HTML]{C0C0C0}\textbf{Theme}}                      & \multicolumn{1}{l|}{\cellcolor[HTML]{C0C0C0}\textbf{F1}} & \multicolumn{1}{l|}{\cellcolor[HTML]{C0C0C0}\textbf{F2}} \\ \hline
\cellcolor[HTML]{CBCEFB}A.1 Network Speed       & 339                      & 28,406                   & \cellcolor[HTML]{CBCEFB}A.2 Server Count                   & 68                       & 4,607                    & \multicolumn{1}{l|}{\cellcolor[HTML]{C0C0C0}A.3.4 Issues with country selection} & \multicolumn{1}{l|}{\cellcolor[HTML]{C0C0C0}}            & \multicolumn{1}{l|}{\cellcolor[HTML]{C0C0C0}}            \\ \hline
\cellcolor[HTML]{C0C0C0}A.1.1 Network Speed     & \cellcolor[HTML]{C0C0C0} & \cellcolor[HTML]{C0C0C0} & \cellcolor[HTML]{C0C0C0}A.2.1 Low server count             & \cellcolor[HTML]{C0C0C0} & \cellcolor[HTML]{C0C0C0} & \multicolumn{1}{l|}{\cellcolor[HTML]{CBCEFB}A.4 Network Protocol}                & \multicolumn{1}{l|}{11}                                  & \multicolumn{1}{l|}{102}                                 \\ \hline
\rowcolor[HTML]{C0C0C0} 
A.1.1.1 Speed decreased with time               &                          &                          & A2.2 High server count                                     &                          &                          & \multicolumn{1}{l|}{\cellcolor[HTML]{C0C0C0}A.4.1 Port Fowarding}                & \multicolumn{1}{l|}{\cellcolor[HTML]{C0C0C0}}            & \multicolumn{1}{l|}{\cellcolor[HTML]{C0C0C0}}            \\ \hline
A.1.1.2 Fast network speed                      & \cellcolor[HTML]{C0C0C0} & \cellcolor[HTML]{C0C0C0} & \cellcolor[HTML]{CBCEFB}A.3 Server switching               & 27                       & 1,806                    & \multicolumn{1}{l|}{\cellcolor[HTML]{C0C0C0}A.4.2 Support for IPv6}              & \multicolumn{1}{l|}{\cellcolor[HTML]{C0C0C0}}            & \multicolumn{1}{l|}{\cellcolor[HTML]{C0C0C0}}            \\ \hline
A.1.1.3 Slow network speed                      & \cellcolor[HTML]{C0C0C0} & \cellcolor[HTML]{C0C0C0} & \cellcolor[HTML]{C0C0C0}A.3.1 Ease of Switching Server     & \cellcolor[HTML]{C0C0C0} & \cellcolor[HTML]{C0C0C0} & \multicolumn{1}{l|}{\cellcolor[HTML]{C0C0C0}A.4.3 Wireguard}                     & \multicolumn{1}{l|}{\cellcolor[HTML]{C0C0C0}}            & \multicolumn{1}{l|}{\cellcolor[HTML]{C0C0C0}}            \\ \hline
A.1.1.4 No Buffering                            & \cellcolor[HTML]{C0C0C0} & \cellcolor[HTML]{C0C0C0} & \cellcolor[HTML]{C0C0C0}A.3.1.1 Easy Switching Server      & \cellcolor[HTML]{C0C0C0} & \cellcolor[HTML]{C0C0C0} & \multicolumn{1}{l|}{\cellcolor[HTML]{C0C0C0}A.4.4 Issues with TCP/UDP}           & \multicolumn{1}{l|}{\cellcolor[HTML]{C0C0C0}}            & \multicolumn{1}{l|}{\cellcolor[HTML]{C0C0C0}}            \\ \hline
\cellcolor[HTML]{C0C0C0}A.1.2 Ping Value        & \cellcolor[HTML]{C0C0C0} & \cellcolor[HTML]{C0C0C0} & A.3.1.2 Switching Server Complex                           & \cellcolor[HTML]{C0C0C0} & \cellcolor[HTML]{C0C0C0} &                                                                                  &                                                          &                                                          \\ \cline{1-6}
\cellcolor[HTML]{C0C0C0}A.1.2.1 High Ping Value & \cellcolor[HTML]{C0C0C0} & \cellcolor[HTML]{C0C0C0} & \cellcolor[HTML]{C0C0C0}A.3.2 Manual switching required    & \cellcolor[HTML]{C0C0C0} & \cellcolor[HTML]{C0C0C0} &                                                                                  &                                                          &                                                          \\ \cline{1-6}
A.1.2.2 Low Ping Value                          & \cellcolor[HTML]{C0C0C0} & \cellcolor[HTML]{C0C0C0} & \cellcolor[HTML]{C0C0C0}A.3.3 Issues with server selection & \cellcolor[HTML]{C0C0C0} & \cellcolor[HTML]{C0C0C0} &                                                                                  &                                                          &                                                          \\ \cline{1-6}
\end{tabular}
\end{table*}

\begin{table*}[]
\centering
\footnotesize
\color{\mytablecolor}
\caption{The hierarchy of themes derived from open coding of reviews in the category of ``User-friendliness''. Here, F1 refers to the number of reviews in which this theme appeared in our manual analysis, and F2 refers to the number of reviews in which this theme was identified by the automated approach.}
\label{tab:user-freind-frequency}
\begin{tabular}{|l|l|l|l|l|l|lll}
\hline
\rowcolor[HTML]{C0C0C0} 
\textbf{Theme}                                                                                     & \textbf{F1}              & \textbf{F2}              & \textbf{Theme}                                                                                  & \textbf{F1}              & \textbf{F2}              & \multicolumn{1}{l|}{\cellcolor[HTML]{C0C0C0}\textbf{Theme}}         & \multicolumn{1}{l|}{\cellcolor[HTML]{C0C0C0}\textbf{F1}} & \multicolumn{1}{l|}{\cellcolor[HTML]{C0C0C0}\textbf{F2}} \\ \hline
\cellcolor[HTML]{CBCEFB}B.1 Assistance                                                             & 151                      & 15,222                   & \cellcolor[HTML]{C0C0C0}B.3.2 General UI issues                                                 & \cellcolor[HTML]{C0C0C0} & \cellcolor[HTML]{C0C0C0} & \multicolumn{1}{l|}{\cellcolor[HTML]{CBCEFB}B.6 Reliability}        & \multicolumn{1}{l|}{44}                                  & \multicolumn{1}{l|}{4,895}                               \\ \hline
\cellcolor[HTML]{C0C0C0}B.1.1 Technical support                                                    & \cellcolor[HTML]{C0C0C0} & \cellcolor[HTML]{C0C0C0} & B.3.2.1 Bad UI                                                                                  & \cellcolor[HTML]{C0C0C0} & \cellcolor[HTML]{C0C0C0} & \multicolumn{1}{l|}{\cellcolor[HTML]{C0C0C0}B.6.1 Unreliable Owner} & \multicolumn{1}{l|}{\cellcolor[HTML]{C0C0C0}}            & \multicolumn{1}{l|}{\cellcolor[HTML]{C0C0C0}}            \\ \hline
B.1.1.1 Good tech support                                                                          & \cellcolor[HTML]{C0C0C0} & \cellcolor[HTML]{C0C0C0} & B.3.2.2 Buggy UI                                                                                & \cellcolor[HTML]{C0C0C0} & \cellcolor[HTML]{C0C0C0} & \multicolumn{1}{l|}{\cellcolor[HTML]{C0C0C0}B.6.2 Reliable app}     & \multicolumn{1}{l|}{\cellcolor[HTML]{C0C0C0}}            & \multicolumn{1}{l|}{\cellcolor[HTML]{C0C0C0}}            \\ \hline
B.1.1.2 Bad tech support                                                                           & \cellcolor[HTML]{C0C0C0} & \cellcolor[HTML]{C0C0C0} & B.3.2.3 Good UI                                                                                 & \cellcolor[HTML]{C0C0C0} & \cellcolor[HTML]{C0C0C0} & \multicolumn{1}{l|}{\cellcolor[HTML]{C0C0C0}B.6.3 Unreliable app}   & \multicolumn{1}{l|}{\cellcolor[HTML]{C0C0C0}}            & \multicolumn{1}{l|}{\cellcolor[HTML]{C0C0C0}}            \\ \hline
\cellcolor[HTML]{C0C0C0}B.1.2 Customer support                                                     & \cellcolor[HTML]{C0C0C0} & \cellcolor[HTML]{C0C0C0} & \cellcolor[HTML]{C0C0C0}B.3.3 General UX issues                                                 & \cellcolor[HTML]{C0C0C0} & \cellcolor[HTML]{C0C0C0} & \multicolumn{1}{l|}{\cellcolor[HTML]{CBCEFB}B.7 Reputation}         & \multicolumn{1}{l|}{33}                                  & \multicolumn{1}{l|}{1,955}                               \\ \hline
B.1.2.1 Good customer support                                                                      & \cellcolor[HTML]{C0C0C0} & \cellcolor[HTML]{C0C0C0} & B.3.3.1 Login/out Issues                                                                        & \cellcolor[HTML]{C0C0C0} & \cellcolor[HTML]{C0C0C0} & \multicolumn{1}{l|}{\cellcolor[HTML]{C0C0C0}B.7.1 Branding}         & \multicolumn{1}{l|}{\cellcolor[HTML]{C0C0C0}}            & \multicolumn{1}{l|}{\cellcolor[HTML]{C0C0C0}}            \\ \hline
B.1.2.2 Bad customer support                                                                       & \cellcolor[HTML]{C0C0C0} & \cellcolor[HTML]{C0C0C0} & B.3.3.2 App Freezes                                                                             & \cellcolor[HTML]{C0C0C0} & \cellcolor[HTML]{C0C0C0} & \multicolumn{1}{l|}{\cellcolor[HTML]{C0C0C0}B.7.2 Online reviews}   & \multicolumn{1}{l|}{\cellcolor[HTML]{C0C0C0}}            & \multicolumn{1}{l|}{\cellcolor[HTML]{C0C0C0}}            \\ \hline
B.1.2.3 Doesn't care about users                                                                   & \cellcolor[HTML]{C0C0C0} & \cellcolor[HTML]{C0C0C0} & \begin{tabular}[c]{@{}l@{}}B.3.3.3 Server info\\ unavailable\end{tabular}                       & \cellcolor[HTML]{C0C0C0} & \cellcolor[HTML]{C0C0C0} & \multicolumn{1}{l|}{B.7.2.1 Bad online reviews}                     & \multicolumn{1}{l|}{\cellcolor[HTML]{C0C0C0}}            & \multicolumn{1}{l|}{\cellcolor[HTML]{C0C0C0}}            \\ \hline
\cellcolor[HTML]{CBCEFB}B.2 Services Provided                                                      & 119                      & 21,689                   & B.3.3.4 Annoying notifications                                                                  & \cellcolor[HTML]{C0C0C0} & \cellcolor[HTML]{C0C0C0} & \multicolumn{1}{l|}{B.7.2.2 Good online reviews}                    & \multicolumn{1}{l|}{\cellcolor[HTML]{C0C0C0}}            & \multicolumn{1}{l|}{\cellcolor[HTML]{C0C0C0}}            \\ \hline
\cellcolor[HTML]{C0C0C0}B.2.1 Amount of data available                                             & \cellcolor[HTML]{C0C0C0} & \cellcolor[HTML]{C0C0C0} & B.3.3.5 More widgets required                                                                   & \cellcolor[HTML]{C0C0C0} & \cellcolor[HTML]{C0C0C0} & \multicolumn{1}{l|}{B.7.2.3 App overrated}                          & \multicolumn{1}{l|}{\cellcolor[HTML]{C0C0C0}}            & \multicolumn{1}{l|}{\cellcolor[HTML]{C0C0C0}}            \\ \hline
B.2.1.1 More data available                                                                        & \cellcolor[HTML]{C0C0C0} & \cellcolor[HTML]{C0C0C0} & B.3.3.6 Takes time to start up                                                                  & \cellcolor[HTML]{C0C0C0} & \cellcolor[HTML]{C0C0C0} & \multicolumn{1}{l|}{B.7.2.4 Used by influencers}                    & \multicolumn{1}{l|}{\cellcolor[HTML]{C0C0C0}}            & \multicolumn{1}{l|}{\cellcolor[HTML]{C0C0C0}}            \\ \hline
B.2.1.2 Less data available                                                                        & \cellcolor[HTML]{C0C0C0} & \cellcolor[HTML]{C0C0C0} & \begin{tabular}[c]{@{}l@{}}B.3.3.7 App stops other background \\ apps from working\end{tabular} & \cellcolor[HTML]{C0C0C0} & \cellcolor[HTML]{C0C0C0} & \multicolumn{1}{l|}{\cellcolor[HTML]{C0C0C0}B.7.3 History}          & \multicolumn{1}{l|}{\cellcolor[HTML]{C0C0C0}}            & \multicolumn{1}{l|}{\cellcolor[HTML]{C0C0C0}}            \\ \hline
\cellcolor[HTML]{C0C0C0}B.2.2 Support for multiple devices                                         & \cellcolor[HTML]{C0C0C0} & \cellcolor[HTML]{C0C0C0} & B.3.3.8 Constant turn on/off reqd                                                               & \cellcolor[HTML]{C0C0C0} & \cellcolor[HTML]{C0C0C0} & \multicolumn{1}{l|}{B.7.3.1 Shady past}                             & \multicolumn{1}{l|}{\cellcolor[HTML]{C0C0C0}}            & \multicolumn{1}{l|}{\cellcolor[HTML]{C0C0C0}}            \\ \hline
\begin{tabular}[c]{@{}l@{}}B.2.2.1 Support many multiple devices \\ simultaneously\end{tabular}    & \cellcolor[HTML]{C0C0C0} & \cellcolor[HTML]{C0C0C0} & \cellcolor[HTML]{CBCEFB}B.4 User friendliness                                                   & 57                       & 22,402                   & \multicolumn{1}{l|}{B.7.3.2 Flawless past}                          & \multicolumn{1}{l|}{\cellcolor[HTML]{C0C0C0}}            & \multicolumn{1}{l|}{\cellcolor[HTML]{C0C0C0}}            \\ \hline
\begin{tabular}[c]{@{}l@{}}B.2.2.2 Doesn't support multiple devices \\ simultaneously\end{tabular} & \cellcolor[HTML]{C0C0C0} & \cellcolor[HTML]{C0C0C0} & \cellcolor[HTML]{C0C0C0}B.4.1 Ease of setup                                                     & \cellcolor[HTML]{C0C0C0} & \cellcolor[HTML]{C0C0C0} &                                                                     &                                                          &                                                          \\ \cline{1-6}
\begin{tabular}[c]{@{}l@{}}B.2.2.3 Support less multiple devices \\ simultaneously\end{tabular}    & \cellcolor[HTML]{C0C0C0} & \cellcolor[HTML]{C0C0C0} & \cellcolor[HTML]{C0C0C0}B.4.2 Ease of use                                                       & \cellcolor[HTML]{C0C0C0} & \cellcolor[HTML]{C0C0C0} &                                                                     &                                                          &                                                          \\ \cline{1-6}
\cellcolor[HTML]{C0C0C0}B.2.3 Cross OS support                                                     & \cellcolor[HTML]{C0C0C0} & \cellcolor[HTML]{C0C0C0} & \cellcolor[HTML]{C0C0C0}B.4.3 Accessibility                                                     & \cellcolor[HTML]{C0C0C0} & \cellcolor[HTML]{C0C0C0} &                                                                     &                                                          &                                                          \\ \cline{1-6}
\cellcolor[HTML]{C0C0C0}B.2.4 Compatibility with environment                                       & \cellcolor[HTML]{C0C0C0} & \cellcolor[HTML]{C0C0C0} & \cellcolor[HTML]{CBCEFB}B.5 Decient                                                             & 42                       & 8,694                    &                                                                     &                                                          &                                                          \\ \cline{1-6}
B.2.4.1 Incompatible with environment                                                              & \cellcolor[HTML]{C0C0C0} & \cellcolor[HTML]{C0C0C0} & \cellcolor[HTML]{C0C0C0}B.5.1 Removal of app reviews                                            & \cellcolor[HTML]{C0C0C0} & \cellcolor[HTML]{C0C0C0} &                                                                     &                                                          &                                                          \\ \cline{1-6}
B.2.4.2 Compatible with environment                                                                & \cellcolor[HTML]{C0C0C0} & \cellcolor[HTML]{C0C0C0} & \cellcolor[HTML]{C0C0C0}B.5.2 Financial fraud                                                   & \cellcolor[HTML]{C0C0C0} & \cellcolor[HTML]{C0C0C0} &                                                                     &                                                          &                                                          \\ \cline{1-6}
\cellcolor[HTML]{C0C0C0}B.2.5 Issues with third-party apps                                         & \cellcolor[HTML]{C0C0C0} & \cellcolor[HTML]{C0C0C0} & \begin{tabular}[c]{@{}l@{}}B.5.2.1 Not working even after \\ payment\end{tabular}               & \cellcolor[HTML]{C0C0C0} & \cellcolor[HTML]{C0C0C0} &                                                                     &                                                          &                                                          \\ \cline{1-6}
B.2.5.1 Hotspot not working                                                                        & \cellcolor[HTML]{C0C0C0} & \cellcolor[HTML]{C0C0C0} & B.5.2.2 Money taken without permission                                                          & \cellcolor[HTML]{C0C0C0} & \cellcolor[HTML]{C0C0C0} &                                                                     &                                                          &                                                          \\ \cline{1-6}
B.2.5.2 Stops email from working                                                                   & \cellcolor[HTML]{C0C0C0} & \cellcolor[HTML]{C0C0C0} & B.5.2.3 Takes advantage of People                                                               & \cellcolor[HTML]{C0C0C0} & \cellcolor[HTML]{C0C0C0} &                                                                     &                                                          &                                                          \\ \cline{1-6}
B.2.5.3 Issues with WIFI                                                                           & \cellcolor[HTML]{C0C0C0} & \cellcolor[HTML]{C0C0C0} & B.5.2.4 Increasing price without info                                                           & \cellcolor[HTML]{C0C0C0} & \cellcolor[HTML]{C0C0C0} &                                                                     &                                                          &                                                          \\ \cline{1-6}
B.2.5.4 Office 365 web not working                                                                 & \cellcolor[HTML]{C0C0C0} & \cellcolor[HTML]{C0C0C0} & B.5.2.5 Difficult to unsubscribe                                                                & \cellcolor[HTML]{C0C0C0} & \cellcolor[HTML]{C0C0C0} &                                                                     &                                                          &                                                          \\ \cline{1-6}
\cellcolor[HTML]{C0C0C0}B.2.6 Support for always-on-VPN                                            & \cellcolor[HTML]{C0C0C0} & \cellcolor[HTML]{C0C0C0} & \cellcolor[HTML]{C0C0C0}B.5.3 False claims                                                      & \cellcolor[HTML]{C0C0C0} & \cellcolor[HTML]{C0C0C0} &                                                                     &                                                          &                                                          \\ \cline{1-6}
\cellcolor[HTML]{CBCEFB}B.3 UI/UX                                                                  & 129                      & 19,406                   & B.5.3.1 Claims free but isn't                                                                   & \cellcolor[HTML]{C0C0C0} & \cellcolor[HTML]{C0C0C0} &                                                                     &                                                          &                                                          \\ \cline{1-6}
\cellcolor[HTML]{C0C0C0}B.3.1 Mobile app                                                           & \cellcolor[HTML]{C0C0C0} & \cellcolor[HTML]{C0C0C0} & B.5.3.2 Does what it claims                                                                     & \cellcolor[HTML]{C0C0C0} & \cellcolor[HTML]{C0C0C0} &                                                                     &                                                          &                                                          \\ \cline{1-6}
B.3.1.1 Bad mobile app                                                                             & \cellcolor[HTML]{C0C0C0} & \cellcolor[HTML]{C0C0C0} & \cellcolor[HTML]{C0C0C0}B.5.4 User data usage                                                   & \cellcolor[HTML]{C0C0C0} & \cellcolor[HTML]{C0C0C0} &                                                                     &                                                          &                                                          \\ \cline{1-6}
B.3.1.2 Good Mobile App                                                                            & \cellcolor[HTML]{C0C0C0} & \cellcolor[HTML]{C0C0C0} & B.5.4.1 Exploits user data package                                                              & \cellcolor[HTML]{C0C0C0} & \cellcolor[HTML]{C0C0C0} &                                                                     &                                                          &                                                          \\ \cline{1-6}
B.3.1.3 App oversimplified                                                                         & \cellcolor[HTML]{C0C0C0} & \cellcolor[HTML]{C0C0C0} & B.5.4.2 Doesn't exploit user data package                                                       & \cellcolor[HTML]{C0C0C0} & \cellcolor[HTML]{C0C0C0} &                                                                     &                                                          &                                                          \\ \cline{1-6}
\end{tabular}
\end{table*}

\begin{table*}[]
\color{\mytablecolor}
\centering
\footnotesize
\caption{The hierarchy of themes derived from open coding of reviews in the category of ``Security \& Privacy''. Here, F1 refers to the number of reviews in which this theme appeared in our manual analysis, and F2 refers to the number of reviews in which this theme was identified by the automated approach.}
\label{tab:security-frequency}
\begin{tabular}{|l|l|l|l|l|l|lll}
\hline
\rowcolor[HTML]{C0C0C0} 
\multicolumn{1}{|c|}{\cellcolor[HTML]{C0C0C0}\textbf{Themes}} & \multicolumn{1}{c|}{\cellcolor[HTML]{C0C0C0}\textbf{F1}} & \multicolumn{1}{c|}{\cellcolor[HTML]{C0C0C0}\textbf{F2}} & \multicolumn{1}{c|}{\cellcolor[HTML]{C0C0C0}\textbf{Themes}}                       & \multicolumn{1}{c|}{\cellcolor[HTML]{C0C0C0}\textbf{F1}} & \multicolumn{1}{c|}{\cellcolor[HTML]{C0C0C0}\textbf{F2}} & \multicolumn{1}{c|}{\cellcolor[HTML]{C0C0C0}\textbf{Themes}}                                                 & \multicolumn{1}{c|}{\cellcolor[HTML]{C0C0C0}\textbf{F1}} & \multicolumn{1}{c|}{\cellcolor[HTML]{C0C0C0}\textbf{F2}} \\ \hline
\cellcolor[HTML]{CBCEFB}E.1 Security                          & 152                                                      & 15,564                                                   & \begin{tabular}[c]{@{}l@{}}E.1.5.1 Unpreferred headquarter \\ country\end{tabular} & \cellcolor[HTML]{C0C0C0}                                 & \cellcolor[HTML]{C0C0C0}                                 & \multicolumn{1}{l|}{\cellcolor[HTML]{C0C0C0}E.2.2 Usage access Permission}                                   & \multicolumn{1}{l|}{\cellcolor[HTML]{C0C0C0}}            & \multicolumn{1}{l|}{\cellcolor[HTML]{C0C0C0}}            \\ \hline
\cellcolor[HTML]{C0C0C0}E.1.1 Encryption standard             & \cellcolor[HTML]{C0C0C0}                                 & \cellcolor[HTML]{C0C0C0}                                 & E.1.5.2 Preferred headquarter country                                              & \cellcolor[HTML]{C0C0C0}                                 & \cellcolor[HTML]{C0C0C0}                                 & \multicolumn{1}{l|}{\begin{tabular}[c]{@{}l@{}}E.2.2.1 Usage access permission \\ required\end{tabular}}     & \multicolumn{1}{l|}{\cellcolor[HTML]{C0C0C0}}            & \multicolumn{1}{l|}{\cellcolor[HTML]{C0C0C0}}            \\ \hline
E.1.1.1 Good Encryption                                       & \cellcolor[HTML]{C0C0C0}                                 & \cellcolor[HTML]{C0C0C0}                                 & \cellcolor[HTML]{C0C0C0}E.1.6 User activity logs                                   & \cellcolor[HTML]{C0C0C0}                                 & \cellcolor[HTML]{C0C0C0}                                 & \multicolumn{1}{l|}{\begin{tabular}[c]{@{}l@{}}E.2.2.2 Usage access permission not \\ required\end{tabular}} & \multicolumn{1}{l|}{\cellcolor[HTML]{C0C0C0}}            & \multicolumn{1}{l|}{\cellcolor[HTML]{C0C0C0}}            \\ \hline
E.1.1.2 Substandard Encryption                                & \cellcolor[HTML]{C0C0C0}                                 & \cellcolor[HTML]{C0C0C0}                                 & E.1.6.1 Sells users activity logs                                                  & \cellcolor[HTML]{C0C0C0}                                 & \cellcolor[HTML]{C0C0C0}                                 & \multicolumn{1}{l|}{\cellcolor[HTML]{C0C0C0}E.2.3 Privacy policy of VPN}                                     & \multicolumn{1}{l|}{\cellcolor[HTML]{C0C0C0}}            & \multicolumn{1}{l|}{\cellcolor[HTML]{C0C0C0}}            \\ \hline
\cellcolor[HTML]{C0C0C0}E.1.2 App security                    & \cellcolor[HTML]{C0C0C0}                                 & \cellcolor[HTML]{C0C0C0}                                 & E1.6.2 Doesn't sells user logs                                                     & \cellcolor[HTML]{C0C0C0}                                 & \cellcolor[HTML]{C0C0C0}                                 & \multicolumn{1}{l|}{E.2.3.1 Good privacy policy}                                                             & \multicolumn{1}{l|}{\cellcolor[HTML]{C0C0C0}}            & \multicolumn{1}{l|}{\cellcolor[HTML]{C0C0C0}}            \\ \hline
E.1.2.1 Virus in app                                          & \cellcolor[HTML]{C0C0C0}                                 & \cellcolor[HTML]{C0C0C0}                                 & E.1.6.3 Keeps user activity logs                                                   & \cellcolor[HTML]{C0C0C0}                                 & \cellcolor[HTML]{C0C0C0}                                 & \multicolumn{1}{l|}{E.2.3.2 Bad privacy policy}                                                              & \multicolumn{1}{l|}{\cellcolor[HTML]{C0C0C0}}            & \multicolumn{1}{l|}{\cellcolor[HTML]{C0C0C0}}            \\ \hline
E.1.2.2 Hides IP address                                      & \cellcolor[HTML]{C0C0C0}                                 & \cellcolor[HTML]{C0C0C0}                                 & E.1.6.4 No user activity  logs                                                     & \cellcolor[HTML]{C0C0C0}                                 & \cellcolor[HTML]{C0C0C0}                                 & \multicolumn{1}{l|}{\cellcolor[HTML]{C0C0C0}E.2.4 SSH issues}                                                & \multicolumn{1}{l|}{\cellcolor[HTML]{C0C0C0}}            & \multicolumn{1}{l|}{\cellcolor[HTML]{C0C0C0}}            \\ \hline
E.1.2.3 Doesn't hide IP                                       & \cellcolor[HTML]{C0C0C0}                                 & \cellcolor[HTML]{C0C0C0}                                 & \cellcolor[HTML]{C0C0C0}E.1.7 Kill switch                                          & \cellcolor[HTML]{C0C0C0}                                 & \cellcolor[HTML]{C0C0C0}                                 & \multicolumn{1}{l|}{\cellcolor[HTML]{C0C0C0}E.2.5 Obfuscated servers present}                                & \multicolumn{1}{l|}{\cellcolor[HTML]{C0C0C0}}            & \multicolumn{1}{l|}{\cellcolor[HTML]{C0C0C0}}            \\ \hline
E.1.2.4 Very secure app                                       & \cellcolor[HTML]{C0C0C0}                                 & \cellcolor[HTML]{C0C0C0}                                 & E.1.7.1 Kill switch works                                                          & \cellcolor[HTML]{C0C0C0}                                 & \cellcolor[HTML]{C0C0C0}                                 &                                                                                                              &                                                          &                                                          \\ \cline{1-6}
E.1.2.4 Insecure app                                          & \cellcolor[HTML]{C0C0C0}                                 & \cellcolor[HTML]{C0C0C0}                                 & E.1.7.2 Buggy kill switch                                                          & \cellcolor[HTML]{C0C0C0}                                 & \cellcolor[HTML]{C0C0C0}                                 &                                                                                                              &                                                          &                                                          \\ \cline{1-6}
\cellcolor[HTML]{C0C0C0}E.1.3 Server security                 & \cellcolor[HTML]{C0C0C0}                                 & \cellcolor[HTML]{C0C0C0}                                 & E.1.7.3 No kill switch                                                             & \cellcolor[HTML]{C0C0C0}                                 & \cellcolor[HTML]{C0C0C0}                                 &                                                                                                              &                                                          &                                                          \\ \cline{1-6}
\cellcolor[HTML]{C0C0C0}E.1.4 DNS leaks                       & \cellcolor[HTML]{C0C0C0}                                 & \cellcolor[HTML]{C0C0C0}                                 & \cellcolor[HTML]{CBCEFB}E.2 Privacy                                                & 44                                                       & 5,861                                                    &                                                                                                              &                                                          &                                                          \\ \cline{1-6}
\cellcolor[HTML]{C0C0C0}E.1.5 Headquarter country of VPN      & \cellcolor[HTML]{C0C0C0}                                 & \cellcolor[HTML]{C0C0C0}                                 & \cellcolor[HTML]{C0C0C0}E.2.1 RTC leaks                                            & \cellcolor[HTML]{C0C0C0}                                 & \cellcolor[HTML]{C0C0C0}                                 &                                                                                                              &                                                          &                                                          \\ \cline{1-6}
\end{tabular}
\end{table*}

\begin{table*}[]
\color{\mytablecolor}
\centering
\footnotesize
\caption{The hierarchy of themes derived from open coding of reviews in the category of ``Finance''. Here, F1 refers to the number of reviews in which this theme appeared in our manual analysis, and F2 refers to the number of reviews in which this theme was identified by the automated approach.}
\label{tab:finance-frequency}
\begin{tabular}{|l|l|l|lll}
\hline
\rowcolor[HTML]{C0C0C0} 
\multicolumn{1}{|c|}{\cellcolor[HTML]{C0C0C0}\textbf{Themes}}      & \multicolumn{1}{c|}{\cellcolor[HTML]{C0C0C0}\textbf{F1}} & \multicolumn{1}{c|}{\cellcolor[HTML]{C0C0C0}\textbf{F2}} & \multicolumn{1}{c|}{\cellcolor[HTML]{C0C0C0}\textbf{Themes}}                                                                                & \multicolumn{1}{c|}{\cellcolor[HTML]{C0C0C0}\textbf{F1}} & \multicolumn{1}{c|}{\cellcolor[HTML]{C0C0C0}\textbf{F2}} \\ \hline
\cellcolor[HTML]{CBCEFB}F.1 Payment                                & 25                                                       & 3,792                                                    & \multicolumn{1}{l|}{\cellcolor[HTML]{CBCEFB}F.4 Premium version of app}                                                                     & \multicolumn{1}{l|}{14}                                  & \multicolumn{1}{l|}{2,260}                               \\ \hline
\rowcolor[HTML]{C0C0C0} 
\cellcolor[HTML]{C0C0C0}F.1.1 Payment medium                       &                                                          &                                                          & \multicolumn{1}{l|}{\cellcolor[HTML]{C0C0C0}F.4.1 Premium not good}                                                                         & \multicolumn{1}{l|}{\cellcolor[HTML]{C0C0C0}}            & \multicolumn{1}{l|}{\cellcolor[HTML]{C0C0C0}}            \\ \hline
F.1.1.1 Bad payment system                                         & \cellcolor[HTML]{C0C0C0}                                 & \cellcolor[HTML]{C0C0C0}                                 & \multicolumn{1}{l|}{\cellcolor[HTML]{C0C0C0}F.4.2 Premium not worth it}                                                                     & \multicolumn{1}{l|}{\cellcolor[HTML]{C0C0C0}}            & \multicolumn{1}{l|}{\cellcolor[HTML]{C0C0C0}}            \\ \hline
F.1.1.2 Accepts cryptocurrency payment                             & \cellcolor[HTML]{C0C0C0}                                 & \cellcolor[HTML]{C0C0C0}                                 & \multicolumn{1}{l|}{\cellcolor[HTML]{C0C0C0}F.4.3 Worth the money}                                                                          & \multicolumn{1}{l|}{\cellcolor[HTML]{C0C0C0}}            & \multicolumn{1}{l|}{\cellcolor[HTML]{C0C0C0}}            \\ \hline
\cellcolor[HTML]{C0C0C0}F.1.2 Subscription options                 & \cellcolor[HTML]{C0C0C0}                                 & \cellcolor[HTML]{C0C0C0}                                 & \multicolumn{1}{l|}{\cellcolor[HTML]{CBCEFB}F.5 Free version of app}                                                                        & \multicolumn{1}{l|}{41}                                  & \multicolumn{1}{l|}{4,435}                               \\ \hline
F.1.2.1 More subscription options                                  & \cellcolor[HTML]{C0C0C0}                                 & \cellcolor[HTML]{C0C0C0}                                 & \multicolumn{1}{l|}{\cellcolor[HTML]{C0C0C0}\begin{tabular}[c]{@{}l@{}}F.5.1 Payment information required for \\ free app\end{tabular}}     & \multicolumn{1}{l|}{\cellcolor[HTML]{C0C0C0}}            & \multicolumn{1}{l|}{\cellcolor[HTML]{C0C0C0}}            \\ \hline
F.1.2.2 Less flexible plans                                        & \cellcolor[HTML]{C0C0C0}                                 & \cellcolor[HTML]{C0C0C0}                                 & \multicolumn{1}{l|}{\cellcolor[HTML]{C0C0C0}F.5.2 Limited features without premium}                                                         & \multicolumn{1}{l|}{\cellcolor[HTML]{C0C0C0}}            & \multicolumn{1}{l|}{\cellcolor[HTML]{C0C0C0}}            \\ \hline
\cellcolor[HTML]{CBCEFB}F.2 Price of VPN                           & 182                                                      & 15,862                                                   & \multicolumn{1}{l|}{\cellcolor[HTML]{C0C0C0}F.5.3 Free version of app is good}                                                              & \multicolumn{1}{l|}{\cellcolor[HTML]{C0C0C0}}            & \multicolumn{1}{l|}{\cellcolor[HTML]{C0C0C0}}            \\ \hline
\rowcolor[HTML]{C0C0C0} 
\cellcolor[HTML]{C0C0C0}F.2.1 High pricing                         &                                                          &                                                          & \multicolumn{1}{l|}{\cellcolor[HTML]{C0C0C0}F.5.4 Free version of App is bad}                                                               & \multicolumn{1}{l|}{\cellcolor[HTML]{C0C0C0}}            & \multicolumn{1}{l|}{\cellcolor[HTML]{C0C0C0}}            \\ \hline
\rowcolor[HTML]{C0C0C0} 
\cellcolor[HTML]{C0C0C0}F.2.2 Cheap VPN                            &                                                          &                                                          & \multicolumn{1}{l|}{\cellcolor[HTML]{C0C0C0}\begin{tabular}[c]{@{}l@{}}F.5.5 Payment information not required for \\ free app\end{tabular}} & \multicolumn{1}{l|}{\cellcolor[HTML]{C0C0C0}}            & \multicolumn{1}{l|}{\cellcolor[HTML]{C0C0C0}}            \\ \hline
\cellcolor[HTML]{CBCEFB}F.3 Refund                                 & 32                                                       & 2,757                                                    & \multicolumn{1}{l|}{\cellcolor[HTML]{C0C0C0}F.5.6 No free trial provided}                                                                   & \multicolumn{1}{l|}{\cellcolor[HTML]{C0C0C0}}            & \multicolumn{1}{l|}{\cellcolor[HTML]{C0C0C0}}            \\ \hline
\rowcolor[HTML]{C0C0C0} 
\cellcolor[HTML]{C0C0C0}{\color[HTML]{222222} F.3.1 Refund Issues} &                                                          &                                                          & \multicolumn{1}{l|}{\cellcolor[HTML]{C0C0C0}F.5.7 Free trial available}                                                                     & \multicolumn{1}{l|}{\cellcolor[HTML]{C0C0C0}}            & \multicolumn{1}{l|}{\cellcolor[HTML]{C0C0C0}}            \\ \hline
\cellcolor[HTML]{C0C0C0}F.3.2 Good refund policy                   & \cellcolor[HTML]{C0C0C0}                                 & \cellcolor[HTML]{C0C0C0}                                 &                                                                                                                                             &                                                          &                                                          \\ \cline{1-3}
\end{tabular}%
\end{table*}

\begin{table*}[]
\color{\mytablecolor}
\centering
\footnotesize
\caption{The hierarchy of themes derived from open coding of reviews in the category of ``Dynamic data services''. Here, F1 refers to the number of reviews in which this theme appeared in our manual analysis, and F2 refers to the number of reviews in which this theme was identified by the automated approach.}
\label{tab:dds-frequency}
\begin{tabular}{|l|l|l|l|l|l|}
\hline
\rowcolor[HTML]{C0C0C0} 
\multicolumn{1}{|c|}{\cellcolor[HTML]{C0C0C0}\textbf{Themes}} & \multicolumn{1}{c|}{\cellcolor[HTML]{C0C0C0}\textbf{F1}} & \multicolumn{1}{c|}{\cellcolor[HTML]{C0C0C0}\textbf{F2}} & \multicolumn{1}{c|}{\cellcolor[HTML]{C0C0C0}\textbf{Themes}} & \multicolumn{1}{c|}{\cellcolor[HTML]{C0C0C0}\textbf{F1}} & \multicolumn{1}{c|}{\cellcolor[HTML]{C0C0C0}\textbf{F2}} \\ \hline
\cellcolor[HTML]{CBCEFB}G.1 OTT                               & 51                                                       & 6,530                                                    & \cellcolor[HTML]{C0C0C0}G.3.1 Too many ads                   & \cellcolor[HTML]{C0C0C0}                                 & \cellcolor[HTML]{C0C0C0}                                 \\ \hline
\cellcolor[HTML]{CBCEFB}G.2 Games                             & 8                                                        & 1,410                                                    & \cellcolor[HTML]{C0C0C0}G.3.2 Less Ads                       & \cellcolor[HTML]{C0C0C0}                                 & \cellcolor[HTML]{C0C0C0}                                 \\ \hline
\cellcolor[HTML]{C0C0C0}G.2.1 Good for games                  & \cellcolor[HTML]{C0C0C0}                                 & \cellcolor[HTML]{C0C0C0}                                 & \cellcolor[HTML]{CBCEFB}G.4 Streaming                        & 34                                                       & 2,504                                                    \\ \hline
\rowcolor[HTML]{C0C0C0} 
\cellcolor[HTML]{C0C0C0}G.2.2 Bad for games                   &                                                          &                                                          & \cellcolor[HTML]{C0C0C0}G.4.1 Good for streaming             &                                                          &                                                          \\ \hline
\cellcolor[HTML]{CBCEFB}G.3 Ads                               & 23                                                       & 5,798                                                    & \cellcolor[HTML]{C0C0C0}G.4.2 Bad for streaming              & \cellcolor[HTML]{C0C0C0}                                 & \cellcolor[HTML]{C0C0C0}                                 \\ \hline
\end{tabular}
\end{table*}

\begin{table*}[]
\color{\mytablecolor}
\centering
\footnotesize
\caption{The hierarchy of themes derived from open coding of reviews in the category of ``Software''. Here, F1 refers to the number of reviews in which this theme appeared in our manual analysis, and F2 refers to the number of reviews in which this theme was identified by the automated approach.}
\label{tab:software-frequency}
\begin{tabular}{|l|l|l|l|l|l|lll}
\hline
\rowcolor[HTML]{C0C0C0} 
\multicolumn{1}{|c|}{\cellcolor[HTML]{C0C0C0}\textbf{Themes}}                                         & \multicolumn{1}{c|}{\cellcolor[HTML]{C0C0C0}\textbf{F1}} & \multicolumn{1}{c|}{\cellcolor[HTML]{C0C0C0}\textbf{F2}} & \multicolumn{1}{c|}{\cellcolor[HTML]{C0C0C0}\textbf{Themes}}                                             & \multicolumn{1}{c|}{\cellcolor[HTML]{C0C0C0}\textbf{F1}} & \multicolumn{1}{c|}{\cellcolor[HTML]{C0C0C0}\textbf{F2}} & \multicolumn{1}{c|}{\cellcolor[HTML]{C0C0C0}\textbf{Themes}}                                                               & \multicolumn{1}{c|}{\cellcolor[HTML]{C0C0C0}\textbf{F1}} & \multicolumn{1}{c|}{\cellcolor[HTML]{C0C0C0}\textbf{F2}} \\ \hline
\cellcolor[HTML]{CBCEFB}H.1 Updates                                                                   & 59                                                       & 11,360                                                   & H.2.2.6 Issues with P2P downloads                                                                        & \cellcolor[HTML]{C0C0C0}                                 & \cellcolor[HTML]{C0C0C0}                                 & \multicolumn{1}{l|}{\cellcolor[HTML]{CBCEFB}H.5 Resource consumption}                                                      & \multicolumn{1}{l|}{40}                                  & \multicolumn{1}{l|}{3,163}                               \\ \hline
\cellcolor[HTML]{C0C0C0}{\color[HTML]{3C4043} H.1.1 Updates make app worse}                           & \cellcolor[HTML]{C0C0C0}                                 & \cellcolor[HTML]{C0C0C0}                                 & H.2.2.7 Has adblocker                                                                                    & \cellcolor[HTML]{C0C0C0}                                 & \cellcolor[HTML]{C0C0C0}                                 & \multicolumn{1}{l|}{\cellcolor[HTML]{C0C0C0}H.5.1 Battery consumption}                                                     & \multicolumn{1}{l|}{\cellcolor[HTML]{C0C0C0}}            & \multicolumn{1}{l|}{\cellcolor[HTML]{C0C0C0}}            \\ \hline
\cellcolor[HTML]{C0C0C0}H.1.2 Bugs retained after updates                                             & \cellcolor[HTML]{C0C0C0}                                 & \cellcolor[HTML]{C0C0C0}                                 & H.2.2.8 No adblocking                                                                                    & \cellcolor[HTML]{C0C0C0}                                 & \cellcolor[HTML]{C0C0C0}                                 & \multicolumn{1}{l|}{H.5.1.1 Consumes battery}                                                                              & \multicolumn{1}{l|}{\cellcolor[HTML]{C0C0C0}}            & \multicolumn{1}{l|}{\cellcolor[HTML]{C0C0C0}}            \\ \hline
\cellcolor[HTML]{C0C0C0}H.1.3 Updates improving the app                                               & \cellcolor[HTML]{C0C0C0}                                 & \cellcolor[HTML]{C0C0C0}                                 & \begin{tabular}[c]{@{}l@{}}H.2.2.9 Particular VPN specific \\ feature not working\end{tabular}           & \cellcolor[HTML]{C0C0C0}                                 & \cellcolor[HTML]{C0C0C0}                                 & \multicolumn{1}{l|}{H.5.1.2 Less Battery drain}                                                                            & \multicolumn{1}{l|}{\cellcolor[HTML]{C0C0C0}}            & \multicolumn{1}{l|}{\cellcolor[HTML]{C0C0C0}}            \\ \hline
\cellcolor[HTML]{C0C0C0}H.1.4 Infrequent updates                                                      & \cellcolor[HTML]{C0C0C0}                                 & \cellcolor[HTML]{C0C0C0}                                 & H.2.2.10 Can not use private DNS                                                                         & \cellcolor[HTML]{C0C0C0}                                 & \cellcolor[HTML]{C0C0C0}                                 & \multicolumn{1}{l|}{\cellcolor[HTML]{C0C0C0}\begin{tabular}[c]{@{}l@{}}H.5.2 System resources \\ consumption\end{tabular}} & \multicolumn{1}{l|}{\cellcolor[HTML]{C0C0C0}}            & \multicolumn{1}{l|}{\cellcolor[HTML]{C0C0C0}}            \\ \hline
\cellcolor[HTML]{CBCEFB}H.2 Features                                                                  & 100                                                      & 11,726                                                   & \cellcolor[HTML]{CBCEFB}{\color[HTML]{222222} H.3 Devices}                                               & 17                                                       & 1,188                                                    & \multicolumn{1}{l|}{\begin{tabular}[c]{@{}l@{}}H.5.2.1 High system resource \\ consumption\end{tabular}}                   & \multicolumn{1}{l|}{\cellcolor[HTML]{C0C0C0}}            & \multicolumn{1}{l|}{\cellcolor[HTML]{C0C0C0}}            \\ \hline
\cellcolor[HTML]{C0C0C0}H.2.1 Number of features                                                      & \cellcolor[HTML]{C0C0C0}                                 & \cellcolor[HTML]{C0C0C0}                                 & \cellcolor[HTML]{C0C0C0}H.3.1 Routers not supported                                                      & \cellcolor[HTML]{C0C0C0}                                 & \cellcolor[HTML]{C0C0C0}                                 & \multicolumn{1}{l|}{\begin{tabular}[c]{@{}l@{}}H.5.2.2 Low system resource \\ consumption\end{tabular}}                    & \multicolumn{1}{l|}{\cellcolor[HTML]{C0C0C0}}            & \multicolumn{1}{l|}{\cellcolor[HTML]{C0C0C0}}            \\ \hline
H.2.1.1 Not many features                                                                             & \cellcolor[HTML]{C0C0C0}                                 & \cellcolor[HTML]{C0C0C0}                                 & \cellcolor[HTML]{C0C0C0}H.3.2 Works with rooted device                                                   & \cellcolor[HTML]{C0C0C0}                                 & \cellcolor[HTML]{C0C0C0}                                 & \multicolumn{1}{l|}{\cellcolor[HTML]{CBCEFB}H.6 Self sufficiency software}                                                 & \multicolumn{1}{l|}{15}                                  & \multicolumn{1}{l|}{277}                                 \\ \hline
\begin{tabular}[c]{@{}l@{}}H.2.1.2 Too many unwanted \\ features\end{tabular}                         & \cellcolor[HTML]{C0C0C0}                                 & \cellcolor[HTML]{C0C0C0}                                 & \cellcolor[HTML]{C0C0C0}H.3.3 Works well with routers                                                    & \cellcolor[HTML]{C0C0C0}                                 & \cellcolor[HTML]{C0C0C0}                                 & \multicolumn{1}{l|}{\cellcolor[HTML]{CBCEFB}H.7 Software stability}                                                        & \multicolumn{1}{l|}{19}                                  & \multicolumn{1}{l|}{2,125}                               \\ \hline
H.2.1.3 More features required                                                                        & \cellcolor[HTML]{C0C0C0}                                 & \cellcolor[HTML]{C0C0C0}                                 & \cellcolor[HTML]{C0C0C0}\begin{tabular}[c]{@{}l@{}}H.3.4 Not working with Android \\ TV\end{tabular}     & \cellcolor[HTML]{C0C0C0}                                 & \cellcolor[HTML]{C0C0C0}                                 & \multicolumn{1}{l|}{\cellcolor[HTML]{C0C0C0}H.7.1 Good Stability}                                                          & \multicolumn{1}{l|}{\cellcolor[HTML]{C0C0C0}}            & \multicolumn{1}{l|}{\cellcolor[HTML]{C0C0C0}}            \\ \hline
H.2.1.4 Good number of features                                                                       & \cellcolor[HTML]{C0C0C0}                                 & \cellcolor[HTML]{C0C0C0}                                 & \cellcolor[HTML]{C0C0C0}\begin{tabular}[c]{@{}l@{}}H.3.5 Does not works with Fire \\ TV\end{tabular}     & \cellcolor[HTML]{C0C0C0}                                 & \cellcolor[HTML]{C0C0C0}                                 & \multicolumn{1}{l|}{\cellcolor[HTML]{C0C0C0}H.7.2 App crashes}                                                             & \multicolumn{1}{l|}{\cellcolor[HTML]{C0C0C0}}            & \multicolumn{1}{l|}{\cellcolor[HTML]{C0C0C0}}            \\ \hline
\rowcolor[HTML]{C0C0C0} 
\cellcolor[HTML]{C0C0C0}\begin{tabular}[c]{@{}l@{}}H.2.2 Presence/Absense of \\ features\end{tabular} &                                                          &                                                          & \cellcolor[HTML]{C0C0C0}H.3.6 Issues with Xbox                                                           &                                                          &                                                          & \multicolumn{1}{l|}{\cellcolor[HTML]{C0C0C0}H.7.3 Unstable app}                                                            & \multicolumn{1}{l|}{\cellcolor[HTML]{C0C0C0}}            & \multicolumn{1}{l|}{\cellcolor[HTML]{C0C0C0}}            \\ \hline
H.2.2.1 Works with torrent                                                                            & \cellcolor[HTML]{C0C0C0}                                 & \cellcolor[HTML]{C0C0C0}                                 & \cellcolor[HTML]{CBCEFB}H.4 Bypassing capabilities                                                       & 65                                                       & 9,566                                                    & \multicolumn{1}{l|}{\cellcolor[HTML]{C0C0C0}H.7.4 App does not crashes}                                                    & \multicolumn{1}{l|}{\cellcolor[HTML]{C0C0C0}}            & \multicolumn{1}{l|}{\cellcolor[HTML]{C0C0C0}}            \\ \hline
H.2.2.2 Issues with torrent                                                                           & \cellcolor[HTML]{C0C0C0}                                 & \cellcolor[HTML]{C0C0C0}                                 & \cellcolor[HTML]{C0C0C0}\begin{tabular}[c]{@{}l@{}}H.4.1 Couldn't bypass blocked \\ service\end{tabular} & \cellcolor[HTML]{C0C0C0}                                 & \cellcolor[HTML]{C0C0C0}                                 &                                                                                                                            &                                                          &                                                          \\ \cline{1-6}
\begin{tabular}[c]{@{}l@{}}H.2.2.3 Works good with email \\ services\end{tabular}                     & \cellcolor[HTML]{C0C0C0}                                 & \cellcolor[HTML]{C0C0C0}                                 & \cellcolor[HTML]{C0C0C0}H.4.2 Bypasses blocked services                                                  & \cellcolor[HTML]{C0C0C0}                                 & \cellcolor[HTML]{C0C0C0}                                 &                                                                                                                            &                                                          &                                                          \\ \cline{1-6}
H.2.2.4 P2P not available                                                                             & \cellcolor[HTML]{C0C0C0}                                 & \cellcolor[HTML]{C0C0C0}                                 & \cellcolor[HTML]{C0C0C0}H.4.3 VPN blocks website                                                         & \cellcolor[HTML]{C0C0C0}                                 & \cellcolor[HTML]{C0C0C0}                                 &                                                                                                                            &                                                          &                                                          \\ \cline{1-6}
H.2.2.5 P2P available                                                                                 & \cellcolor[HTML]{C0C0C0}                                 & \cellcolor[HTML]{C0C0C0}                                 & \cellcolor[HTML]{C0C0C0}H.4.4 VPN blocked by government                                                  & \cellcolor[HTML]{C0C0C0}                                 & \cellcolor[HTML]{C0C0C0}                                 &                                                                                                                            &                                                          &                                                          \\ \cline{1-6}
\cellcolor[HTML]{C0C0C0}H.3.1 Routers not supported                                                   & \cellcolor[HTML]{C0C0C0}                                 & \cellcolor[HTML]{C0C0C0}                                 & \cellcolor[HTML]{C0C0C0}H.4.5 Detected as proxy by websites                                              & \cellcolor[HTML]{C0C0C0}                                 & \cellcolor[HTML]{C0C0C0}                                 &                                                                                                                            &                                                          &                                                          \\ \cline{1-6}
\end{tabular}
\end{table*}

\begin{table}[]
\color{\mytablecolor}
\footnotesize
\centering
\caption{The hierarchy of themes derived from open coding of reviews in the category of ``Connection''.}
\label{tab:connection-heirarchy}
\begin{tabular}{|l|}
\hline
\rowcolor[HTML]{C0C0C0} 
\multicolumn{1}{|c|}{\cellcolor[HTML]{C0C0C0}\textbf{Themes}} \\ \hline
\rowcolor[HTML]{C0C0C0} 
C.1 Connection speed                                          \\ \hline
C.1.1 Fast connection speed                                   \\ \hline
C.1.2 Slow connection speed                                   \\ \hline
\rowcolor[HTML]{C0C0C0} 
C.2 Connection stability                                      \\ \hline
\rowcolor[HTML]{C0C0C0} 
C.3 Split tunneling                                           \\ \hline
\rowcolor[HTML]{C0C0C0} 
C.4 Ease of reconnection                                      \\ \hline
\rowcolor[HTML]{C0C0C0} 
C.5 Auto-connect feature                                      \\ \hline
C.5.1 Bad auto-connect feature                                \\ \hline
C.5.2 Good auto-connect feature                               \\ \hline
\end{tabular}%
\end{table}

\section{Classification of reviews into actual switch, potential switch and irrelevant}\label{sec:review-model-appendix}
\newtext{To classify reviews into three classes (actual switch, potential switch and irrelevant), we used the following models: DeBERTa\cite{deberta}, FLAN T5\cite{flant5}, GAN BERT\cite{gan-bert}, BERT\cite{bertbase}, and LLAMA-2\cite{llama2}. Since the task is a sequence classification task and hence, we used different architectures of models described in Table \ref{tab:model-architectures} along with the specification of model sizes. The hyperparameter tuning was done using Optuna \cite{optuna}, which is a framework for optimization for hyperparameters. The performance of tuned models is shown in Table \ref{tab:val-performance}.}

\noindent\newtext{\textbf{Error Analysis of DeBERTa: }Since we achieved the best accuracy for the downstream task using the fine-tuned DeBERTa model. Class-wise performance of the DeBERTa model is shown in Table \ref{tab: val-table-deberta}. We also tried to get insights into the performance of the model using error analysis and observed that the model mostly commits mistakes between the "Potential" and "Actual" shift classes, which can be attributed to fine linguistic boundaries between the reviews in the classes.}

\begin{table}[]
\centering
{\color{\mytablecolor}
\footnotesize
\caption{Class-wise performance of DeBERTa on the validation dataset.}
\label{tab: val-table-deberta}
\vspace{1mm}
\begin{tabular}{|
>{\columncolor[HTML]{C0C0C0}}l |l|l|l|}
\hline
\textbf{}                & \cellcolor[HTML]{C0C0C0}\textbf{Precision} & \cellcolor[HTML]{C0C0C0}\textbf{Recall} & \cellcolor[HTML]{C0C0C0}\textbf{F1-Score} \\ \hline
\textbf{Actual Shift}    & 0.91                                       & 0.79                                    & 0.85                                      \\ \hline
\textbf{Irrelevant}      & 0.77                                       & 0.90                                    & 0.83                                      \\ \hline
\textbf{Potential Shift} & 0.85                                       & 0.88                                    & 0.87                                      \\ \hline
\end{tabular}
}
\end{table}

\begin{table}[!t]
\footnotesize
\centering
\caption{Specification and architecture of models used in review classification and theme identification tasks.}
\label{tab:model-architectures}
\vspace{1mm}
\resizebox{\columnwidth}{!}{%
\begin{tabular}{|
>{\columncolor[HTML]{C0C0C0}}c |
>{\columncolor[HTML]{FFFFFF}}c |
>{\columncolor[HTML]{FFFFFF}}c |
>{\columncolor[HTML]{FFFFFF}}c |}
\hline
\textbf{Model}    & \cellcolor[HTML]{C0C0C0}\textbf{Model Specification}                                                                                                                       & \cellcolor[HTML]{C0C0C0}\textbf{Review Classification} & \cellcolor[HTML]{C0C0C0}\textbf{Theme Identification}                                                   \\ \hline
\textbf{Flan T5}  & Flan-T5 base                                                                                                                                                               & Sequence Classification                                & Seq2Seq                                                                                                 \\ \hline
\textbf{LLAMA-2}  & Llama-2-7B                                                                                                                                                                 & Causal LM                                              & Causal LM                                                                                               \\ \hline
\textbf{GAN BERT} & \begin{tabular}[c]{@{}c@{}}Generator and Descriminator\\ are ANNs with 1 hidden layer \\ of size 512\\ BERT used for encoding \\ sequences is BERT-Base model\end{tabular} & Sequence Classsification                               & \begin{tabular}[c]{@{}c@{}}Sequence Classification \\ (With multiple classification heads)\end{tabular} \\ \hline
\textbf{DeBERTa}  & DeBERTa-base                                                                                                                                                               & Sequence Classsification                               & \begin{tabular}[c]{@{}c@{}}Sequence Classification \\ (With multiple classification heads)\end{tabular} \\ \hline
\textbf{BERT}     & BERT-base                                                                                                                                                                  & Sequence Classsification                               & \begin{tabular}[c]{@{}c@{}}Sequence Classification \\ (With multiple classification heads)\end{tabular} \\ \hline
\textbf{BART}     & BART-base                                                                                                                                                                  & -                                                      & Seq2Seq                                                                                                 \\ \hline
\end{tabular}%
}
\end{table}

\begin{table}[]
\footnotesize
\centering
{\color{\mytablecolor}
\caption{Performance of fine-tuned model (For theme classification) on the validation dataset. The fraction of correctly identified themes means the average number of themes correctly identified for a review and vice versa. The fraction of unidentified themes refers to the average number of themes which the model failed to identify for a review.}
\label{tab:theme-model-performance}
\vspace{1mm}
\begin{tabular}{|l|l|l|l|}
\hline
\rowcolor[HTML]{C0C0C0} 
\textbf{Model} & \textbf{Correct Themes} & \textbf{Incorrect Themes} & \textbf{Missing Themes} \\ \hline
Flan T5        & 0.79                    & 0.17                       & 0.21                    \\ \hline
DeBERTa        & 0.76                    & 0.25                       & 0.24                    \\ \hline
LLAMA 2        & 0.61                    & 0.41                       & 0.39                    \\ \hline
\textbf{BART}  & \textbf{0.81}           & \textbf{0.21}              & \textbf{0.19}           \\ \hline
Bert           & 0.75                    & 0.12                       & 0.25                    \\ \hline
GAN Bert       & 0.74                    & 0.24                       & 0.26                    \\ \hline
\end{tabular}
}
\end{table}

\begin{table}[]
\centering
\footnotesize
\caption{Performance of fine-tuned BART model (For theme classification) on validation dataset and additional random sample (with manual ground truth labels). The fraction of correctly identified themes means the average number of themes correctly identified for a review and vice versa. The fraction of unidentified themes refers to the average number of themes which the model failed to identify for a review.}
\label{tab:classModelTheme}
\vspace{1mm}
\begin{tabular}{|l|l|l|}
\hline
                                                                                  & \textbf{\begin{tabular}[c]{@{}l@{}}\% reviews in \\ validation dataset\end{tabular}} & \textbf{\begin{tabular}[c]{@{}l@{}}\% Reviews in additional \\ random sample\end{tabular}} \\ \hline
\textbf{\begin{tabular}[c]{@{}l@{}}Correctly identified \\ themes\end{tabular}}   & 81\%                                                                         & 77\%                                                                            \\ \hline
\textbf{\begin{tabular}[c]{@{}l@{}}Incorrectly identified \\ themes\end{tabular}} & 21\%                                                                         & 12\%                                                                             \\ \hline
\textbf{\begin{tabular}[c]{@{}l@{}}Unidentified \\ themes\end{tabular}}           & 19\%                                                                         & 23\%                                                                            \\ \hline
\end{tabular}
\end{table}

\section{Theme identification model}\label{sec:theme-identification-appendix}
\newtext{For the identification of themes in reviews, we used the following models: DeBERTa\cite{deberta}, FLAN T5\cite{flant5}, GAN BERT\cite{gan-bert}, BERT\cite{bertbase}, BART\cite{bart}, and LLAMA-2\cite{llama2}. Since the task can be modelled as both sequence multi-label classification or sequence generation task, hence, we used different architectures of models described in Table \ref{tab:model-architectures} along with the specification of model sizes. The hyperparameter tuning was done using Optuna \cite{optuna}, which is a framework for optimization for hyperparameters. }

\noindent\newtext{\textbf{Accuracy of models: }We found that BART model outperformed other models into consideration, identifying more than 80\% of themes correctly in the reviews as shown in Table~\ref{tab:theme-model-performance} which reports the accuracy of the models on validation set. The result of the accuracy of BART model against additional manual ground truth is reported in Table~\ref{tab:classModelTheme}.}

\noindent\newtext{\textbf{Error Analysis of BART: }Since the fine-tuned BART model predicted the highest fraction of themes correctly, we did further analysis to look into the performance of the model. Class-wise performance of the BART model is shown in Table \ref{tab:bart-classwise}. We also tried to get insights into the performance of the model using error analysis and observed that the model mostly commits mistakes in predicting less frequently occurring classes like Reputation/Review. We also noted that the model committed mistakes in understanding the contextual meaning of phrases and made predictions using the phrases themselves, which we found a bit amusing as LLMs are known for understanding the context and then making decisions.}

\noindent\newtext{\textbf{Error Analysis of LLAMA-2: }In both the tasks, viz Theme Identification and Review classification, we noted a poor performance by LLAMA-2 despite, it being the biggest model in our basket. While evaluating the fine-tuned LLAMA-2 model, we noted model hallucinated a lot\cite{hallucination}, and hence the wrong predictions. We experimented with multiple prompts and hyperparameter tuning, but we couldn't achieve performance up to the mark of other models.}

\begin{table*}[!b]
\footnotesize
\centering
\caption{List of top 10 named entities occurring in individual websites. Names shown in bold refer to different VPNs.}
\label{tab:ner}
\vspace{1mm}
\begin{tabular}{|c|c|c|c|c|c|}
\hline
\rowcolor[HTML]{E7E7E7} 
\textbf{\begin{tabular}[c]{@{}c@{}}Website\\ A\end{tabular}} & \textbf{\begin{tabular}[c]{@{}c@{}}Website\\ B\end{tabular}} & \textbf{\begin{tabular}[c]{@{}c@{}}Website\\ C\end{tabular}} & \textbf{\begin{tabular}[c]{@{}c@{}}Website\\ D\end{tabular}} & \textbf{\begin{tabular}[c]{@{}c@{}}Website\\ E\end{tabular}} & \textbf{\begin{tabular}[c]{@{}c@{}}Website\\ F\end{tabular}}   \\ \hline
Netflix                                                      & Netflix                                                      & Netflix                                                      & Netflix                                                      & \textbf{Surfshark}                                           & \textbf{Surfshark}                                             \\ \hline
DNS                                                          & \textbf{Surfshark}                                           & China                                                        & BBC                                                          & Netflix                                                      & Android                                                        \\ \hline
Russia                                                       & DNS                                                          & DNS                                                          & DNS                                                          & DNS                                                          & DNS                                                            \\ \hline
\textbf{Proton}                                              & BBC                                                          & Android                                                      & MBPS                                                         & MBPS                                                         & India                                                          \\ \hline
\textbf{Norton}                                              & \textbf{Nord}                                                & \textbf{Surfshark}                                           & \textbf{Surfshark}                                           & \textbf{Proton}                                              & Netflix                                                        \\ \hline
China                                                        & Android                                                      & \textbf{Norton}                                              & Android                                                      & China                                                        & Taiwan                                                         \\ \hline
\textbf{Surfshark}                                           & Russia                                                       & Amazon                                                       & ITV                                                          & \textbf{Nord}                                                & \textbf{Nord}                                                  \\ \hline
Android                                                      & MacOS                                                        & \textbf{Mullvad}                                             & Hong Kong                                                    & Russia                                                       & \textbf{\begin{tabular}[c]{@{}c@{}}Google  One\end{tabular}} \\ \hline
\textbf{Mullvad}                                             & \textbf{NordVPN}                                             & Russia                                                       & China                                                        & Mac                                                          & MacOS                                                          \\ \hline
\begin{tabular}[c]{@{}c@{}}MacOS\\ Windows\end{tabular}      & Disney                                                       & Germany                                                      & India                                                        & Germany                                                      & Russia                                                         \\ \hline
\end{tabular}
\end{table*}

\begin{table*}[!t]
\centering
\footnotesize
{\color{\mytablecolor}
\caption{Class-wise performance of BART model on theme identification task on validation dataset}
\label{tab:bart-classwise}
\vspace{1mm}
\begin{tabular}{|c|c|c|c|c|c|}
\hline
\rowcolor[HTML]{C0C0C0} 
\textbf{Theme}       & \textbf{Accuracy} & \textbf{F1-Score} & \textbf{Theme}       & \textbf{Accuracy} & \textbf{F1-Score} \\ \hline
Bypassing capability & 0.95              & 0.73              & Ads                  & 1                 & 1                 \\ \hline
Network speed        & 0.88              & 0.86              & Price                & 0.93              & 0.88              \\ \hline
Premium              & 0.99              & 0.89              & UI/UX                & 0.93              & 0.85              \\ \hline
Free version         & 0.97              & 0.69              & Server count         & 0.95              & 0.75              \\ \hline
Deceit               & 0.96              & 0.69              & Assistance provided  & 0.93              & 0.82              \\ \hline
Security             & 0.89              & 0.8               & Refund               & 0.97              & 0.69              \\ \hline
Service provided     & 0.9               & 0.74              & User friendliness    & 0.99              & 0.94              \\ \hline
Payment              & 0.98              & 0.74              & Reliability          & 0.99              & 0.94              \\ \hline
Reputation/Review    & 0.99              & \textbf{0.49}     & Features             & 0.94              & \textbf{0.68}     \\ \hline
Updates              & 1                 & 1                 & Streaming            & 0.98              & 0.75              \\ \hline
Geography            & 0.97              & 0.89              & Games                & 0.99              & 0.79              \\ \hline
Connection           & 0.9               & 0.81              & Resource consumption & 0.98              & 0.74              \\ \hline
OTT                  & 0.94              & 0.68              & Devices              & 0.99              & 0.83              \\ \hline
Privacy              & 0.97              & \textbf{0.69}     & Network  Protocol    & 0.99              & 0.85              \\ \hline
Not Self Sufficient  & 0.98              & 0.91              & Network Protocol     & 0.99              & 0.89              \\ \hline
\end{tabular}
}
\end{table*}

\section{Desired features of VPNs as indicated by VPN-switching reviews}\label{sec:desired-features-appendix}

The top 10 most-frequent features (themes) are in Table~\ref{tab:most-frequent-indiv-feature}. Also, VPN-specific features refers to functionalities like kill switch, always on VPN, split tunnelling, etc. One interesting observation from Table~\ref{tab:most-frequent-indiv-feature} is that even though VPNs were introduced as Privacy Enhancing Techniques, privacy is not among our corpus's top 10 most frequently mentioned features.

Table~\ref{tab:comms} further presents the exact set of 10 communities of co-occurred features in VPN-switching reviews. Notably, in these communities, popular features appear with non-popular ones and provide a roadmap for catering to specific user bases with customized VPN solutions.

\begin{table*}[]
\footnotesize
\centering
{\color{\mytablecolor}
\caption{Communities of desired features for VPN apps based on co-occurance in same review. Community names marked in bold (C1, C2, C3, C5, C6, C8 and C10) either have ‘Security’ or ‘Privacy’ as member features.}
\label{tab:comms}
\vspace{1mm}
\begin{tabular}{|c|p{12cm}|}
\hline
\rowcolor[HTML]{E7E7E7} 
\textbf{Community} & \multicolumn{1}{c|}{\cellcolor[HTML]{E7E7E7}\textbf{Set of themes included in the community}}                                                                                                                                    \\ \hline
\rowcolor[HTML]{FFFFFF} 
\textbf{C1}        & \begin{tabular}[c]{@{}l@{}}Connection, UI/UX, Network speed, Security, Resource consumption, Geography, Service provided,\\ User-friendliness, Not Self Sufficient, Reliability\end{tabular} \\ \hline
\rowcolor[HTML]{FFFFFF} 
\textbf{C2}                 & Network Traffic, Ads, OTT, Privacy                                                                                                                                                                     \\ \hline
\rowcolor[HTML]{FFFFFF} 
\textbf{C3}        & \begin{tabular}[c]{@{}l@{}}Updates, Payment, OTT, Refund, Features, Deceit, Bypassing capability, Privacy, Assistance provided  \\ Reputation/Review, Split tunnelling\end{tabular}                                    \\ \hline
\rowcolor[HTML]{FFFFFF} 
C4                 & Devices, Games, Free version, OTT                                                                                                                                                                                           \\ \hline
\rowcolor[HTML]{FFFFFF} 
\textbf{C5}        & Premium, Streaming, Network Speed, Free version, Protocol, Security                                                                                                                                                  \\ \hline
\rowcolor[HTML]{FFFFFF} 
\textbf{C6}        & Security, Privacy, Switching server, Geography, Premium                                                                                                                                                                     \\ \hline
\rowcolor[HTML]{FFFFFF} 
C7                 & Connection, Network speed, Premium, Free version, Protocol                                                                                                                                                             \\ \hline
\rowcolor[HTML]{FFFFFF} 
\textbf{C8}        & Protocol, Bypassing capability, Network, Updates, Privacy                                                                                                                                                                     \\ \hline
\rowcolor[HTML]{FFFFFF} 
C9                 & \begin{tabular}[c]{@{}l@{}}Bypassing capability, Devices, OTT, Reputation/Review, Server count, Service provided, Switching server\end{tabular}                                                                     \\ \hline
\rowcolor[HTML]{FFFFFF} 
\textbf{C10}       & \begin{tabular}[c]{@{}l@{}}Refund, Reputation/Review, Privacy, Service provided, Split tunnelling, UI/UX, Updates, User-friendliness\end{tabular}                                                                  \\ \hline
\end{tabular}
}
\end{table*}

\begin{table}[]
\scriptsize
\centering
{\color{\mytablecolor}
\caption{Most frequently occurring themes in the reviews}
\label{tab:most-frequent-indiv-feature}
\vspace{1mm}
\begin{tabular}{|l|c|}
\hline
\rowcolor[HTML]{E7E7E7} 
\multicolumn{1}{|c|}{\cellcolor[HTML]{E7E7E7}Theme} & Count of reviews \\ \hline
Connection                                       & 41,020           \\ \hline
Network Speed                                               & 28,406           \\ \hline
User-friendliness                                    & 22,402           \\ \hline
Service provided                                               & 21,689           \\ \hline
UI/UX                                            & 19,406           \\ \hline
Price                                  & 15,862           \\ \hline
Security                                                 & \textbf{15,564}           \\ \hline
Assistance provided                                           & 15,258           \\ \hline
Geography                             & 12,487           \\ \hline
(VPN-specific) Features                                             & 11,726           \\ \hline
\end{tabular}
}
\end{table}

\section{Terms used in Content Analysis for reference}\label{sec:refernce-terms-in-content}

The terms which have been used to study bias in the content of blogs collected from VPN recommendation websites have been presented in Table \ref{tab:ref-terms}. We have used general terms like "Best" and "Worst", which are not related to the identified themes, along with terms related to themes like "Logs" and "Cheapest". To further study that if VPN recommendation websites are only doing positive publicity of selected VPN apps or negative also, we have used negative terms like "Bloated", "Slow", and "Costly".

\begin{table}[]
\scriptsize
\centering
\caption{Keywords used in the word-association analysis to understand bias towards specific VPNs and the coverage of themes in VPN blogs.}
\label{tab:ref-terms}
\vspace{1mm}
\begin{tabular}{|c|ll|}
\hline
\rowcolor[HTML]{E7E7E7} 
\cellcolor[HTML]{E7E7E7}                                                                                                                   & \multicolumn{2}{c|}{\cellcolor[HTML]{E7E7E7}\textbf{\begin{tabular}[c]{@{}c@{}}Related to themes\\ (Second set of keywords)\end{tabular}}}                                                                                                                                                                              \\ \cline{2-3} 
\rowcolor[HTML]{E7E7E7} 
\multirow{-2}{*}{\cellcolor[HTML]{E7E7E7}\textbf{\begin{tabular}[c]{@{}c@{}}Not related to themes\\ (First set of keywords)\end{tabular}}} & \multicolumn{1}{l|}{\cellcolor[HTML]{E7E7E7}\textbf{Negative terms}}                                                                                             & \textbf{Positive terms}                                                                                                                              \\ \hline
\multicolumn{1}{|l|}{\begin{tabular}[c]{@{}l@{}}vpn, best, worst, china, \\ speed, useless, \\ useful, ad\end{tabular}}                    & \multicolumn{1}{l|}{\begin{tabular}[c]{@{}l@{}}costly, data leak, slow, \\ slowest, complicated \\ interface,\\ bad gui, less features, \\ bloated\end{tabular}} & \begin{tabular}[c]{@{}l@{}}cheapest, secured, \\ fastest, logs, \\ simple gui, \\ most  secure, free\\ worth it, price,\\ many features\end{tabular} \\ \hline
\end{tabular}
\end{table}

\section{Additional detailed results of our VPN blog  analysis}\label{sec:blog_appendix}

\noindent \textbf{Description of identified topics in VPN blogs:} We have represented topics identified by applying Topic Modelling on the complete corpus along with the manually assigned title in Table \ref{tab:topics-complete-corpus}. Table~\ref{tab:topics-review} enlists all the topics identified by applying topic modelling on "Review" blogs, and Table~\ref{tab:topics-best-pick} enlists all the topics identified by applying topic modelling on "Best-Pick" blogs. 

\vspace{1mm}

\noindent \textbf{Frequency of top named entities in VPN blogs:} Table~\ref{tab:ner} provides the top 10 named entities occurring in individual websites. The Names shown in bold refer to different VPNs. Clearly, across websites, a few VPN apps are most frequently mentioned. 

\vspace{1mm}

\noindent \textbf{Keywords used for our word-association analysis:} Table~\ref{tab:ref-terms} provides two sets of keywords we used for our word-association analysis. The first set of keywords is not related to themes. The second set of keywords is related to different themes identified during VPN-switching. 

\vspace{1mm}

\noindent \newtext{\textbf{Themes identified in blogs corpus by BART model:} Table~\ref{tab:themes-blogs} enlists top-5 themes identified by our fine-tuned BART model for each category in blogs corpus. As discussed in Section~\ref{sec:blogs}, we can see while some important themes are consistently covered by these blogging websites, there is a lack of coverage of other important issues like Connection or User-Friendliness, which are the most frequently appearing themes in our review corpus.}

\begin{table}[]
\footnotesize
\centering
{\color{\mytablecolor}
\caption{Top themes (by frequency) identified in the blogs corpus}
\label{tab:themes-blogs}
\vspace{1mm}
\begin{tabular}{|c|c|c|}
\hline
\rowcolor[HTML]{C0C0C0} 
\textbf{Overall Dataset} & \textbf{VPN Best Pick Blogs} & \multicolumn{1}{l|}{\cellcolor[HTML]{C0C0C0}\textbf{VPN Review Blogs}} \\ \hline
Features                 & Network Speed                & Features                                                               \\ \hline
Security                 & Price                        & Network Speed                                                          \\ \hline
Price                    & Security                     & Streaming                                                              \\ \hline
Network Speed            & Features                     & Price                                                                  \\ \hline
Streaming                & Streaming                    & Security                                                               \\ \hline
\end{tabular}%
}
\end{table}

\begin{table}[]
\footnotesize
\centering
\caption{Topics identified by applying "Topic Modelling" to the complete corpus and title assigned to the topics using manual analysis. Topics marked with "-" seemed irrelevant in reference to our work.}
\label{tab:topics-complete-corpus}
\vspace{1mm}
\begin{tabular}{|c|c|}
\hline
\rowcolor[HTML]{E7E7E7} 
\multicolumn{1}{|l|}{\cellcolor[HTML]{E7E7E7}\textbf{Terms occurring in the topic}}                 & \textbf{\begin{tabular}[c]{@{}c@{}}Title Assigned\\ to the Topic\end{tabular}}        \\ \hline
\begin{tabular}[c]{@{}c@{}}VPN, BBC, ITV, Streaming, TV,\\ Hulu, Prime\end{tabular}                 & \begin{tabular}[c]{@{}c@{}}Related\\ to OTT\end{tabular}                              \\ \hline
\begin{tabular}[c]{@{}c@{}}VPN, Privacy, DNS, Logging, \\ Address, Information, IP\end{tabular}     & \cellcolor[HTML]{FFFFFF}\begin{tabular}[c]{@{}c@{}}Privacy\\ Related\end{tabular}     \\ \hline
\begin{tabular}[c]{@{}c@{}}Tests, Support, Open, Times, \\ Recording, Technical, Setup\end{tabular} & \cellcolor[HTML]{FFFFFF}-                                                             \\ \hline
\begin{tabular}[c]{@{}c@{}}Help, Expert, Setup, Everyday\\ Standard, Basis\end{tabular}             & \cellcolor[HTML]{FFFFFF}-                                                             \\ \hline
\begin{tabular}[c]{@{}c@{}}Security, Privacy, Avast, Read,\\ Response, Year, Right\end{tabular}     & \cellcolor[HTML]{FFFFFF}\begin{tabular}[c]{@{}c@{}}Security\\ \& Privacy\end{tabular} \\ \hline
\begin{tabular}[c]{@{}c@{}}Free, Access, Internet, Traffic,\\ Also, Data\end{tabular}               & \cellcolor[HTML]{FFFFFF}-                                                             \\ \hline
\begin{tabular}[c]{@{}c@{}}Netflix, Streaming, Torrenting, \\ Express, Best, Access\end{tabular}    & \cellcolor[HTML]{FFFFFF}\begin{tabular}[c]{@{}c@{}}Related to\\ OTT\end{tabular}      \\ \hline
\begin{tabular}[c]{@{}c@{}}Free, Unblock, Support, One,\\ Month, Year\end{tabular}                  & \cellcolor[HTML]{FFFFFF}-                                                             \\ \hline
\begin{tabular}[c]{@{}c@{}}VPN, Ranked, Open, Torrent,\\ Windows, VPNs\end{tabular}                 & \cellcolor[HTML]{FFFFFF}-                                                             \\ \hline
\begin{tabular}[c]{@{}c@{}}VPN, Servers, MBPS, Server, \\ Countries, Also\end{tabular}              & \cellcolor[HTML]{FFFFFF}-                                                             \\ \hline
\end{tabular}

\end{table}

\begin{table}[]
\footnotesize
\centering
\caption{Topics identified by applying "Topic Modelling" to the review blogs and titles assigned to the topics using manual analysis. Topics marked with "-" seemed irrelevant in reference to our work.}
\label{tab:topics-review}
\vspace{1mm}
\begin{tabular}{|c|c|}
\hline
\rowcolor[HTML]{E7E7E7} 
\multicolumn{1}{|l|}{\cellcolor[HTML]{E7E7E7}\textbf{Terms occurring in the topic}}               & \textbf{\begin{tabular}[c]{@{}c@{}}Title Assigned\\ to the Topic\end{tabular}}        \\ \hline
\begin{tabular}[c]{@{}c@{}}VPN, Free, IP, Traffic, Encryption\\ Switch, China, Data\end{tabular}  & -                                                                                     \\ \hline
\begin{tabular}[c]{@{}c@{}}Streaming, US, Netflix, BBC, ITV,\\ Test, UK, Video\end{tabular}       & \cellcolor[HTML]{FFFFFF}\begin{tabular}[c]{@{}c@{}}Related to\\ OTT\end{tabular}      \\ \hline
\begin{tabular}[c]{@{}c@{}}Server, Speed, Limited, Options,\\ Tests, Also\end{tabular}            & \cellcolor[HTML]{FFFFFF}-                                                             \\ \hline
\begin{tabular}[c]{@{}c@{}}Security, VPN, Safe, Overall, \\ Privacy, Server\end{tabular}          & \cellcolor[HTML]{FFFFFF}\begin{tabular}[c]{@{}c@{}}Related to\\ Security\end{tabular} \\ \hline
\begin{tabular}[c]{@{}c@{}}Free, Server, Country, Astrill, Links, \\ Tested, However\end{tabular} & \cellcolor[HTML]{FFFFFF}-                                                             \\ \hline
\begin{tabular}[c]{@{}c@{}}Advice, Recorded, Live, Recorded,\\ Table, Technical\end{tabular}      & \cellcolor[HTML]{FFFFFF}-                                                             \\ \hline
\begin{tabular}[c]{@{}c@{}}Android, IOS, Windows, Devices,\\ Switch, App, Apps\end{tabular}       & \cellcolor[HTML]{FFFFFF}\begin{tabular}[c]{@{}c@{}}Related to\\ Devices\end{tabular}  \\ \hline
\begin{tabular}[c]{@{}c@{}}Privacy, Policy, Logging, Testing,\\ VPNs, May, Recommend\end{tabular} & \cellcolor[HTML]{FFFFFF}\begin{tabular}[c]{@{}c@{}}Related to\\ Privacy\end{tabular}  \\ \hline
\begin{tabular}[c]{@{}c@{}}VPN, VPNs, HBO, Prime, Netflix,\\ ITV\end{tabular}                     & \cellcolor[HTML]{FFFFFF}\begin{tabular}[c]{@{}c@{}}Related to \\ OTT\end{tabular}     \\ \hline
\begin{tabular}[c]{@{}c@{}}VPN, Server, Speed, Countries,\\ Download, Also, Network\end{tabular}  & \cellcolor[HTML]{FFFFFF}-                                                             \\ \hline
\end{tabular}

\end{table}

\begin{table}[]
\footnotesize
\centering
\caption{Topics identified by applying "Topic Modelling" to the best-pick blogs and titles assigned to the topics using manual analysis. Topics marked with "-" seemed irrelevant in reference to our work.}
\label{tab:topics-best-pick}
\vspace{1mm}
\begin{tabular}{|c|c|}
\hline
\rowcolor[HTML]{E7E7E7} 
\multicolumn{1}{|l|}{\cellcolor[HTML]{E7E7E7}\textbf{Terms occurring in the topic}}              & \textbf{\begin{tabular}[c]{@{}c@{}}Title Assigned\\ to the Topic\end{tabular}}        \\ \hline
\begin{tabular}[c]{@{}c@{}}VPN, Service, Products, Free,\\ Also, Content, Best, Get\end{tabular} & -                                                                                     \\ \hline
\begin{tabular}[c]{@{}c@{}}Free, VPNs, Money, Gurantee,\\ Back, Also\end{tabular}                & \cellcolor[HTML]{FFFFFF}-                                                             \\ \hline
\begin{tabular}[c]{@{}c@{}}VPN, Best, NordVPN, Surfshark,\\ ExpressVPN, Well, Tab\end{tabular}   & \cellcolor[HTML]{FFFFFF}-                                                             \\ \hline
\begin{tabular}[c]{@{}c@{}}Free, New, Opens, Tab, Month,\\ Best, Get, Services\end{tabular}      & \cellcolor[HTML]{FFFFFF}-                                                             \\ \hline
\begin{tabular}[c]{@{}c@{}}Streaming, Netflix, Amazon, \\ IPlayer, ITV, BBC\end{tabular}         & \cellcolor[HTML]{FFFFFF}\begin{tabular}[c]{@{}c@{}}Related to\\ OTT\end{tabular}      \\ \hline
\begin{tabular}[c]{@{}c@{}}Netflix, Content, Meets, Access,\\ Prime\end{tabular}                 & \cellcolor[HTML]{FFFFFF}\begin{tabular}[c]{@{}c@{}}Related to\\ OTT\end{tabular}      \\ \hline
\begin{tabular}[c]{@{}c@{}}Speeds, Connection, MBPS,\\ Servers, Using, VPNs\end{tabular}         & \cellcolor[HTML]{FFFFFF}\begin{tabular}[c]{@{}c@{}}Related to\\ Speed\end{tabular}    \\ \hline
\begin{tabular}[c]{@{}c@{}}Security, Logging, Features,\\ Android, MacOS\end{tabular}            & \cellcolor[HTML]{FFFFFF}\begin{tabular}[c]{@{}c@{}}Related to\\ Security\end{tabular} \\ \hline
\begin{tabular}[c]{@{}c@{}}Reviews, Independent, IP, May,\\ Site, Data, Ensure\end{tabular}      & \cellcolor[HTML]{FFFFFF}-                                                             \\ \hline
\begin{tabular}[c]{@{}c@{}}Affect, Guide, Tom, Future,\\ Writes, Team\end{tabular}               & \cellcolor[HTML]{FFFFFF}-                                                             \\ \hline
\end{tabular}

\end{table}

\end{document}